\documentclass[12pt]{article}
\pdfoutput=1

\usepackage{booktabs}
\usepackage{tabulary}
\usepackage{multirow}
\usepackage{rotating}
\usepackage{epsfig}
\usepackage{psfrag}
\usepackage{latexsym}
\usepackage{indentfirst}
\usepackage{fancyhdr}
\usepackage{dsfont}
\usepackage{amsmath}
\usepackage{amssymb}
\usepackage{amsfonts}
\usepackage{mathrsfs}
\usepackage{amsthm}
\usepackage{pifont}
\usepackage{dsfont}
\usepackage{multirow}
\usepackage{array}
\usepackage{chngpage}
\usepackage{longtable}
\usepackage{cite}
\usepackage{bbold}
\usepackage{color}
\usepackage{braket}
\usepackage{colordvi}
\usepackage{makecell}
\usepackage{fancybox}
\usepackage[footnotesize]{caption2}
\usepackage{graphicx}
\usepackage[center,footnotesize,hang]{subfigure}
\usepackage{bbm}
\usepackage{url}
\usepackage[colorlinks, linkcolor=red, anchorcolor=black, citecolor=green]{hyperref}
\usepackage{multirow}
\usepackage[integrals]{wasysym}
\usepackage{harpoon}
\usepackage{textcomp}  %  for the command \textlbrackdbl
\usepackage{enumitem}
\usepackage{adjustbox}
\usepackage{mathrsfs}  % $\mathscr{ABCDEFGHIJKLMNOPQRSTUVWXYZ}$
\newcommand{\PreserveBackslash}[1]{\let\temp=\\#1\let\\=\temp}
\newcolumntype{C}[1]{>{\PreserveBackslash\centering}p{#1}}
\newcolumntype{R}[1]{>{\PreserveBackslash\raggedleft}p{#1}}
\newcolumntype{L}[1]{>{\PreserveBackslash\raggedright}p{#1}}
\allowdisplaybreaks  % automated pagination

\newcommand{\bq}{\begin{eqnarray}}
\newcommand{\nq}{\end{eqnarray}}

\newcommand{\cleqn}{\setcounter{equation}{0}}

\begin{document}
\title{
\begin{flushright}
\hfill\mbox{{\small\tt USTC-ICTS/PCFT-20-22}}\\[5mm]
\begin{minipage}{0.2\linewidth}
\normalsize
\end{minipage}
\end{flushright}
{\Large \bf Half-integral weight modular forms and application to neutrino mass models  \\[2mm]} }
\date{}

\author{
Xiang-Gan~Liu$^{1,2}$\footnote{E-mail: {\tt
hepliuxg@mail.ustc.edu.cn}}, \
Chang-Yuan~Yao$^{3}$\footnote{E-mail: {\tt
yaocy@nankai.edu.cn}},  \
Bu-Yao~Qu$^{1,2}$\footnote{E-mail: {\tt
qubuyao@mail.ustc.edu.cn}},  \
Gui-Jun~Ding$^{1,2}$\footnote{E-mail: {\tt
dinggj@ustc.edu.cn}}
\\*[20pt]
\centerline{
\begin{minipage}{\linewidth}
\begin{center}
$^1${\it\small Peng Huanwu Center for Fundamental Theory, Hefei, Anhui 230026, China} \\[2mm]
$^2${\it \small
Interdisciplinary Center for Theoretical Study and  Department of Modern Physics,\\
University of Science and Technology of China, Hefei, Anhui 230026, China}\\[2mm]
$^3${\it \small
School of Physics, Nankai University, Tianjin 300071, China}\\
\end{center}
\end{minipage}}
\\[10mm]}
\maketitle
\thispagestyle{empty}

\begin{abstract}

We generalize the modular invariance approach to include the half-integral weight modular forms. Accordingly the modular group should be extended to its metaplectic covering group for consistency. We introduce the well-defined half-integral weight modular forms for congruence subgroup $\Gamma(4N)$ and show that they can be decomposed into the irreducible multiplets of finite metaplectic group $\widetilde{\Gamma}_{4N}$.
We construct concrete expressions of the half-integral/integral modular forms for $\Gamma(4)$ up to weight 6 and arrange them into the irreducible representations of $\widetilde{\Gamma}_4$. We present three typical models with $\widetilde{\Gamma}_4$ modular symmetry for neutrino masses and mixing, and the phenomenological predictions of each model are analyzed numerically.

\end{abstract}
\newpage

%%%%%%%%%%%%%%%%%%%%%%%%%%%%%%%%%%%%%%%%%%%%%
\section{\label{sec:introduction}Introduction}

How to understand the mass hierarchies and flavor mixing patterns of quark and lepton is still one of the greatest challenges in particle physics. The origin of the large mass hierarchies among quark and charged lepton, the tiny but non-zero neutrino masses, and observed drastically different patterns of quark and lepton flavor mixing can not be explained by the Standard Model (SM). In many scenarios beyond the SM, flavor symmetry is still a very interesting and promising approach to solve these mysteries. Especially in recent years, the attempt to explain the large mixing angles in lepton sector with some discrete non-abelian flavor groups has made good progress. However,
generally a large number of scalar fields so-called flavons transforming nontrivially under discrete flavor symmetry are necessary to spontaneously break the flavor symmetry group. Moreover, auxiliary symmetries such as the product of cyclic groups are generally introduced to forbid the dangerous terms and to achieve the desired vacuum alignment in the neutrino and charged lepton sectors. In short, the flavor symmetry breaking sector has to be intelligently designed and the structure is complex in traditional discrete flavor symmetry models.

Recently modular invariance has been suggested as the origin of flavor symmetry~\cite{Feruglio:2017spp}. Notice that modular symmetry naturally appears in torus and orbifold compactifications of string theory. Some recent related work about the modular symmetry on $T^2$ and $T^2 \times T^2$ with magnetic fluxes can been seen in~\cite{Kikuchi:2020frp,Kikuchi:2020nxn}, where zero-modes wavefunctions behave as weight $1/2$ and $1$ modular forms. The modular invariance as flavor symmetry is a bottom-up approach~\cite{Feruglio:2017spp}, flavons are not absolutely necessary, and flavor symmetry can be uniquely broken by the vacuum expectation value of complex modulus $\tau$. Therefore the above mentioned issue of vacuum alignment is drastically simplified although a moduli stabilization mechanism is needed. In this approach, the Yukawa couplings are modular forms which are holomorphic functions of modulus $\tau$, and the superpotential is completely determined by modular invariance in the limit of supersymmetry while the K\"ahler potential is not fixed by modular symmetry~\cite{Chen:2019ewa}. In a top-down approach  motivated from string theory~\cite{Nilles:2020nnc,Nilles:2020kgo,Nilles:2020tdp}, the modular and traditional flavor symmetries are combined to form the eclectic flavor groups. The K\"ahler potential as well as the representation and weight assignment for the matter fields are severely restricted in this scheme although the order of eclectic flavor group is larger.

The finite modular group $\Gamma_2 \cong S_3$\cite{Kobayashi:2018vbk,Kobayashi:2018wkl,Kobayashi:2019rzp,Okada:2019xqk}, $\Gamma_3\cong A_4$~\cite{Feruglio:2017spp,Criado:2018thu,Kobayashi:2018vbk,Kobayashi:2018scp,deAnda:2018ecu,Okada:2018yrn,Kobayashi:2018wkl,Novichkov:2018yse,Nomura:2019jxj,Okada:2019uoy,Nomura:2019yft,Ding:2019zxk,Okada:2019mjf,Nomura:2019lnr,Kobayashi:2019xvz,Asaka:2019vev,Gui-JunDing:2019wap,Zhang:2019ngf,Nomura:2019xsb,Wang:2019xbo,Kobayashi:2019gtp,King:2020qaj,Okada:2020rjb,Nomura:2020opk}, $\Gamma_4\cong S_4$~\cite{Penedo:2018nmg,Novichkov:2018ovf,deMedeirosVarzielas:2019cyj,Kobayashi:2019mna,King:2019vhv,Criado:2019tzk,Wang:2019ovr,Gui-JunDing:2019wap,Wang:2020dbp}, $\Gamma_5\cong A_5$~\cite{Novichkov:2018nkm,Ding:2019xna,Criado:2019tzk} and $\Gamma_7\cong PSL(2,\mathbb{Z}_7)$~\cite{Ding:2020msi} have been considered. The quark masses and CKM parameters together with the lepton masses and mixing can be explained by using modular symmetry~\cite{Lu:2019vgm,Okada:2018yrn,Okada:2019uoy,Okada:2020rjb}.
Modular symmetry has been also discussed in $SU(5)$ grand unification theory~\cite{deAnda:2018ecu,Kobayashi:2019rzp}. Notably, the dynamics of modular symmetry could be tested at present and future neutrino oscillation experiments~\cite{Ding:2020yen}. The modular symmetry can be consistently combined with the generalized CP symmetry~\cite{Novichkov:2019sqv,Baur:2019kwi,Acharya:1995ag,Dent:2001cc,Giedt:2002ns}.
Multiple modular symmetries with direct product has been proposed~\cite{deMedeirosVarzielas:2019cyj,King:2019vhv}. A comprehensive discussion about flavor symmetry, CP symmetry and modular invariance in string theory was recently given in~\cite{Baur:2019kwi,Baur:2019iai}. The modular invariance approach is generalized to include the odd weight modular forms which can be arranged into irreducible representations of the homogeneous finite modular group $\Gamma'_{N}$~\cite{Liu:2019khw}. $\Gamma'_N$ is the double covering of the inhomogeneous finite modular group $\Gamma_N$. A simultaneous description of quark and lepton sectors can be achieved in the modular symmetries $\Gamma'_3\cong T'$~\cite{Liu:2019khw,Lu:2019vgm} and $\Gamma'_4\cong S'_4$~\cite{Liu:2020akv,Novichkov:2020eep}. It is notable that quite predictive flavor models can be constructed with $S'_4$~\cite{Liu:2020akv}.
The modular symmetry has the merits of both abelian flavor symmetry and discrete non-abelian flavor symmetry, it can naturally generates texture zeros in fermion mass matrix~\cite{Lu:2019vgm} after including odd weight modular forms, and the modular weight can play the role of Froggatt-Nielsen charge to generate the fermion mass hierarchies~\cite{King:2020qaj}.

In this work, we further extend the modular invariance approach to include half-integral weight modular forms. In order to consistently discuss the action of the modular transformations of the half-integral weight modular forms, one should consider the metaplectic covering of the classical modular group $\text{SL}_2(\mathbb{Z})$. Accordingly the framework of modular invariance is extended to the metaplectic modular invariance. The more general modular forms of rational weights can be studied in a similar way. It is known that the half-integral weight modular forms can be defined for the principal congruence subgroup $\Gamma(4N)$.
We find that the half-integral weight modular forms for $\Gamma(4N)$ can be arranged into irreducible multiplets of the finite metaplectic modular group $\widetilde{\Gamma}_{4N}$ which is the quadruple covering of the inhomogeneous finite modular group $\Gamma_{4N}$ or the double covering of the homogeneous finite modular group $\Gamma'_{4N}$. In this work, we focus on the lowest level case of $\Gamma(4)$, and use the corresponding modular forms of half-integral weight to construct lepton mass models.

The rest of the paper is organized as follows. In section~\ref{sec:half integral MF}, we introduce the metaplectic group and give the definition of the half-integral weight modular forms. We show that the half-integral weight modular forms of $\Gamma(4N)$ arrange themselves into different irreducible multiplets of the finite metaplectic group $\widetilde{\Gamma}_{4N}$. We also present some useful known results of rational weight modular forms by mathematician, the explicit expressions of the lowest rational weight modular forms for certain congruence subgroup $\Gamma(N)$, and the corresponding finite metaplectic group $\widetilde{\Gamma}_{N}$ are summarized in table~\ref{tab:MFspace and finite group}. In section~\ref{sec:MF level 4}, we construct the half-integral and integral weight modular forms for $\Gamma(4)$ up to weight 6 in terms of Jacobi theta functions, and organize them into irreducible representations of finite metaplectic group $\widetilde{\Gamma}_4=\widetilde{S}_4$. In section~\ref{sec:modular building}, we generalize the modular invariant theory to include the half-integral weight modular forms. Moreover, we present three phenomenologically viable models for lepton masses and flavor mixing based on the finite metaplectic group $\widetilde{\Gamma}_4\equiv \widetilde{S}_4$. Appendix~\ref{sec:app-1} contains the analytical formulas of the Kronecker symbol and two-cocycle relevant to the definition of half-integral weight modular forms. The multiplier systems of the rational weight modular forms are given in Appendix~\ref{sec:app-multiplier}.
The group theory of $\widetilde{S}_4$ and the Clebsch-Gordan (CG) coefficients in our working basis are presented in Appendix~\ref{sec:app-S4QC}. We give another representation basis of $\widetilde{S}_4$
and the corresponding forms of CG coefficients and modular forms in Appendix~\ref{sec:app-another-basis}.

\section{\label{sec:half integral MF}Modular symmetry, metaplectic group and half-integral weight modular forms }

The full modular group $\text{SL}_2(\mathbb{Z})$ is the group of $2\times2$ matrices with integer entries and determinant 1:
\begin{equation}
\text{SL}_{2}(\mathbb{Z})=\left\{\begin{pmatrix}
a  &  b \\
c  &  d
\end{pmatrix}\bigg|a,b,c,d\in\mathbb{Z}, ad-bc=1
\right\}\,.
\end{equation}
It is quite common to use the notation $\Gamma$ for $\text{SL}_2(\mathbb{Z})$. It is well-known that $\text{SL}_2(\mathbb{Z})$ is finitely generated, and its generators are used usually chosen to be $S$ and $T$ with
\begin{equation}
S=\begin{pmatrix}
0  ~&~  1 \\
-1  ~&~  0
\end{pmatrix},~~~~T=\begin{pmatrix}
1   ~&~  1  \\
0   ~&~  1
\end{pmatrix}\,,
\end{equation}
which satisfy the relations
\begin{equation}
S^4=(ST)^3=I\,.
\end{equation}
Here $I$ is the two dimensional unit matrix. Let $N$ be a positive integer, the principal congruence subgroup $\Gamma(N)$ of level N is defined as
\begin{equation}
\Gamma(N)=\left\{
\begin{pmatrix}
a  &  b \\
c  & d
\end{pmatrix}\in \text{SL}_2(\mathbb{Z})\bigg| a,d=1(\text{mod}~N),~ b,c=0(\text{mod}~N)
\right\}\,,
\end{equation}
which implies that $\Gamma(N)$ is a normal subgroup of finite index in $\text{SL}_2(\mathbb{Z})$, and obviously we have $\Gamma(1)=\text{SL}_{2}(\mathbb{Z})$. We denote by $\mathcal{H}$ the upper half plane, i.e. the set of complex numbers $\tau$ with $\texttt{Im}(\tau)>0$. We can view elements of $\text{SL}_2(\mathbb{Z})$ as acting in the following way on $\mathcal{H}$:
\begin{equation}
\gamma=\begin{pmatrix}
a  &  b \\
c  &  d
\end{pmatrix}\in SL_2(\mathbb{Z}), ~~~~ \gamma\tau=\gamma(\tau)=\frac{a\tau+b}{c\tau+d}\,.
\end{equation}
The modular form $f(\tau)$ of weight $k$ and level $N$ is a holomorphic function of the complex modulus $\tau$ and it satisfies the transformation formula
\begin{equation}
f\left(\frac{a\tau+b}{c\tau+d}\right)=(c\tau+d)^{k}f(\tau)~~~~\text{for~all}~~\begin{pmatrix}
a  &  b \\
c   & d
\end{pmatrix}\in\Gamma(N)~\text{and}~\tau\in\mathcal{H}\,.
\end{equation}
It has been shown that the modular forms of integral weight $k$ and level $N$ can be arranged into different irreducible representations of the homogeneous finite modular group $\Gamma'_N \equiv \Gamma/\Gamma(N)$ up to the factor $(c\tau+d)^k$ in \cite{Liu:2019khw}. In the present work, we intend to include half-integral weight modular forms such that the square root of the $c\tau+d$ appears in the transformation formula. It is crucial to deal with the two branches for the square root in a systematic way.
The most common choice and the one we will always use is to choose the principal branch of the square root, i.e., for a complex number $z$, $z^{1/2}$ always means $-\pi/2 < \texttt{Arg}(z^{1/2}) \leq \pi/2$, in particular if $z< 0$ is real, $z^{1/2}$ is a pure positive imaginary number such as $(-1)^{1/2}=i$. Therefore $(z_1 z_2)^{1/2}$ is equal to $z_1^{1/2} z_2^{1/2}$ only up to a sign $\pm 1$, i.e., $(z_1z_2)^{1/2}= z_1^{1/2} z_2^{1/2}$ for $-\pi < \texttt{Arg}(z_1) + \texttt{Arg}(z_2) \leq \pi$ and $(z_1z_2)^{1/2}= -z_1^{1/2} z_2^{1/2}$ otherwise. For an (even or odd) integer $k$, $z^{k/2}$ always refer to $(z^{1/2})^k$. Note that this is not always equal to $(z^k)^{1/2}$ for $k$ odd. It is non-trivial to define the half-integral weight modular forms, and $J_{k/2}(\gamma,\tau)\equiv(c\tau+d)^{k/2}$ is not the automorphy factor anymore, and certain multiplier is generally needed. For instance, the half-integral $k/2$ weight modular form $f(\tau)$ can be consistently defined for the principal congruence subgroup $\Gamma(4N)$, it is a holomorphic function of $\tau$ and satisfies the following condition,
\begin{equation}
\label{eq:MF-Gamma4N}f(h\tau)=v^k(h)(c\tau+d)^{k/2}f_i(\tau)= v^k(h)J_{k/2}(h,\tau)f_i(\tau),~~~~ h=\begin{pmatrix}
a & b \\ c & d
\end{pmatrix} \in \Gamma(4N)\,,
\end{equation}
where $v(h)=(\frac{c}{d})$ is the Kronecker symbol, it is $1$ or $-1$ here and more details can be found in the Appendix~\ref{sec:app-1}. Notice that $v^k(h)$ and $J_{k/2}(h, \tau)$ satisfy the following identities~\cite{stromberg2013weil,cohen2017modular},
\begin{subequations}
\begin{eqnarray}
\label{eq:J-k/2-prop}&&J_{k/2}(\gamma_1 \gamma_2,\tau)= \zeta_{k/2}^{-1}(\gamma_1,\gamma_2) J_{k/2}(\gamma_1,\gamma_2 \tau) J_{k/2}(\gamma_2, \tau), \quad \gamma_{1,2} \in \Gamma \,,\\
\label{eq:vh-prop}&&v^k(h_1 h_2)=\zeta_{k/2}(h_1,h_2)v^k(h_1)v^k(h_2),~~~h_{1,2} \in \Gamma(4N)\,,
\end{eqnarray}
\end{subequations}
where $\zeta_{k/2}(\gamma_1,\gamma_2)=\zeta^k_{1/2}(\gamma_1,\gamma_2)\in\{1, e^{\pi i k}\}$, and the explicit expression of $\zeta_{1/2}(\gamma_1,\gamma_2)$ is given in Eq.~\eqref{eq:zeta-value}. Note that $\zeta_{k/2}(\gamma_1,\gamma_2)$ is always equal to 1 for any values of $\gamma_1$ and $\gamma_2$ if $k$ is even. We denote the factor $\widetilde{J}_{k/2}(h,\tau)\equiv v^k(h)(c\tau+d)^{k/2}$, using Eqs.~(\ref{eq:J-k/2-prop}, \ref{eq:vh-prop}) it is easy to check $\widetilde{J}_{k/2}$ satisfies the cocycle relation
\begin{equation}
\widetilde{J}_{k/2}(h_1h_2,\tau)=\widetilde{J}_{k/2}(h_1,h_2\tau)\widetilde{J}_{k/2}(h_2,\tau)\,, \quad h_{1,2}\in \Gamma(4N)\,.
\end{equation}
This means that $\widetilde{J}_{k/2}(h,\tau)$ is the correct automorphy factor for $\Gamma(4N)$, this generalized automorphy factor eliminates the ambiguity caused by half-integral weight, and the half-integral weight modular form defined in Eq.~\eqref{eq:MF-Gamma4N} really makes sense.

\subsection{Metaplectic group}

In order to discuss the action of the full modular group on the half-integral modular forms, one has to consider the metaplectic (twofold) cover group $\text{Mp}_2(\mathbb{Z})$ of $\text{SL}_{2}(\mathbb{Z})$~\cite{shimura1973modular}. For notation simplicity, we shall denote $\text{Mp}_2(\mathbb{Z})$ as $\widetilde{\Gamma}$ in the following. The elements of $\widetilde{\Gamma}$ can be written in the form~\cite{shimura1973modular}:
\begin{equation}
\widetilde{\Gamma} = \Big\{ \widetilde{\gamma}=(\gamma, \phi(\gamma,\tau)) ~\Big|~ \gamma=\begin{pmatrix}
a  &  b  \\
c   &  d
\end{pmatrix} \in \Gamma,~~\phi(\gamma,\tau)^2=(c\tau+d) \Big\}\,,
\end{equation}
which implies $\phi(\gamma,\tau)=\pm (c\tau+d)^{1/2}=\epsilon J_{1/2}(\gamma,\tau)$ with $\epsilon=\pm1$. The multiplication law of $\text{Mp}_2(\mathbb{Z})$ is given by
\begin{equation}
(\gamma_1,\phi(\gamma_1,\tau))(\gamma_2, \phi(\gamma_2,\tau))=(\gamma_1\gamma_2, \phi(\gamma_1, \gamma_2\tau)\phi(\gamma_2,\tau) )\,,
\end{equation}
or equivalently
\begin{equation}
(\gamma_1,\epsilon_1J_{1/2}(\gamma_1,\tau))(\gamma_2, \epsilon_2J_{1/2}(\gamma_2,\tau))=(\gamma_1\gamma_2, \epsilon_1\epsilon_2\zeta_{1/2}(\gamma_1,\gamma_2)J_{1/2}(\gamma_1\gamma_2,\tau))\,,
\end{equation}
where $\epsilon_1, \epsilon_2\in\{\pm1\}$. Obviously each element $\gamma\in\Gamma$ corresponds to two elements $\widetilde{\gamma}=(\gamma, \pm J_{1/2}(\gamma,\tau))$ of the metaplectic group $\widetilde{\Gamma}$. Lets us consider the natural projection mapping $P:(\gamma, \pm J_{1/2}(\gamma,\tau)) \mapsto\gamma$, then it is easy to see the kernel $\text{Ker}(P)=(1, \pm 1) \cong\{ \pm 1\}$, therefore $\widetilde{\Gamma}$ can be viewed as the central extension of the modular group $\Gamma$ by the group $\{\pm1\}$.

Using the generators $S$ and $T$ of $\text{SL}_2(\mathbb{Z})$, it is easy to see that the metaplectic group $\widetilde{\Gamma}$ can be generated by $\widetilde{S}$ and $\widetilde{T}$~\cite{bruinier2010weil,stromberg2013weil}:
\begin{equation}
\widetilde{S}= \left(\begin{pmatrix}
0 & 1\\ -1 & 0 \end{pmatrix} , -\sqrt{-\tau} \right), ~\quad~ \widetilde{T}=\left(\begin{pmatrix}
1 & 1\\ 0 & 1 \end{pmatrix} , 1 \right)\,,
\end{equation}
where $\sqrt{-\tau}$ denotes the principal branch of the square root, possessing positive real part. Notice that $\widetilde{S}$ is of order 8 while $\widetilde{T}$ is of infinite order, and we have
\begin{equation}
\widetilde{S}\widetilde{T}=\left(\begin{pmatrix}
 0 & 1\\ -1 & -1 \end{pmatrix} , -\sqrt{-\tau-1} \right) ,\quad \widetilde{S}^2 \equiv \widetilde{R} = \left(\begin{pmatrix}
 -1 & 0\\ 0 & -1 \end{pmatrix}, -i\right)\,,
\end{equation}
which are of orders 3 and 4, respectively. Hence the generators $\widetilde{S}$ and $\widetilde{T}$ fulfill the relations
\begin{equation}
\label{eq:STtilde-rules}\widetilde{S}^8=(\widetilde{S}\widetilde{T})^3=1\,,
\end{equation}
or equivalently
\begin{equation}
\label{eq:STtilde-rules-equiv}\widetilde{S}^2 =\widetilde{R},~~~ (\widetilde{S}\widetilde{T})^3=\widetilde{R}^4=1,~~~\widetilde{T}\widetilde{R}=\widetilde{R}\widetilde{T}\,,
\end{equation}
Because $\widetilde{R}$ is commutable with both generators $\widetilde{S}$ and $\widetilde{T}$, $\widetilde{R}$ generates the center of $\widetilde{\Gamma}$. Notice the identities $\widetilde{R}^2=\left(\begin{pmatrix}
1 & 0\\ 0 & 1 \end{pmatrix}, -1\right)$ and $\widetilde{R}^2(\gamma, J_{1/2}(\gamma,\tau)) =(\gamma, -J_{1/2}(\gamma,\tau))$, therefore the modular group $\text{SL}_{2}(\mathbb{Z})$ is isomorphic to the quotient of $\text{Mp}_{2}(\mathbb{Z})$ over the $Z_2$ subgroup $Z^{\widetilde{R}^2}_2=\{1, \widetilde{R}^2\}$,
\begin{equation}
\text{Mp}_{2}(\mathbb{Z})/Z^{\widetilde{R}^2}_2\cong\text{SL}(2,\mathbb{Z})\,.
\end{equation}
A well-known metaplectic congruence subgroup is~\cite{bruinier2010weil,stromberg2013weil}:
\begin{equation}
\label{eq:meta-congruence}\widetilde{\Gamma}(4N) = \Big\{ \widetilde{h}=(h, v(h)J_{1/2}(h,\tau) ~|~ h \in \Gamma(4N) \Big\}\,,
\end{equation}
where $v(h)=(\frac{c}{d})$ is the Kronecker symbol. $\widetilde{\Gamma}(4N)$ is an infinite normal subgroup of $\widetilde{\Gamma}$ and it is isomorphic to the principal congruence subgroup $\Gamma(4N)$.
Likewise the finite metaplectic group is the quotient group $\widetilde{\Gamma}_{4N} \equiv \widetilde{\Gamma}/\widetilde{\Gamma}(4N)$. It is easy to check
\begin{equation}
\widetilde{T}^{4N}=\left(\begin{pmatrix}
1 ~&~ 4N \\ 0 ~&~ 1 \end{pmatrix} ,~ 1 \right) \in \widetilde{\Gamma}(4N)\,.
\end{equation}
Consequently the relation
\begin{equation}
\label{eq:T-relation}\widetilde{T}^{4N}=1
\end{equation}
is generally fulfilled in the group $\widetilde{\Gamma}_{4N}$. In the present work, we focus on the lowest case $N=1$. The finite metaplectic group $\widetilde{\Gamma}_{4}$ denoted as $\widetilde{S}_4$ is a group of order 96 with group ID $[96, 67]$ in \texttt{GAP}~\cite{GAP}. The conjugacy classes and the irreducible representations of $\widetilde{S}_4$ are given Appendix~\ref{sec:app-S4QC}. For larger $N$, the relations in Eqs.~(\ref{eq:STtilde-rules}, \ref{eq:T-relation}) or equivalently Eqs.~(\ref{eq:STtilde-rules-equiv}, \ref{eq:T-relation}) are not sufficient and addition relations are needed to render the group $\widetilde{\Gamma}_{4N}$ finite. For instance, for the case of $N=2$,
the multiplication rules of $\widetilde{\Gamma}_{8}$ for the generators $S$ and $T$ are\footnote{The relations can also be written as $\widetilde{S}^8=(\widetilde{S}\widetilde{T})^3=\widetilde{T}^8=\widetilde{S}^5\widetilde{T}^6\widetilde{S}\widetilde{T}^4\widetilde{S}\widetilde{T}^2\widetilde{S}\widetilde{T}^4=1,~
\widetilde{T}\widetilde{S}^2=\widetilde{S}^2\widetilde{T}$. }
\begin{equation}
\widetilde{S}^2 =\widetilde{R},~~~ (\widetilde{S}\widetilde{T})^3=\widetilde{R}^4=\widetilde{T}^8=\widetilde{R}^2\widetilde{S}\widetilde{T}^6\widetilde{S}\widetilde{T}^4\widetilde{S}\widetilde{T}^2\widetilde{S}\widetilde{T}^4=1,
~~~\widetilde{T}\widetilde{R}=\widetilde{R}\widetilde{T}\,.
\end{equation}
Thus $\widetilde{\Gamma}_{8}$ is a group of order 768 and its group \texttt{ID} in \texttt{GAP} is [768,\, 1085324].

\subsection{\label{subsec:half-weight-MF-decom}Half-integral weight modular forms }

For an element $\widetilde{\gamma}=(\gamma, \phi(\gamma,\tau))$, we define the weight-$k/2$ slash operator $|[\widetilde{\gamma}]_{k/2}$ on the modular function $f(\tau)$ as~\cite{shimura1973modular}:
\begin{equation}
f(\tau)|[\widetilde{\gamma}]_{k/2}=f(\gamma \tau) \phi^{-k}(\gamma,\tau)\,.
\end{equation}
The slash operator has the property,
\begin{equation}
\label{eq:slash-prop}f(\tau)| [\widetilde{\gamma}_1]_{k/2} | [\widetilde{\gamma}_2]_{k/2} = f(\tau) | [\widetilde{\gamma}_1\widetilde{\gamma}_2]_{k/2} ,\quad \widetilde{\gamma}_{1,2} \in \widetilde{\Gamma} \,.
\end{equation}
The modular forms of the metaplectic congruence subgroup $\widetilde{\Gamma}(4N)$ is a holomorphic modular function invariant under the action of slash operator $|[\widetilde{h}]_{k/2}$, i.e.
\begin{equation}
\label{eq:MF-def-Gamma4N}f(\tau)| [\widetilde{h}]_{k/2} = f(\tau)~~~\text{or}~~~f(h\tau)=\phi^{k}(h,\tau)f(\tau),~~~\widetilde{h}\in\widetilde{\Gamma}(4N)\,.
\end{equation}
This is actually the same as the condition in Eq.~\eqref{eq:MF-Gamma4N} which the half-integral weight modular forms of $\Gamma(4N)$ should satisfy. It can be seen that $\widetilde{\Gamma}(4N)$ is the more natural group acting on the half-integral weight modular forms of $\Gamma(4N)$.

The weight $k/2$ modular forms of $\widetilde{\Gamma}(4N)$
span a linear space $\mathcal{M}_{k/2}(\widetilde{\Gamma}(4N))$ of finite dimension $n=\texttt{dim}\mathcal{M}_{k/2}(\widetilde{\Gamma}(4N))$. Let us denote a multiplet of linearly independent modular form $f(\tau) \equiv (f_1(\tau), f_2(\tau), \dots , f_n(\tau))^T$, $\widetilde{\gamma}=(\gamma,~\epsilon J_{1/2} (\gamma,\tau))$ and $\widetilde{h}=(h, v(h)J_{1/2}(h,\tau))$ stand for a generic element of $\widetilde{\Gamma}$ and $\widetilde{\Gamma}(4N)$ respectively. It is straightforward to check that the following identity is fulfilled,
\begin{equation}
f(\tau)|[\widetilde{\gamma}]_{k/2} | [\widetilde{h}]_{k/2}=f(\tau)| [\widetilde{\gamma}]_{k/2}\,.
\end{equation}
This means that the function $f(\tau)| [\widetilde{\gamma}]_{k/2}$ is invariant under the action of slash operator $|[\widetilde{h}]_{k/2}$, in other word, $f(\tau)| [\widetilde{\gamma}]_{k/2}=f(\gamma\tau) \phi^{-k}(\gamma,\tau)$ should be a modular form of $\widetilde{\Gamma}(4N)$, therefore $f(\tau)| [\widetilde{\gamma}]_{k/2}$ can be written as a linear combination of $f_i(\tau)$:
\begin{equation}
\label{eq:madter-eq-decomp}f(\tau)| [\widetilde{\gamma}]_{k/2} = \rho(\widetilde{\gamma}) f(\tau)~~~\text{or}~~~f(\gamma\tau)=\phi^k(\gamma,\tau)\rho(\widetilde{\gamma})f(\tau)\,,
\end{equation}
where $\rho(\widetilde{\gamma})$ is a $n\times n$ dimensional matrix depending on $\widetilde{\gamma}$. Using the identity $f(\tau)| [\widetilde{\gamma}_1]_{k/2}[\widetilde{\gamma}_2]_{k/2} = f(\tau) | [\widetilde{\gamma}_1\widetilde{\gamma}_2]_{k/2}$ in Eq.~\eqref{eq:slash-prop}, we can obtain\footnote{Analogously we find that $\rho$ forms a projective representation of the modular group $\Gamma$: $\rho(\gamma_1\gamma_2)=\zeta_{k/2}(\gamma_1,\gamma_2)\rho(\gamma_1)\rho(\gamma_2)$ for $\gamma_{1,2} \in \Gamma$. That is to say, the projection representation is lifted to the linear representation by extending the $\Gamma$ to the metaplectic group $\widetilde{\Gamma}$. }
\begin{equation}
\rho(\widetilde{\gamma}_1)\rho(\widetilde{\gamma}_2)=\rho(\widetilde{\gamma}_1\widetilde{\gamma}_2)\,.
\end{equation}
Hence $\rho$ is a linear representation of the metaplectic group $\widetilde{\Gamma}$. For $\widetilde{\gamma}=\widetilde{h}=(h, v(h)J_{1/2}(h,\tau))\in \widetilde{\Gamma}(4N)$, Eq.~\eqref{eq:madter-eq-decomp} gives us
\begin{equation}
f(\tau)|[\widetilde{h}]_{k/2} = \rho(\widetilde{h})f(\tau)\,.
\end{equation}
Comparing with the definition of modular form $f(\tau)| [\widetilde{h}]_{k/2} = f(\tau)$ in Eq.~\eqref{eq:MF-def-Gamma4N}, we obtain $\rho(\widetilde{h})=1$. As a consequence, $\rho(\widetilde{\gamma})$ is actually a linear representation of the quotient group $\widetilde{\Gamma}_{4N} \equiv \widetilde{\Gamma}/\widetilde{\Gamma}(4N)$.
The finite representation $\rho$ can always be decomposed into a direct sum of irreducible unitary representations of $\widetilde{\Gamma}_{4N}$ such that the modular forms of half-integral weights can be arranged into different irreducible representations of the finite group $\widetilde{\Gamma}_{4N}$.
Furthermore, applying Eq.~\eqref{eq:madter-eq-decomp} for $\widetilde{\gamma}=\widetilde{R}$, we obtain
\begin{equation}
f(\widetilde{R}\tau)=f(\tau)= (-i)^k \rho(\widetilde{R})f(\tau)\,,
\end{equation}
which implies
\begin{equation}
\rho(\widetilde{R})=i^{k},~~~~\begin{cases}
k ~\texttt{odd}\,~: ~\rho^4(\widetilde{R})=1\,, \\
k ~\texttt{even}: ~\rho^2(\widetilde{R})=1\,.
\end{cases}
\end{equation}
Taking into account with the general relations in Eqs.~(\ref{eq:STtilde-rules}, \ref{eq:T-relation}), we can know
\begin{equation}
\label{eq:rep-matrices-relations}\hskip-0.1in\begin{cases}
k ~\texttt{odd}\,~: ~~\rho^2(\widetilde{S})=\rho(\widetilde{R}),~~\rho^4(\widetilde{R})=\rho^3(\widetilde{S}\widetilde{T})=\rho^{4N}(\widetilde{T})=1,~~\rho(\widetilde{T})\rho(\widetilde{R})=\rho(\widetilde{R})\rho(\widetilde{T})\,, \\
k ~\texttt{even}: ~~\rho^2(\widetilde{S})=\rho(\widetilde{R}),~~\rho^2(\widetilde{R})=\rho^3(\widetilde{S}\widetilde{T})=\rho^{4N}(\widetilde{T})=1,~~\rho(\widetilde{T})\rho(\widetilde{R})=\rho(\widetilde{R})\rho(\widetilde{T})\,.
\end{cases}
\end{equation}
Notice that the representation matrices of the generators $\widetilde{S}$ and $\widetilde{T}$ satisfy the same relations as those of the homogeneous finite modular group $\Gamma'_{4N}$~\cite{Liu:2019khw}. The equations in Eq.~\eqref{eq:rep-matrices-relations} show explicitly that the half-integral weight modular form can be decomposed into irreducible representation of finite metaplectic group $\widetilde{\Gamma}_{4N}$, and the integral weight modular forms are arranged into irreducible multiplets of $\Gamma'_{4N}$.

\subsection{\label{subsec:rational-weight-MF}Rational weight modular forms}

Analogously modular forms of rational weight $r$ can be defined for certain congruence subgroup. Similar to Eq.~\eqref{eq:MF-Gamma4N}, a multiplier system $v(\gamma)$ is needed such that $v(\gamma)(c\tau+d)^{r}$ is the correct automorphy factor satisfying the cocycle relation, and the ambiguity of multi-valued branches caused by the rational power is properly eliminated. It is also a big challenge to explicitly construct the basis of the linear space of the rational weight modular forms. It is remarkable that some mathematicians have found out the multiplier system $v(\gamma)$ for the principal congruence subgroup $\Gamma(N)$ with odd integer $N\geq5$,
the explicit expression of $v(\gamma)$ is given in Appendix~\ref{sec:app-multiplier}, and the corresponding modular forms of rational weights are constructed~\cite{ibukiyama2000modular,ibukiyama2020graded}. We will describe the main results below.

First of all, for any odd integer $5 \leq N \leq 13$, the ring of the modular forms of rational weight $r=(N-3)/(2N)$ for the principal congruence subgroup $\Gamma(N)$ can be constructed from the holomorphic functions $f^{(N)}_n(\tau)$~\cite{ibukiyama2000modular},
\begin{equation}
\label{eq:rationalMF}
f^{(N)}_n(\tau)=\theta_{(\frac{n}{2N},\frac{1}{2})}(N\tau)/\eta(\tau)^{\frac{3}{N}}\,,
\end{equation}
where $n$ are odd integers with $1 \leq n \leq N-2$, and the theta constant with characteristic $(m',m'')$ is defined by
\begin{equation}
\label{eq:theta_constans}
\theta_{(m',m'')}(\tau)=\sum_{m\in \mathbb{Z}} e^{2\pi i \tau (\frac{1}{2}(m+m')^2\tau+(m+m')m'')}\,,
\end{equation}
and Dedekind eta function is
\begin{equation}
\eta(\tau)=e^{\pi i \tau/12} \prod_{n=1}^{\infty} (1-e^{2\pi i n\tau})\,.
\end{equation}
Consequently there are $(N-1)/2$ linearly independent modular forms $f^{(N)}_1, f^{(N)}_3, \dots f^{(N)}_{N-2}$ of rational weight $r=(N-3)/(2N)$, and the graded rings of modular forms $\mathcal{M}(\Gamma(N)) = \bigoplus_{m\geq 1} \mathcal{M}_{m\frac{N-3}{2N}}(\Gamma(N))$ can be generated by the tensor products of these lowest weight modular forms.
The dimension formula of $ \mathcal{M}_{m\frac{N-3}{2N}}(\Gamma(N))$ for any odd integer $N \geq 5$ and any integer $m > \frac{4(N-6)}{N-3}$ is given by~\cite{ibukiyama2000modular,ibukiyama2020graded}
\begin{equation}
\dim \mathcal{M}_{m\frac{(N-3)}{2N}}(\Gamma(N)) =\frac{N^2\left[m(N-3)-2(N-6)\right]}{48}\prod_{p | N} (1-\frac{1}{p^2})\,,
\end{equation}
where the product is over the prime divisors $p$ of $N$. As shown in section~\ref{subsec:half-weight-MF-decom}, we expect that rational weight modular forms can be organized into different irreducible multiplets of the finite metaplectic congruence subgroup. We summarize the dimension formula, modular forms of rational weight $r=(N-3)/(2N)$ and the corresponding finite metaplectic congruence subgroup in table~\ref{tab:MFspace and finite group}.
We also include the half-integral weight case which we are concerned with. We would like to mention that the theory of modular forms with real weight and even complex weight are also developed~\cite{bruggeman2018holomorphic}, and then the modular group $\text{SL}_2(\mathbb{Z})$ should be extended to the universal covering groups. Some concrete examples are given in~\cite{aoki2017jacobi,manin2018modular}.

\begin{table}[t!]
\centering
\resizebox{1.0\textwidth}{!}{
\begin{tabular}{|c|c|c|c|c|c|c|}
\hline\hline
$N$ & weight $r$ & $\texttt{dim}\mathcal{M}_{r}(\Gamma(N)) $  & $\mathcal{M}_{r}(\Gamma(N))|_{k=1}$ &  $\widetilde{\Gamma}_{N}$ & $|\widetilde{\Gamma}_{N}|$ & $\texttt{GAP~ID}$   \rule[-2ex]{0pt}{5ex} \\ \hline
4 & $k/2$& $k+1$ & $\Big\{\theta_3(0|2\tau),~ \theta_2(0|2\tau) \Big\}$ & $\widetilde{S}_4$ & 96 & [96,67]  \rule[-2ex]{0pt}{5ex} \\ \hline
5 & $k/5$ &$k+1$ & $\Big\{f^{(5)}_1(\tau), ~f^{(5)}_3(\tau) \Big\}$ &  $Z_5\times \Gamma'_5$   &  600 & [600,54]  \rule[-2ex]{0pt}{5ex} \\ \hline
7 & $2k/7$ & $\begin{cases} 4k-2 ~\,(\text{for}~ k \geq 2) \\ 3~\, (\text{for}~ k=1) \end{cases}$  & $\Big\{f^{(7)}_1(\tau), ~f^{(7)}_3(\tau), ~f^{(7)}_5(\tau) \Big\}$ & $Z_7\times \Gamma_7$ &  1176 & [1176,212]   \rule[-2ex]{0pt}{5ex} \\ \hline
9 & $k/3$ & $\begin{cases} 9k-9 ~\,(\text{for}~ k \geq 3) \\ 10~\, (\text{for}~ k=2) \\ 4 ~~~ (\text{for}~ k=1) \end{cases}$ & $\Big\{f^{(9)}_1(\tau), ~f^{(9)}_3(\tau), ~f^{(9)}_5(\tau),~f^{(9)}_7(\tau) \Big\}$ &  $\widetilde{\Gamma}_9$ & 1944 & [1944,2976]  \rule[-2ex]{0pt}{5ex}\\ \hline \hline
\end{tabular}}
\caption{The dimension formula of $\texttt{dim}\mathcal{M}_{r}(\Gamma(N))$ for $N=4,5,7,9$, the linear space $\mathcal{M}_{r}(\Gamma(N))|_{k=1}$ of the lowest fractional weight modular forms,
and the finite metaplectic group $\widetilde{\Gamma}_{N}$. Notice that the functions $f^{(N)}_n(\tau)$ and the theta constants $\theta_{2,3}(0|2\tau)$ are defined in Eq.~\eqref{eq:rationalMF} and Eq.~\eqref{eq:theta-function-def} respectively.
\label{tab:MFspace and finite group} }
\end{table}

\section{\label{sec:MF level 4}Half-integral/integral weight modular forms of level 4}

Half weight modular forms of level $4N$ have been studied extensively in math since Shimura's original work~\cite{Shimura1973},
a general construction of the modular space $\mathcal{M}_{1/2}(\Gamma(4N))$ using theta functions associated with lattices has been given in the literature \cite{Modulformen1984Ueber,gritsenko2019theta}. In particular for the simplest case of level $4N=4$, the linear space of the half weight modular forms can be generated by the following two Jacobi theta functions\footnote{These two modular forms also appeared in~\cite{Novichkov:2020eep}.}
\begin{equation}
\label{eq:e12-half-weight}\mathcal{M}_{1/2}(\Gamma(4))=\{e_1(\tau)\equiv\theta_3(0|2\tau),~~ e_2(\tau)\equiv \theta_2(0|2\tau) \},
\end{equation}
with
\begin{align}
\nonumber&\theta_2(0|2\tau)=\sum_{m \in \mathbb{Z}} e^{2\pi i \tau  (m+\frac{1}{2})^2 } =2q^{1/4}(1+q^2+q^6+q^{12}+\dots)\,, \\
\nonumber&\theta_3(0|2\tau)=\sum_{m \in \mathbb{Z}} e^{2\pi i \tau  m^2}=1+2q+2q^4+2q^9+2q^{16}+\dots \,, \\
\label{eq:theta-function-def}&\theta_4(0|2\tau)=\sum_{m \in \mathbb{Z}} (-1)^me^{2\pi i \tau m^2 }=1-2q+2q^4-2q^9+2q^{16}+\dots \,.
\end{align}
Using the basic transformation properties of Jacobi theta function~\cite{olver2010nist}, we can obtain the following transformation rules under the action of the generators $S$ and $T$,
\begin{align}
\nonumber&\theta_3(0|2\tau) \xrightarrow{T} \theta_3(0|2\tau) ,~~\quad~~ \theta_2(0|2\tau) \xrightarrow{T} i \theta_2(0|2\tau) \,,\\
\nonumber&\theta_3(0|2\tau) \xrightarrow{S} \theta_3(0|-\frac{2}{\tau})=(-i\frac{\tau}{2})^{1/2}\theta_3(0|\frac{\tau}{2})\,,\\
\label{eq:thetaconstant}&\theta_2(0|2\tau) \xrightarrow{S} \theta_2(0|-\frac{2}{\tau})=(-i\frac{\tau}{2})^{1/2}\theta_4(0|\frac{\tau}{2})\,.
\end{align}
From definition of the theta function in Eq.~\eqref{eq:theta-function-def}, we know
\begin{align}
\nonumber
\theta_3(0|\frac{\tau}{2})&=\sum_{m\in\mathbb{Z}}e^{\pi i (\frac{\tau}{2} m^2)}=\sum_{n\in \mathbb{Z}} e^{2\pi i\tau n^2}+\sum_{n\in \mathbb{Z}}e^{2\pi i\tau (n+\frac{1}{2})^2}=\theta_3(0|2\tau)+\theta_2(0|2\tau)\,,\\
\theta_4(0|\frac{\tau}{2})&=\sum_{m\in\mathbb{Z}}(-1)^m e^{\pi i (\frac{\tau}{2} m^2)}=\sum_{n\in \mathbb{Z}} e^{2\pi i\tau n^2} - \sum_{n\in \mathbb{Z}}e^{2\pi i\tau (n+\frac{1}{2})^2}=\theta_3(0|2\tau) - \theta_2(0|2\tau)\,.
\end{align}
As a consequence, under the action of $S$ and $T$, the basis vectors $e_1(\tau)$ and $e_2(\tau)$ transform as
\begin{align}
\nonumber
&~~~~~~~~~~~~~~~~~~~~~e_1(\tau) \xrightarrow{T} e_1(\tau) ,~~\quad~~ e_2(\tau) \xrightarrow{T} i e_2(\tau) \,, \\
&e_1(\tau) \xrightarrow{S} (-i\frac{\tau}{2})^{1/2}(e_1(\tau)+e_2(\tau)),\quad e_2(\tau) \xrightarrow{S} (-i\frac{\tau}{2})^{1/2}(e_1(\tau)-e_2(\tau))\,.
\end{align}
We find these two linearly independent modular forms $e_1(\tau)$ and $e_2(\tau)$ arrange themselves into a doublet denoted as
\begin{equation}
Y^{(\frac{1}{2})}_{\mathbf{\hat{2}}}(\tau)=
\begin{pmatrix}
e_1(\tau) \\
-e_2(\tau)
\end{pmatrix}
\equiv\begin{pmatrix}
\vartheta_1  \\ \vartheta_2
\end{pmatrix} \,,
\end{equation}
which transforms in the two-dimensional irreducible representation $\mathbf{\hat{2}}$ of $\widetilde{\Gamma}_4 \equiv \widetilde{S_4}$,
\begin{equation}
Y^{(\frac{1}{2})}_{\mathbf{\hat{2}}}(-1/\tau)=-\sqrt{-\tau}\,\rho_{\mathbf{\hat{2}}}(\widetilde{S})Y^{(\frac{1}{2})}_{\mathbf{\hat{2}}}(\tau),~~~~
Y^{(\frac{1}{2})}_{\mathbf{\hat{2}}}(\tau+1)=\rho_{\mathbf{\hat{2}}}(\widetilde{T})Y^{(\frac{1}{2})}_{\mathbf{\hat{2}}}(\tau)\,,
\end{equation}
where the unitary representation matrices $\rho_{\mathbf{\hat{2}}}(\widetilde{S})$ and $\rho_{\mathbf{\hat{2}}}(\widetilde{T})$ are given in table~\ref{tab:Rep_baseB}. All higher (half-integral and integral) weight modular forms can be constructed from the tensor products of $Y^{(\frac{1}{2})}_{\mathbf{\hat{2}}}(\tau)$ by using the CG coefficients of group $\widetilde{S}_4$. For instance, we find the weight 1 modular forms make up a triplet $\mathbf{\hat{3}'}$ of $\widetilde{S}_4$,
\begin{equation}
Y_{\mathbf{\hat{3}'}}^{(1)}=\frac{1}{\sqrt{2}}\left(Y_{\mathbf{\hat{2}}}^{(\frac{1}{2})}Y_{\mathbf{\hat{2}}}^{(\frac{1}{2})}\right)_{\mathbf{\hat{3}'_{s}}}=\begin{pmatrix}
\sqrt{2}\vartheta_1\vartheta_2\\
\vartheta_1^2\\
-\vartheta_2^2\\
\end{pmatrix}\,,
\end{equation}
where we have multiplied an overall constant $1/\sqrt{2}$ to make the resulting expression relatively simple. The nontrivial constraint $Y^{(1)2}_1+2Y^{(1)}_2Y^{(1)}_3=0 $ in Ref.~\cite{Liu:2020akv} are now trivial. Notice another tensor product $(Y^{(\frac{1}{2})}_{\mathbf{\hat{2}}} Y^{(\frac{1}{2})}_{\mathbf{\hat{2}}})_\mathbf{\hat{1}'}=0$ because of the antisymmetric CG coefficient for the contraction $\mathbf{\hat{2}}\otimes\mathbf{\hat{2}}\rightarrow\mathbf{\hat{1}'}$.
It is straightforward to check that $Y^{(1)}_{\mathbf{\hat{3}'}}$ is the same as the original weight one modular forms given in~\cite{Novichkov:2020eep} up to a permutation, the discrepancy arises from the different convention for the representation matrices of the generators $S$ and $T$. In a similar fashion, we can obtain higher weights modular forms and decompose them into different irreducible multiplets of $\widetilde{S}_4$. In the following, we present linearly independent half-integral and integral weight modular forms up to weight 6, and we normalize the overall constant for simplicity. There are four linearly independent modular forms of weight $3/2$, and they can be arranged into a quartet representation $\mathbf{4'}$ of $\widetilde{S}_4$,
\begin{equation}
Y_{\mathbf{4'}}^{(\frac{3}{2})}=\frac{1}{\sqrt{3}}\left(Y_{\mathbf{\hat{2}}}^{(\frac{1}{2})}Y_{\mathbf{\hat{3}'}}^{(1)}\right)_{\mathbf{4'}}=\begin{pmatrix}
\vartheta_2^3\\
\sqrt{3}\vartheta_1^2\vartheta_2\\
\vartheta_1^3\\
\sqrt{3}\vartheta_1\vartheta_2^2\\
\end{pmatrix}\,.
\end{equation}
The weight $2$ modular forms of level 4 can be decomposed into a doublet $\mathbf{2}$ and a triplet $\mathbf{3}$,
\begin{eqnarray}
\nonumber&&Y_{\mathbf{2}}^{(2)}=\left(Y_{\mathbf{\hat{2}}}^{(\frac{1}{2})}Y_{\mathbf{4'}}^{(\frac{3}{2})}\right)_{\mathbf{2}}=\begin{pmatrix}
\vartheta_1^4+\vartheta_2^4\\
-2\sqrt{3}\vartheta_1^2\vartheta_2^2\\
\end{pmatrix}\,,\\
&&Y_{\mathbf{3}}^{(2)}=\frac{1}{\sqrt{2}}\left(Y_{\mathbf{\hat{2}}}^{(\frac{1}{2})}Y_{\mathbf{4'}}^{(\frac{3}{2})}\right)_{\mathbf{3}}=\begin{pmatrix}
\vartheta_1^4-\vartheta_2^4\\
2\sqrt{2}\vartheta_1^3\vartheta_2\\
2\sqrt{2}\vartheta_1\vartheta_2^3\\
\end{pmatrix}\,.
\end{eqnarray}
At weight $5/2$, we have 6 independent modular forms which transform according to the irreducible representations $\mathbf{\hat{2}}$ and $\mathbf{4}$ of $\widetilde{S}_4$,
\begin{eqnarray}
\nonumber&&Y_{\mathbf{\hat{2}}}^{(\frac{5}{2})}=-\left(Y_{\mathbf{\hat{2}}}^{(\frac{1}{2})}Y_{\mathbf{3}}^{(2)}\right)_{\mathbf{\hat{2}}}=\begin{pmatrix}
\vartheta_1^5-5\vartheta_1\vartheta_2^4\\
\vartheta_2^5-5\vartheta_1^4\vartheta_2\\
\end{pmatrix}\,,\\
&&Y_{\mathbf{4}}^{(\frac{5}{2})}=\left(Y_{\mathbf{\hat{2}}}^{(\frac{1}{2})}Y_{\mathbf{2}}^{(2)}\right)_{\mathbf{4}}=\begin{pmatrix}
\vartheta_1\left(\vartheta_1^4+\vartheta_2^4\right)\\
2\sqrt{3}\vartheta_1^3\vartheta_2^2\\
-\vartheta_2\left(\vartheta_1^4+\vartheta_2^4\right)\\
-2\sqrt{3}\vartheta_1^2\vartheta_2^3\\
\end{pmatrix}\,.
\end{eqnarray}
The weight $3$ modular forms can be arranged into a singlet and two triplets representation of $\widetilde{S}_4$ as follows,
\begin{eqnarray}
\nonumber&&Y_{\mathbf{\hat{1}'}}^{(3)}=-\frac{1}{6}\left(Y_{\mathbf{\hat{2}}}^{(\frac{1}{2})}Y_{\mathbf{\hat{2}}}^{(\frac{5}{2})}\right)_{\mathbf{\hat{1}'_{a}}}=\vartheta_1\vartheta_2\left(\vartheta_1^4-\vartheta_2^4\right)\,, \\
\nonumber&&Y_{\mathbf{\hat{3}}}^{(3)}=-\frac{1}{\sqrt{3}}\left(Y_{\mathbf{\hat{2}}}^{(\frac{1}{2})}Y_{\mathbf{4}}^{(\frac{5}{2})}\right)_{\mathbf{\hat{3}}}=\begin{pmatrix}
4\sqrt{2}\vartheta_1^3\vartheta_2^3\\
\vartheta_1^6+3\vartheta_1^2\vartheta_2^4\\
-\vartheta_2^2\left(3\vartheta_1^4+\vartheta_2^4\right)\\
\end{pmatrix}\,,\\
&&Y_{\mathbf{\hat{3}'}}^{(3)}=-\frac{1}{\sqrt{2}}\left(Y_{\mathbf{\hat{2}}}^{(\frac{1}{2})}Y_{\mathbf{\hat{2}}}^{(\frac{5}{2})}\right)_{\mathbf{\hat{3}'_{s}}}=\begin{pmatrix}
2\sqrt{2}\vartheta_1\vartheta_2\left(\vartheta_1^4+\vartheta_2^4\right)\\
\vartheta_2^6-5\vartheta_1^4\vartheta_2^2\\
5\vartheta_1^2\vartheta_2^4-\vartheta_1^6\\
\end{pmatrix}\,.
\end{eqnarray}
The linear space of modular forms of weight $7/2$ and level 4 has dimension 8, and it can be decomposed into three $\widetilde{S}_4$ multiplets $\mathbf{\widetilde{2}'}$, $\mathbf{\widetilde{2}}$ and $\mathbf{4'}$ as follow,
\begin{eqnarray}
\nonumber&&Y_{\mathbf{\widetilde{2}'}}^{(\frac{7}{2})}=\left(Y_{\mathbf{\hat{2}}}^{(\frac{1}{2})}Y_{\mathbf{\hat{1}'}}^{(3)}\right)_{\mathbf{\widetilde{2}'}}=\vartheta_1\vartheta_2\left(\vartheta_1^4-\vartheta_2^4\right)\begin{pmatrix}
\vartheta_1\\
\vartheta_2\\
\end{pmatrix}\,,\\
\nonumber&&Y_{\mathbf{\widetilde{2}}}^{(\frac{7}{2})}=-\frac{1}{\sqrt{2}}\left(Y_{\mathbf{\hat{2}}}^{(\frac{1}{2})}Y_{\mathbf{\hat{3}}}^{(3)}\right)_{\mathbf{\widetilde{2}}}=\begin{pmatrix}
\vartheta_2^7+7\vartheta_1^4\vartheta_2^3\\
-\vartheta_1^3\left(\vartheta_1^4+7\vartheta_2^4\right)\\
\end{pmatrix}\,,\\
&&Y_{\mathbf{4'}}^{(\frac{7}{2})}=\left(Y_{\mathbf{\hat{2}}}^{(\frac{1}{2})}Y_{\mathbf{\hat{3}}}^{(3)}\right)_{\mathbf{4'}}=\begin{pmatrix}
5\vartheta_1^4\vartheta_2^3-\vartheta_2^7\\
\sqrt{3}\vartheta_1^2\vartheta_2\left(\vartheta_1^4+3\vartheta_2^4\right)\\
5\vartheta_1^3\vartheta_2^4-\vartheta_1^7\\
\sqrt{3}\vartheta_1\vartheta_2^2\left(3\vartheta_1^4+\vartheta_2^4\right)\\
\end{pmatrix}\,.
\end{eqnarray}
At weight $4$, we find the following modular multiplets,
\begin{eqnarray}
\nonumber&&Y_{\mathbf{1}}^{(4)}=-\left(Y_{\mathbf{\hat{2}}}^{(\frac{1}{2})}Y_{\mathbf{\widetilde{2}}}^{(\frac{7}{2})}\right)_{\mathbf{1}}=
\vartheta_1^8+14\vartheta_1^4\vartheta_2^4+\vartheta_2^8\,,\\
\nonumber&&Y_{\mathbf{2}}^{(4)}=-\left(Y_{\mathbf{\hat{2}}}^{(\frac{1}{2})}Y_{\mathbf{4'}}^{(\frac{7}{2})}\right)_{\mathbf{2}}=\begin{pmatrix}
\vartheta_1^8-10\vartheta_1^4\vartheta_2^4+\vartheta_2^8\\
4\sqrt{3}\vartheta_1^2\vartheta_2^2\left(\vartheta_1^4+\vartheta_2^4\right)\\
\end{pmatrix}\,,\\
\nonumber&&Y_{\mathbf{3}}^{(4)}=\left(Y_{\mathbf{\hat{2}}}^{(\frac{1}{2})}Y_{\mathbf{\widetilde{2}}}^{(\frac{7}{2})}\right)_{\mathbf{3}}=\begin{pmatrix}
\vartheta_2^8-\vartheta_1^8\\
\sqrt{2}\vartheta_2\left(\vartheta_1^7+7\vartheta_1^3\vartheta_2^4\right)\\
\sqrt{2}\vartheta_1\left(\vartheta_2^7+7\vartheta_1^4\vartheta_2^3\right)\\
\end{pmatrix}\,,\\
&&Y_{\mathbf{3'}}^{(4)}=\frac{1}{\sqrt{2}}\left(Y_{\mathbf{\hat{2}}}^{(\frac{1}{2})}Y_{\mathbf{\widetilde{2}'}}^{(\frac{7}{2})}\right)_{\mathbf{3'}}=\vartheta_1\vartheta_2\left(\vartheta_1^4-\vartheta_2^4\right)\begin{pmatrix}
\sqrt{2}\vartheta_1\vartheta_2\\
-\vartheta_2^2\\
\vartheta_1^2\\
\end{pmatrix}\,.
\end{eqnarray}
We have 10 linearly independent weight $9/2$ modular forms which arrange themselves into a $\widetilde{S}_4$ doublet $\mathbf{\hat{2}}$ and two quartets transforming in the representation $\mathbf{4}$,
\begin{eqnarray}
\nonumber&&Y_{\mathbf{\hat{2}}}^{(\frac{9}{2})}=\left(Y_{\mathbf{\hat{2}}}^{(\frac{1}{2})}Y_{\mathbf{1}}^{(4)}\right)_{\mathbf{\hat{2}}}=\left(\vartheta_1^8+14\vartheta_1^4\vartheta_2^4+\vartheta_2^8\right)\begin{pmatrix}
\vartheta_1\\
\vartheta_2\\
\end{pmatrix}\,, \\
\nonumber&&Y_{\mathbf{4}I}^{(\frac{9}{2})}=\left(Y_{\mathbf{\hat{2}}}^{(\frac{1}{2})}Y_{\mathbf{2}}^{(4)}\right)_{\mathbf{4}}=\begin{pmatrix}
\vartheta_1\left(\vartheta_1^8-10\vartheta_1^4\vartheta_2^4+\vartheta_2^8\right)\\
-4\sqrt{3}\vartheta_1^3\vartheta_2^2\left(\vartheta_1^4+\vartheta_2^4\right)\\
-\vartheta_2\left(\vartheta_1^8-10\vartheta_1^4\vartheta_2^4+\vartheta_2^8\right)\\
4\sqrt{3}\vartheta_1^2\vartheta_2^3\left(\vartheta_1^4+\vartheta_2^4\right)\\
\end{pmatrix}\,, \\
&&Y_{\mathbf{4}II}^{(\frac{9}{2})}=-\frac{1}{\sqrt{2}}\left(Y_{\mathbf{\hat{2}}}^{(\frac{1}{2})}Y_{\mathbf{3}}^{(4)}\right)_{\mathbf{4}}=\begin{pmatrix}
\vartheta_1^9-7\vartheta_1^5\vartheta_2^4-2\vartheta_1\vartheta_2^8\\
-\sqrt{3}\vartheta_2^2\left(\vartheta_1^7+7\vartheta_1^3\vartheta_2^4\right)\\
-\vartheta_2^9+7\vartheta_1^4\vartheta_2^5+2\vartheta_1^8\vartheta_2\\
\sqrt{3}\vartheta_1^2\left(\vartheta_2^7+7\vartheta_1^4\vartheta_2^3\right)\\
\end{pmatrix}\,.
\end{eqnarray}
At weight 5, we find the following modular multiplets:
\begin{eqnarray}
\nonumber&&Y_{\mathbf{2'}}^{(5)}=\frac{1}{3}\left(Y_{\mathbf{\hat{2}}}^{(\frac{1}{2})}Y_{\mathbf{4}II}^{(\frac{9}{2})}\right)_{\mathbf{2'}}=\vartheta_1\vartheta_2\left(\vartheta_1^4-\vartheta_2^4\right)\begin{pmatrix}
2\sqrt{3}\vartheta_1^2\vartheta_2^2\\
\vartheta_1^4+\vartheta_2^4\\
\end{pmatrix}\,, \\
\nonumber&&Y_{\mathbf{\hat{3}}}^{(5)}=-\frac{1}{\sqrt{3}}\left(Y_{\mathbf{\hat{2}}}^{(\frac{1}{2})}Y_{\mathbf{4}I}^{(\frac{9}{2})}\right)_{\mathbf{\hat{3}}}=\begin{pmatrix}
-8\sqrt{2}\vartheta_1^3\vartheta_2^3\left(\vartheta_1^4+\vartheta_2^4\right)\\
\vartheta_1^2\left(\vartheta_1^8-14\vartheta_1^4\vartheta_2^4-3\vartheta_2^8\right)\\
\vartheta_2^2\left(3\vartheta_1^8+14\vartheta_1^4\vartheta_2^4-\vartheta_2^8\right)\\
\end{pmatrix}\,, \\
\nonumber&&Y_{\mathbf{\hat{3}'}I}^{(5)}=-\left(Y_{\mathbf{\hat{2}}}^{(\frac{1}{2})}Y_{\mathbf{4}I}^{(\frac{9}{2})}\right)_{\mathbf{\hat{3}'}}=\begin{pmatrix}
2\sqrt{2}\vartheta_1\vartheta_2\left(\vartheta_1^8-10\vartheta_1^4\vartheta_2^4+\vartheta_2^8\right)\\
\vartheta_2^2\left(13\vartheta_1^8+2\vartheta_1^4\vartheta_2^4+\vartheta_2^8\right)\\
-\vartheta_1^2\left(\vartheta_1^8+2\vartheta_1^4\vartheta_2^4+13\vartheta_2^8\right)\\
\end{pmatrix}\,, \\
&&Y_{\mathbf{\hat{3}'}II}^{(5)}=\left(Y_{\mathbf{\hat{2}}}^{(\frac{1}{2})}Y_{\mathbf{4}II}^{(\frac{9}{2})}\right)_{\mathbf{\hat{3}'}}=\left(\vartheta_1^8+14\vartheta_1^4\vartheta_2^4+\vartheta_2^8\right)\begin{pmatrix}
\sqrt{2}\vartheta_1\vartheta_2\\
-\vartheta_2^2\\
\vartheta_1^2\\
\end{pmatrix}\,.
\end{eqnarray}
There are 12 independent weight $11/2$ modular form of level 4, and they decompose as $\mathbf{\widetilde{2}}\oplus\mathbf{\widetilde{2}'}\oplus\mathbf{4'}\oplus\mathbf{4'}$ under $\widetilde{S}_4$,
\begin{eqnarray}
\nonumber&&Y_{\mathbf{\widetilde{2}}}^{(\frac{11}{2})}=\frac{1}{\sqrt{2}}\left(Y_{\mathbf{\hat{2}}}^{(\frac{1}{2})}Y_{\mathbf{\hat{3}}}^{(5)}\right)_{\mathbf{\widetilde{2}}}=\begin{pmatrix}
\vartheta_2^3\left(11\vartheta_1^8+22\vartheta_1^4\vartheta_2^4-\vartheta_2^8\right)\\
\vartheta_1^3\left(\vartheta_1^8-22\vartheta_1^4\vartheta_2^4-11\vartheta_2^8\right)\\
\end{pmatrix}\,, \\
\nonumber&&Y_{\mathbf{\widetilde{2}'}}^{(\frac{11}{2})}=-\frac{1}{3\sqrt{2}}\left(Y_{\mathbf{\hat{2}}}^{(\frac{1}{2})}Y_{\mathbf{\hat{3}'}I}^{(5)}\right)_{\mathbf{\widetilde{2}'}}=\vartheta_1\vartheta_2\left(\vartheta_1^4-\vartheta_2^4\right)\begin{pmatrix}
\vartheta_1^5-5\vartheta_1\vartheta_2^4\\
\vartheta_2^5-5\vartheta_1^4\vartheta_2\\
\end{pmatrix}\,, \\
\nonumber&&Y_{\mathbf{4'}I}^{(\frac{11}{2})}=-\frac{1}{\sqrt{3}}\left(Y_{\mathbf{\hat{2}}}^{(\frac{1}{2})}Y_{\mathbf{\hat{3}'}I}^{(5)}\right)_{\mathbf{4'}}=\begin{pmatrix}
\vartheta_2^3\left(13\vartheta_1^8+2\vartheta_1^4\vartheta_2^4+\vartheta_2^8\right)\\
-\sqrt{3}\vartheta_1^2\vartheta_2\left(\vartheta_1^8-14\vartheta_1^4\vartheta_2^4-3\vartheta_2^8\right)\\
\vartheta_1^3\left(\vartheta_1^8+2\vartheta_1^4\vartheta_2^4+13\vartheta_2^8\right)\\
\sqrt{3}\vartheta_1\vartheta_2^2\left(3\vartheta_1^8+14\vartheta_1^4\vartheta_2^4-\vartheta_2^8\right)\\
\end{pmatrix}\,, \\
&&Y_{\mathbf{4'}II}^{(\frac{11}{2})}=\frac{1}{\sqrt{3}}\left(Y_{\mathbf{\hat{2}}}^{(\frac{1}{2})}Y_{\mathbf{\hat{3}'}II}^{(5)}\right)_{\mathbf{4'}}=\left(\vartheta_1^8+14\vartheta_1^4\vartheta_2^4+\vartheta_2^8\right)\begin{pmatrix}
\vartheta_2^3\\
\sqrt{3}\vartheta_1^2\vartheta_2\\
\vartheta_1^3\\
\sqrt{3}\vartheta_1\vartheta_2^2\\
\end{pmatrix}\,.
\end{eqnarray}
The linear space of weight 6 modular form has dimension 13, and they can be arranged into two singlets $\mathbf{1}$, $\mathbf{1'}$, a doublet $\mathbf{2}$, and three triplets $\mathbf{3}\oplus\mathbf{3}\oplus\mathbf{3'}$ under $\widetilde{S}_4$:
\begin{eqnarray}
\nonumber&&Y_{\mathbf{1}}^{(6)}=\left(Y_{\mathbf{\hat{2}}}^{(\frac{1}{2})}Y_{\mathbf{\widetilde{2}}}^{(\frac{11}{2})}\right)_{\mathbf{1}}=
\vartheta_1^{12}-33\vartheta_1^8\vartheta_2^4-33\vartheta_1^4\vartheta_2^8+\vartheta_2^{12}\,,\\
\nonumber&&Y_{\mathbf{1'}}^{(6)}=-\frac{1}{6}\left(Y_{\mathbf{\hat{2}}}^{(\frac{1}{2})}Y_{\mathbf{\widetilde{2}'}}^{(\frac{11}{2})}\right)_{\mathbf{1'}}=
\vartheta_1^2\vartheta_2^2\left(\vartheta_1^4-\vartheta_2^4\right)^2 \,, \\
\nonumber&&Y_{\mathbf{2}}^{(6)}=\left(Y_{\mathbf{\hat{2}}}^{(\frac{1}{2})}Y_{\mathbf{4'}I}^{(\frac{11}{2})}\right)_{\mathbf{2}}=\left(\vartheta_1^8+14\vartheta_1^4\vartheta_2^4+\vartheta_2^8\right)\begin{pmatrix}
\vartheta_1^4+\vartheta_2^4\\
-2\sqrt{3}\vartheta_1^2\vartheta_2^2\\
\end{pmatrix}\,,\\
\nonumber&&Y_{\mathbf{3}I}^{(6)}=\frac{1}{\sqrt{2}}\left(Y_{\mathbf{\hat{2}}}^{(\frac{1}{2})}Y_{\mathbf{4'}I}^{(\frac{11}{2})}\right)_{\mathbf{3}}=\begin{pmatrix}
\vartheta_1^{12}-11\vartheta_1^8\vartheta_2^4+11\vartheta_1^4\vartheta_2^8-\vartheta_2^{12}\\
-\sqrt{2}\vartheta_1^3\vartheta_2\left(\vartheta_1^8-22\vartheta_1^4\vartheta_2^4-11\vartheta_2^8\right)\\
\sqrt{2}\vartheta_1\vartheta_2^3\left(11\vartheta_1^8+22\vartheta_1^4\vartheta_2^4-\vartheta_2^8\right)\\
\end{pmatrix}\,,\\
\nonumber&&Y_{\mathbf{3}II}^{(6)}=\frac{1}{\sqrt{2}}\left(Y_{\mathbf{\hat{2}}}^{(\frac{1}{2})}Y_{\mathbf{4'}II}^{(\frac{11}{2})}\right)_{\mathbf{3}}=\left(\vartheta_1^8+14\vartheta_2^4\vartheta_1^4+\vartheta_2^8\right)\begin{pmatrix}
\vartheta_1^4-\vartheta_2^4\\
2\sqrt{2}\vartheta_1^3\vartheta_2\\
2\sqrt{2}\vartheta_1\vartheta_2^3\\
\end{pmatrix}\,,\\
&&Y_{\mathbf{3'}}^{(6)}=\frac{1}{2\sqrt{3}}\left(Y_{\mathbf{\hat{2}}}^{(\frac{1}{2})}Y_{\mathbf{4'}I}^{(\frac{11}{2})}\right)_{\mathbf{3'}}=\vartheta_1\vartheta_2\left(\vartheta_1^4-\vartheta_2^4\right)\begin{pmatrix}
2\sqrt{2}\vartheta_1\vartheta_2\left(\vartheta_1^4+\vartheta_2^4\right)\\
\vartheta_2^6-5\vartheta_1^4\vartheta_2^2\\
5\vartheta_1^2\vartheta_2^4-\vartheta_1^6\\
\end{pmatrix}\,.
\end{eqnarray}

We summarize the modular forms of level $4$ up to weight 6 in table~\ref{Tab:Level4_MF}. If the complex modulus $\tau$ is stabilized at certain points, it would be invariant under some modular transformation and some residual modular flavor symmetry would be preserved~\cite{Novichkov:2018yse,Gui-JunDing:2019wap}. It is well-known that there are only four fixed points $\tau_S=i,~\tau_{ST}=\omega,~\tau_{TS}=-\omega^2,~\tau_T=+i\infty$ with $\omega=e^{2i\pi/3}$ in the fundamental domain of $\text{SL}_2(\mathbb{Z})$ group~\cite{Novichkov:2018yse,Gui-JunDing:2019wap}. In the following, we give the vacuum alignment of half weight modular form $Y^{(\frac{1}{2})}_\mathbf{\hat{2}}(\tau)$ at these fixed points which could be useful to modular flavor model building~\cite{Gui-JunDing:2019wap},
\begin{align}
\nonumber
& Y^{(\frac{1}{2})}_\mathbf{\hat{2}}(\tau_S)=Y_S\begin{pmatrix}
1 \\ 1-\sqrt{2}
\end{pmatrix}, ~\quad~ Y^{(\frac{1}{2})}_\mathbf{\hat{2}}(\tau_{ST})=Y_{ST}\begin{pmatrix}
\omega \\ -1-e^{\frac{5\pi i}{6}}
\end{pmatrix} \,,\\
& Y^{(\frac{1}{2})}_\mathbf{\hat{2}}(\tau_T)=Y_T\begin{pmatrix}
1 \\ 0
\end{pmatrix}, ~\quad~ Y^{(\frac{1}{2})}_\mathbf{\hat{2}}(\tau_{TS})=Y_{TS}\begin{pmatrix}
\omega^2 \\ -1+e^{\frac{\pi i}{6}}
\end{pmatrix}\,,
\label{eq:MF_vacuum}
\end{align}
with $Y_S=1.00373$,~ $Y_{ST}=-0.49567-0.85852i$,~ $Y_{TS}=-0.49567+0.85852i$ and $Y_{T}=1$. The alignments of higher weight modular forms at fixed points can be easily obtained from Eq.~\eqref{eq:MF_vacuum} and the concrete expressions of higher weight modular forms given above.

\begin{table}[t!]
\centering
\begin{tabular}{|c|c|}
\hline  \hline
Modular weight $k/2$ & Modular forms $Y^{(\frac{k}{2})}_{\mathbf{r}}$ \\[+0.05in] \hline

$k=1$ & $Y^{(\frac{1}{2})}_{\mathbf{\hat{2}}}$\\[+0.05in]  \hline

$k=2$ & $Y^{(1)}_{\mathbf{\hat{3}'}}$\\[+0.05in]  \hline

$k=3$ & $Y^{(\frac{3}{2})}_{\mathbf{4'}}$\\ [+0.05in] \hline

$k=4$ & $Y^{(2)}_{\mathbf{2}}, ~Y^{(2)}_{\mathbf{3}}$\\[+0.05in] \hline

$k=5$ & $Y^{(\frac{5}{2})}_{\mathbf{\hat{2}}}, ~Y^{(\frac{5}{2})}_{\mathbf{4}}$\\[+0.05in] \hline

$k=6$ & $Y^{(3)}_{\mathbf{\hat{1}'}},~ Y^{(3)}_{\mathbf{\hat{3}}},~Y^{(3)}_{\mathbf{\hat{3}'}}$\\[+0.05in] \hline

$k=7$ & $Y^{(\frac{7}{2})}_{\mathbf{\widetilde{2}'}},~ Y^{(\frac{7}{2})}_{\mathbf{\widetilde{2}}},~Y^{(\frac{7}{2})}_{\mathbf{4'}}$\\[+0.05in] \hline

$k=8$ & $Y^{(4)}_{\mathbf{1}},~Y^{(4)}_{\mathbf{2}},~ Y^{(4)}_{\mathbf{3}}, ~Y^{(4)}_{\mathbf{3}'}$\\ [+0.05in]\hline

$k=9$ & $Y^{(\frac{9}{2})}_{\mathbf{\hat{2}}},~Y^{(\frac{9}{2})}_{\mathbf{4}I},~ Y^{(\frac{9}{2})}_{\mathbf{4}II}$\\[+0.05in] \hline

$k=10$ & $Y^{(5)}_{\mathbf{2'}},~ Y^{(5)}_{\mathbf{\hat{3}}}, ~Y^{(5)}_{\mathbf{\hat{3}'}I}~,~Y^{(5)}_{\mathbf{\hat{3}'}II}$\\[+0.05in] \hline

$k=11$ & $Y^{(\frac{11}{2})}_{\mathbf{\widetilde{2}}},~ Y^{(\frac{11}{2})}_{\mathbf{\widetilde{2}'}}, ~Y^{(\frac{11}{2})}_{\mathbf{4'}I},~Y^{(\frac{11}{2})}_{\mathbf{4'}II}$\\[+0.05in] \hline

$k=12$ & $Y^{(6)}_{\mathbf{1}},~Y^{(6)}_{\mathbf{1}'},~Y^{(6)}_{\mathbf{2}},~Y^{(6)}_{\mathbf{3}I}~,~Y^{(6)}_{\mathbf{3}II}~,~Y^{(6)}_{\mathbf{3}'}$\\[+0.05in] \hline \hline
\end{tabular}
\caption{\label{Tab:Level4_MF} Summary of modular forms of level $N=4$ up to weight 6, the subscript $\mathbf{r}$ denotes the transformation property under the finite metaplectic group $\widetilde{S}_4$.  }
\end{table}

\section{\label{sec:modular building}Model for lepton masses and flavor mixing }

As shown in section~\ref{sec:half integral MF}, in order to consistently define half-integral weight modular forms, a multiplier is necessary and the modular group $\text{SL}_{2}(\mathbb{Z})$ should be extended to the metaplectic group $\text{Mp}_{2}(\mathbb{Z})$ which is the double covering of $\text{SL}_{2}(\mathbb{Z})$. As a result, we need to generalize the original modular invariant supersymmetric theory~\cite{Feruglio:2017spp} to metaplectic modular invariant theory.

\subsection{\label{subsec:framework}Metaplectic modular invariant theory}

We adopt the framework of the $\mathcal{N}=1$ global supersymmetry, the most general form of the action is
\begin{equation}
\mathcal{S}=\int d^4x d^2\theta d^2\bar{\theta} K(\Phi_I, \bar{\Phi}_I, \tau, \bar{\tau}) + \int d^4x d^2\theta W(\Phi_I, \tau) + h.c.\,,
\end{equation}
where $K(\Phi_I, \bar{\Phi}_I, \tau, \bar{\tau})$ is the K\"ahler potential, $W(\Phi_I , \tau)$ is the superpotential, and $\Phi_I$ denotes a set of chiral supermultiplets. The metaplectic group acts on the modulus $\tau$ and the superfield $\Phi_I$ in a certain way~\cite{Ferrara:1989bc,Ferrara:1989qb,Feruglio:2017spp}. Analogous to~\cite{Feruglio:2017spp}, we assume the supermultiplet $\Phi_I$ transforms in a representation $\rho_I$ of the finite metaplectic group $\widetilde{\Gamma}_{4N}$ with a weight $-k_I/2$,
\begin{equation}
\label{eq:metaplectic-transf}\tau \rightarrow \widetilde{\gamma}\tau=\dfrac{a\tau+b}{c\tau+d},~~~~\Phi_I \rightarrow \phi^{-k_I}(\gamma,\tau)\rho_I(\widetilde{\gamma})\Phi_I\quad \text{with} ~\widetilde{\gamma}=(\gamma, \phi(\gamma,\tau))\in\widetilde{\Gamma}\,,
\end{equation}
where $\gamma =\begin{pmatrix}
a & b \\ c & d  \end{pmatrix}$ and $\phi(\gamma,\tau)=\epsilon(c\tau+d)^{1/2}$, $\rho_I(\widetilde{\gamma})$ is the unitarity representation matrix of $\widetilde{\gamma}$, and $k_I$ is a generic non-negative integer. The supermultiplets $\Phi_I$ are not modular forms, therefore there is no restriction on the possible value of $k_I$. We can see that the combination of any two metaplectic transformations $\widetilde{\gamma}_1$ and $\widetilde{\gamma}_2$ is also a metaplectic transformation,
\begin{eqnarray}
\nonumber&&\hskip-0.2in \tau \rightarrow \widetilde{\gamma}_1(\widetilde{\gamma}_2\tau)=\gamma_1(\gamma_2\tau)=(\gamma_1\gamma_2)\tau=(\widetilde{\gamma_1}\widetilde{\gamma_2})\tau,\\
&&\hskip-0.2in \Phi_I \rightarrow \phi^{-k_I}(\gamma_1,\gamma_2\tau)\phi^{-k_I}(\gamma_2,\tau)\rho_I(\widetilde{\gamma}_1)\rho_I(\widetilde{\gamma}_2)\Phi_I=\zeta^{-1}_{k_I/2}(\gamma_1,\gamma_2)\phi^{-k_I}(\gamma_1\gamma_2,\tau)
\rho_I(\widetilde{\gamma}_1\widetilde{\gamma}_2)\Phi_I\,.
\end{eqnarray}
If we are still confined to the original modular group $\text{SL}_{2}(\mathbb{Z})$ instead of $\text{Mp}_{2}(\mathbb{Z})$, the combination of two half-integral weight modular transformations would be not equal to a third half-integral weight modular transformation due to presence of the factor $\zeta_{k_I/2}(\gamma_1,\gamma_2)$, so that the modular transformation is not well-defined. For this reason, it is insufficient to simply change the modular weight to a rational or real number when discussing modular transformations of rational or real weight, the classical modular group $\text{SL}_{2}(\mathbb{Z})$ should be enhanced to its metaplectic covering group. The action $\mathcal{S}$ is required invariant under the metaplectic transformation given in Eq.~\eqref{eq:metaplectic-transf}. The K\"ahler potential $K(\Phi_I, \bar{\Phi}_I, \tau, \bar{\tau})$ is a real gauge-invariant function of the chiral supermultiplets $\Phi_I$ and their conjugates. A minimal choice of K\"ahler potential is
\begin{equation}
K(\Phi_I, \bar{\Phi}_I, \tau, \bar{\tau})=-h\Lambda^2\log(-i\tau+i\bar{\tau})+\sum_I(-i\tau+i\bar{\tau})^{-\frac{k_I}{2}}|\Phi_I|^2\,,
\end{equation}
where $h$ is a positive constant. $K(\Phi_I, \bar{\Phi}_I, \tau, \bar{\tau})$ is invariant up to a K\"ahler transformation under the metaplectic transformation. The superpotential $W(\Phi_I , \tau)$ can be expanded in power series of the supermultiplets $\Phi_I$:
\begin{equation}
W(\Phi_I, \tau)= \sum_n Y_{I_1\dots I_n}(\tau) \Phi_{I_1}\dots \Phi_{I_n}\,.
\end{equation}
Invariance of $W(\Phi_I, \tau)$ under the metaplectic transformation in Eq.~\eqref{eq:metaplectic-transf} entails that the function $Y_{I_1\dots I_n}(\tau)$ should be a modular form of weight $k_Y/2$ and level $4N$ transforming in the
representation $\rho_{Y}$ of $\widetilde{\Gamma}_{4N}$,
\begin{equation}
Y_{I_1\dots I_n}(\gamma\tau)=\phi^{k_Y}(\gamma,\tau)\rho_Y(\widetilde{\gamma})Y_{I_1\dots I_n}(\tau), ~\quad~ \widetilde{\gamma}=(\gamma, \phi(\gamma,\tau))\,.
\end{equation}
The modular weight $k_Y/2$ and the representation $\rho_{Y}$ should satisfy the conditions
\begin{eqnarray}
\nonumber&&k_Y=k_{I_1}+\dots +k_{I_n}\,, \\
&&\rho_Y \otimes \rho_{I_1} \otimes \dots \otimes \rho_{I_n} \supset  \mathbf{1}\,,
\end{eqnarray}
where $\mathbf{1}$ refers to invariant singlet of $\widetilde{\Gamma}_{4N}$.

\subsection{Models based on $\widetilde{S}_4$}

In this section, we shall construct lepton models based on the finite  metaplectic modular group $\widetilde{\Gamma}_4\cong \widetilde{S}_4$. In the representations $\mathbf{1}$, $\mathbf{1^{\prime}}$, $\mathbf{\hat{1}}$, $\mathbf{\hat{1}^{\prime}}$, $\mathbf{2}$, $\mathbf{2}'$,  $\mathbf{3}$, $\mathbf{3^{\prime}}$, $\mathbf{\hat{3}}$ and  $\mathbf{\hat{3}^{\prime}}$, the generator $R=\pm1$ and therefore $\widetilde{S}_4$ and the homogeneous finite modular group $S'_4$ are represented by the same set of matrices. As a consequence, all the $S'_4$ modular models obviously can be reproduced from the metaplectic modular group $\widetilde{S}_4$, in particular the successful $S'_4$ models given in our previous work~\cite{Liu:2020akv} can be obtained here. In the following, we shall explore new models beyond $S'_4$, and half-integral weight modular forms would be involved.
The neutrino masses originate from the effective Weinberg operator or the type-I seesaw mechanism, and both scenarios of three right-handed neutrinos and two right-handed neutrinos are considered in the type-I seesaw mechanism. Three models would be presented in the following, the field content and their transformation properties and weights are summarized in table~\ref{tab:model}.

\begin{table}[t!]
\centering
\begin{tabular}{|c|c||c|c|c|c|c|c|c|c|c|} \hline\hline
\multicolumn{2}{|c|}{} & $E_1^c$ & $E_2^c$ & $E_3^c$ & $L$ & $N_1^c$ & $N_2^c$ & $N_3^c$ & $H_{u}$ & $H_{d}$ \rule[1ex]{0pt}{1ex} \\ \hline

\multirow{2}{*}{\textbf{Model I}} & $\widetilde{S}_4$ &  \multicolumn{2}{|c|}{$\mathbf{\widehat{2}}$} & $\mathbf{1}$ &  $\mathbf{\widehat{3}}$ & --- & --- & --- & $\mathbf{1}$ & $\mathbf{1}$ \rule[1ex]{0pt}{1ex} \\ \cline{2-11}
  & $k_I/2$  &\multicolumn{2}{|c|}{$3/2$}  & $0$  & $1$ & --- & ---  & ---  & 0 & 0 \rule[1ex]{0pt}{1ex}\\ \hline

\multirow{2}{*}{\textbf{ Model II}} & $\widetilde{S}_4$ &  \multicolumn{2}{|c|}{$\mathbf{\widetilde{2}}$} & $\mathbf{\widehat{1}'}$ &  $\mathbf{\widehat{3}}$ & \multicolumn{3}{|c|}{$\mathbf{3'}$} & $\mathbf{1}$ & $\mathbf{1}$ \rule[1ex]{0pt}{1ex} \\ \cline{2-11}
  & $k_I/2$  &\multicolumn{2}{|c|}{$3/2$}  & $0$  & $1$ & \multicolumn{3}{|c|}{$1$} & $0$ & $0$ \rule[1ex]{0pt}{1ex}\\ \hline

\multirow{2}{*}{\textbf{ Model III}} & $\widetilde{S}_4$ &  $\mathbf{1}$  &  $\mathbf{1}$ & $\mathbf{\widehat{1}}$ &  $\mathbf{3}$ & \multicolumn{2}{|c|}{$\mathbf{\widetilde{2}}$} & --- & $\mathbf{1}$ & $\mathbf{1}$ \rule[1ex]{0pt}{1ex} \\ \cline{2-11}
  & $k_I/2$  & $1$ & $3$ & $4$  & $1$ & \multicolumn{2}{|c|}{$3/2$} & --- & $0$ & $0$  \rule[1ex]{0pt}{1ex}\\ \hline \hline

\end{tabular}
\caption{\label{tab:model} Transformation properties of the leptonic matter fields under the finite metaplectic group $\widetilde{\Gamma}_4$, and the modular weight $k_I/2$ assignments for each model. The Higgs fields $H_u$ and $H_d$ are assigned to be invariant singlet $\mathbf{1}$ of $\widetilde{S}_4$ with vanishing modular weight.  }
\end{table}

\subsubsection{Model I: neutrino masses from Weinberg operator}
In this model, the neutrino masses are described by the Weinberg operator. The left-handed leptons $L$ are assigned to a triplet $\mathbf{\widehat{3}}$, the first two generations of right-handed charged leptons $E^c_1$ and $E^c_2$ transform as a doublet $\mathbf{\widehat{2}}$ of $\widetilde{S}_4$. For convenience, we use the subscript ``$D$'' to denote the doublet assignment, i.e. $E^c_D\equiv (E^c_1,~ E^c_2)^T $. The third right-handed charged lepton $E^c_{3}$ is invariant under $\widetilde{S}_4$.
The modular weights of the lepton superfields are set to
\begin{equation}
\label{eq:wtI}k_{L}/2=1\,,\quad k_{E^c_D}/2=\frac{3}{2}\,,\quad k_{E^c_3}/2=0\,.
\end{equation}
Thus the modular invariant superpotentials for lepton masses read as follows,
\begin{eqnarray}
\nonumber&&\hskip-0.2in \mathcal{W}_{e}=\alpha (E^{c}_{D}L  Y_{\mathbf{\widehat{2}}}^{\left(\frac{5}{2}\right)})_{\mathbf{1}}H_{d}+\beta ( E^{c}_{D}L Y_{\mathbf{4}}^{\left(\frac{5}{2}\right)})_{\mathbf{1}}H_{d} +\gamma (E^{c}_{3}L Y_{\mathbf{\widehat{3}'}}^{(1)})_{\mathbf{1}}H_{d}\,, \\
&&\hskip-0.2in \mathcal{W}_{\nu}=\frac{g_1}{\Lambda}(LL Y_{\mathbf{2}}^{(2)})_{\mathbf{1}}H_{u}H_{u}+\frac{g_2}{\Lambda }(LL Y_{\mathbf{3}}^{(2)})_{\mathbf{1}}H_{u}H_{u} \,.
\label{eq:We1}
\end{eqnarray}
With the CG coefficients of $\widetilde{S}_4$ group in Appendix~\ref{sec:app-S4QC}, the charged lepton and neutrino mass matrices read as
\begin{eqnarray}
\label{eq:MI}
\nonumber M_{e}&=&\begin{pmatrix}
 \alpha  Y_{\mathbf{\widehat{2}},2}^{\left(\frac{5}{2}\right)}-\sqrt{2} \beta  Y_{\mathbf{4},3}^{\left(\frac{5}{2}\right)} ~& \sqrt{2} \alpha  Y_{\mathbf{\widehat{2}},1}^{\left(\frac{5}{2}\right)}-\beta  Y_{\mathbf{4},1}^{\left(\frac{5}{2}\right)} ~& -\sqrt{3} \beta  Y_{\mathbf{4},2}^{\left(\frac{5}{2}\right)} \\
 \alpha  Y_{\mathbf{\widehat{2}},1}^{\left(\frac{5}{2}\right)}+\sqrt{2} \beta  Y_{\mathbf{4},1}^{\left(\frac{5}{2}\right)} ~& -\sqrt{3} \beta  Y_{\mathbf{4},4}^{\left(\frac{5}{2}\right)} ~& -\sqrt{2} \alpha  Y_{\mathbf{\widehat{2}},2}^{\left(\frac{5}{2}\right)} -\beta  Y_{\mathbf{4},3}^{\left(\frac{5}{2}\right)}\\
 \gamma  Y_{\mathbf{\widehat{3}'},1}^{(1)} ~& \gamma  Y_{\mathbf{\widehat{3}'},3}^{(1)} ~& \gamma  Y_{\mathbf{\widehat{3}'},2}^{(1)} \\
\end{pmatrix}v_{d}\,,\quad\\
M_{\nu }&=&\begin{pmatrix}
 -2 g_1 Y_{\mathbf{2},2}^{(2)} ~& g_2 Y_{\mathbf{3},2}^{(2)} ~& -g_2 Y_{\mathbf{3},3}^{(2)} \\
 g_2 Y_{\mathbf{3},2}^{(2)} ~& \sqrt{3} g_1 Y_{\mathbf{2},1}^{(2)}+g_2 Y_{\mathbf{3},1}^{(2)} ~& g_1 Y_{\mathbf{2},2}^{(2)} \\
 -g_2 Y_{\mathbf{3},3}^{(2)} ~& g_1 Y_{\mathbf{2},2}^{(2)} ~& \sqrt{3} g_1 Y_{\mathbf{2},1}^{(2)}-g_2 Y_{\mathbf{3},1}^{(2)} \\
\end{pmatrix}\frac{v_{u}^2}{\Lambda }\,.
\end{eqnarray}
where $Y_{\mathbf{r},n}^{(w)}$ denotes the $n$th component of weight $w$ modular multiplets $Y_{\mathbf{r}}^{(w)}$. The phases of $\alpha$, $\gamma$ and $g_1$ can be removed by field redefinition while $\beta$ and $g_2$ are generally complex numbers. Thus this model makes use of three real positive parameters $\alpha$, $\gamma$, $g_1$ and two complex parameter $\beta$, $g_{2}$ to describe lepton masses and PMNS matrix. A good agreement between data and predictions is obtained for the following values of the free parameters:
\begin{equation}
\begin{gathered}
\langle\tau\rangle=0.098664+1.0034i\,,\quad |\beta/\alpha|=1.4141\,,\quad \arg(\beta/\alpha)=1.9967\pi\,,\\
\gamma/\alpha=69.8216\,,\quad|g_2/g_1|=0.64756\,,\quad \arg(g_2/g_1)=0.5717\pi\,,\\
\alpha v_d=15.8450~\text{MeV},\quad \frac{g_1v_u^2}{\Lambda}=22.96~\text{meV}\,.
\end{gathered}
\end{equation}
The lepton mixing parameters and neutrino masses are determined to be
\begin{eqnarray}
\nonumber&&\sin^2 \theta_{12}=0.30990\,,~~ \sin^2 \theta_{13}=0.022376\,,~~ \sin^2 \theta_{23}=0.56274\,,~~ \delta_{CP}=1.60547\pi\,,\\
\nonumber&& \alpha_{21}=0.18318\pi\,,~~ \alpha_{31}=0.23584\pi\,,~~ m_{1}=31.523~\text{meV}\,,~~ m_{2}=32.674~\text{meV}\,,\\
&& m_{3}=59.354~\text{meV}\,,~~~~ m_\beta=32.7557~\text{meV} \,,~~~~ m_{\beta\beta}=28.7411~\text{meV}\,.
\end{eqnarray}
It is remarkable that all observables are within the $1\sigma$ experimental range~\cite{Esteban:2018azc}. Here we adopt the particle data group convention for the mixing angles and CP violation phases~\cite{Tanabashi:2018oca}. The lepton mixing matrix in the standard parametrization is written as
\begin{equation}
U=\left(\begin{array}{ccc}
c_{12} c_{13} ~& s_{12} c_{13} ~& s_{13} e^{-i \delta_{C P}} \\
-s_{12} c_{23}-c_{12} s_{13} s_{23} e^{i \delta_{C P}} ~& c_{12} c_{23}-s_{12} s_{13} s_{23} e^{i \delta_{C P}} ~& c_{13} s_{23} \\
s_{12} s_{23}-c_{12} s_{13} c_{23} e^{i \delta_{C P}} ~& -c_{12} s_{23}-s_{12} s_{13} c_{23} e^{i \delta_{C P}} ~& c_{13} c_{23}
\end{array}\right) Q\,,
\end{equation}
where $c_{ij}\equiv\cos \theta_{ij}, s_{ij}\equiv\sin\theta_{ij}, \delta_{CP}$ is the Dirac CP phase, and $Q$ is a diagonal Majorana phase matrix. If all of the three neutrinos have nonzero masses, then the phase matrix is given by $Q=\operatorname{diag}(1, e^{i \frac{\alpha_{21}}{2}}, e^{i \frac{\alpha_{31}}{2}})$. If the lightest neutrino is predicted to be massless, i.e., $m_1=0$ (see model IV), one of the Majorana neutrino phase is unphysical, and we can parameterize the phase matrix as $Q=\operatorname{diag}(1, e^{i \phi}, 1)$, where $\alpha_{21},\alpha_{31}$ or $\phi$ are called Majorana CP phases. From the predicted values of mixing angles and neutrino masses, one can determine the effective neutrino masses $m_{\beta}$ probed by direct kinematic search in beta decay and the effective mass $m_{\beta\beta}$ in neutrinoless double beta decay. Note that $m_{\beta}$ is independent of the CP violation phase and it is defined by
\begin{align}
&m_\beta=\sqrt{m^2_1 \cos^2\theta_{12}\cos^2\theta_{13}+m^2_2\sin^2\theta_{12}\cos^2\theta_{13}+m^2_3\sin^2\theta_{13} }\,.
\end{align}
The latest bound is $m_{\beta}<1.1$ eV at $90\%$ confidence level from KATRIN~\cite{Aker:2019uuj}. Our prediction $m_\beta=32.7557~\text{meV}$ is far below the future sensitivity of KATRIN. The decay rate of the neutrinoless double beta decay is proportional to the square of the effective Majorana mass $m_{\beta\beta}$ defined as
\begin{equation}
m_{\beta\beta}=\left|m_{1} \cos ^{2} \theta_{12} \cos ^{2} \theta_{13}+m_{2} \sin ^{2} \theta_{12} \cos ^{2} \theta_{13} e^{i \alpha_{21}}+m_{3} \sin ^{2} \theta_{13} e^{i\left(\alpha_{31}-2 \delta_{C P}\right)}\right|\,,
\end{equation}
in which all mixing parameters except $\theta_{23}$ are involved. For $m_1=0$, it can be simply written as
\begin{equation}
m_{\beta\beta}=\left|m_{2} \sin ^{2} \theta_{12} \cos ^{2} \theta_{13} e^{i \phi}+m_{3} \sin ^{2} \theta_{13} e^{-i2 \delta_{C P}}\right|\,.
\end{equation}
The current experimental bound from KamLAND-Zen is $m_{\beta\beta} < (61-165)~\text{meV}$~\cite{KamLAND-Zen:2016pfg}. The predicted value $m_{\beta\beta}=28.7411~\text{meV}$ is within the reach of the next generation high sensitive neutrinoless double beta decay experiments.
The most stringent bound on the neutrino mass sum is $\sum_i m_i<120$ meV at $95\%$ confidence level from Plank~\cite{Aghanim:2018eyx}.
In this model, the neutrino mass sum is predicted to be $\sum_i m_i=123.551$ \text{meV} which is slightly larger than the upper limit of Planck.

\subsubsection{Model II: neutrino masses from seesaw mechanism with three right-handed neutrinos}
The neutrino masses arise from type-I seesaw mechanism in this model, and three right-handed neutrinos are introduced to transform as a triplet $\mathbf{3'}$ of $\widetilde{S}_4$. The left-handed leptons $L$ are assigned to a triplet $\mathbf{3}$, the right-handed charged leptons $E^c_D$ and $E^c_{3}$ are assigned to transform as $\mathbf{\widetilde{2}}$ and $\mathbf{\widehat{1}}'$ respectively under $\widetilde{S}_4$.
We take the modular weights of the lepton superfields to be
\begin{equation}
\label{eq:wtIII}k_{L}/2=1\,,\quad k_{E^c_D}/2=\frac{3}{2}\,,\quad k_{E^c_3}/2=4\,,\quad k_{N^c}/2=1\,.
\end{equation}
The superpotentials for lepton masses are given by,
\begin{eqnarray}
\nonumber&&\hskip-0.2in \mathcal{W}_{e}=\alpha (E^{c}_{D}L Y_{\mathbf{\widehat{2}}}^{\left(\frac{5}{2}\right)})_{\mathbf{1}}H_{d}+\beta (E^{c}_{D} L Y_{\mathbf{4}}^{\left(\frac{5}{2}\right)})_{\mathbf{1}}H_{d}+\gamma (E^{c}_{3} L Y_{\mathbf{\widehat{3}}}^{(5)})_{\mathbf{1}}H_{d}\,, \\
&&\hskip-0.2in  \mathcal{W}_{\nu}=g_1(N^{c}L Y_{\mathbf{2}}^{(2)})_{\mathbf{1}}H_{u}+g_2(N^{c}L Y_{\mathbf{3}}^{(2)})_{\mathbf{1}}H_{u}+\Lambda(N^{c}N^{c} Y_{\mathbf{2}}^{(2)})_{\mathbf{1}}\,,
\label{eq:We3}
\end{eqnarray}
which lead to the following charged lepton and neutrino mass matrices
\begin{eqnarray}
\label{eq:MIII}
\nonumber M_{e}&=&\begin{pmatrix}
 \alpha  Y_{\mathbf{\widehat{2}},2}^{\left(\frac{5}{2}\right)}-\sqrt{2} \beta  Y_{\mathbf{4},3}^{\left(\frac{5}{2}\right)} ~& \sqrt{2} \alpha  Y_{\mathbf{\widehat{2}},1}^{\left(\frac{5}{2}\right)}-\beta  Y_{\mathbf{4},1}^{\left(\frac{5}{2}\right)} ~& -\sqrt{3} \beta  Y_{\mathbf{4},2}^{\left(\frac{5}{2}\right)} \\
 \alpha  Y_{\mathbf{\widehat{2}},1}^{\left(\frac{5}{2}\right)}+\sqrt{2} \beta  Y_{\mathbf{4},1}^{\left(\frac{5}{2}\right)} ~& -\sqrt{3} \beta  Y_{\mathbf{4},4}^{\left(\frac{5}{2}\right)} ~& -\sqrt{2} \alpha  Y_{\mathbf{\widehat{2}},2}^{\left(\frac{5}{2}\right)}-\beta  Y_{\mathbf{4},3}^{\left(\frac{5}{2}\right)} \\
 \gamma  Y_{\mathbf{\widehat{3}},1}^{(5)} ~& \gamma  Y_{\mathbf{\widehat{3}},3}^{(5)} ~& \gamma  Y_{\mathbf{\widehat{3}},2}^{(5)}
\end{pmatrix}v_{d}\,,\\
\nonumber  M_{N}&=&\begin{pmatrix}
 2 Y_{\mathbf{2},1}^{(2)} ~& 0 ~& 0 \\
 0 ~& \sqrt{3} Y_{\mathbf{2},2}^{(2)} ~& -Y_{\mathbf{2},1}^{(2)} \\
 0 ~& -Y_{\mathbf{2},1}^{(2)} ~& \sqrt{3} Y_{\mathbf{2},2}^{(2)}
\end{pmatrix}\Lambda\,,  \\
  M_{D}&=&\begin{pmatrix}
 -2 g_1 Y_{\mathbf{2},2}^{(2)} ~& g_2 Y_{\mathbf{3},2}^{(2)} ~& -g_2 Y_{\mathbf{3},3}^{(2)} \\
 g_2 Y_{\mathbf{3},2}^{(2)} ~& \sqrt{3} g_1 Y_{\mathbf{2},1}^{(2)}+g_2 Y_{\mathbf{3},1}^{(2)} ~& g_1 Y_{\mathbf{2},2}^{(2)} \\
 -g_2 Y_{\mathbf{3},3}^{(2)} ~& g_1 Y_{\mathbf{2},2}^{(2)} ~& \sqrt{3} g_1 Y_{\mathbf{2},1}^{(2)}-g_2 Y_{\mathbf{3},1}^{(2)}
\end{pmatrix}v_{u} \,.
\end{eqnarray}
The parameters $\alpha$, $\gamma$ and $g_1$ can be taken real since their phases are unphysical, nevertheless the phases of $\beta$ and $g_2$ can not be absorbed into lepton fields. We get a good agreement between the model and the data by the parameter choice:
\begin{equation}
\begin{gathered}
\langle\tau\rangle=-0.19974+ 1.13171i\,,\quad \gamma=51.0076\,,\quad
|\beta/\alpha|= 1.4142\,,\\
|g_2/g_1|=2.4263\,,~~\arg(\beta/\alpha)=0.004255\pi\,,~~\arg(g_2/g_1)=0.95715\pi\,,\\
~\alpha v_d=22.8482~\text{MeV},~~~~~\frac{g_1v_u^2}{\Lambda}=2.2593~\text{meV}\,.
\end{gathered}
\end{equation}
Accordingly the predictions for the lepton mixing parameters and neutrino masses are given by
\begin{eqnarray}
\nonumber&&\sin^2 \theta_{12}=0.30997\,,~~\sin^2 \theta_{13}=0.022370\,,~~ \sin^2 \theta_{23}=0.56311\,,~~ \delta_{CP}=1.14879\pi\,,\\
\nonumber&&\alpha_{21}=1.1277\pi\,,~~ \alpha_{31}=0.84159\pi\,,~~ m_{1}=6.6056~\text{meV}\,, ~~ m_{2}=10.841~\text{meV}\,,\\
&& m_{3}=50.709~\text{meV}\,,~~~~m_\beta=11.0715~\text{meV} \,,~~~~ m_{\beta\beta}=1.2851~\text{meV}\,.
\end{eqnarray}
which are compatible with the experimental data at $1\sigma$ level~\cite{Esteban:2018azc}. Notice that the neutrino masses are hierarchical normal ordering, the Planck bound $\sum_im_{i}<120$ meV is satisfied, and $m_{\beta}$ and $m_{\beta\beta}$ are quite tiny.
\subsubsection{Model III: neutrino masses from seesaw mechanism with two right-handed neutrinos }

\begin{figure}[ht]
\centering
\includegraphics[width=6.5in]{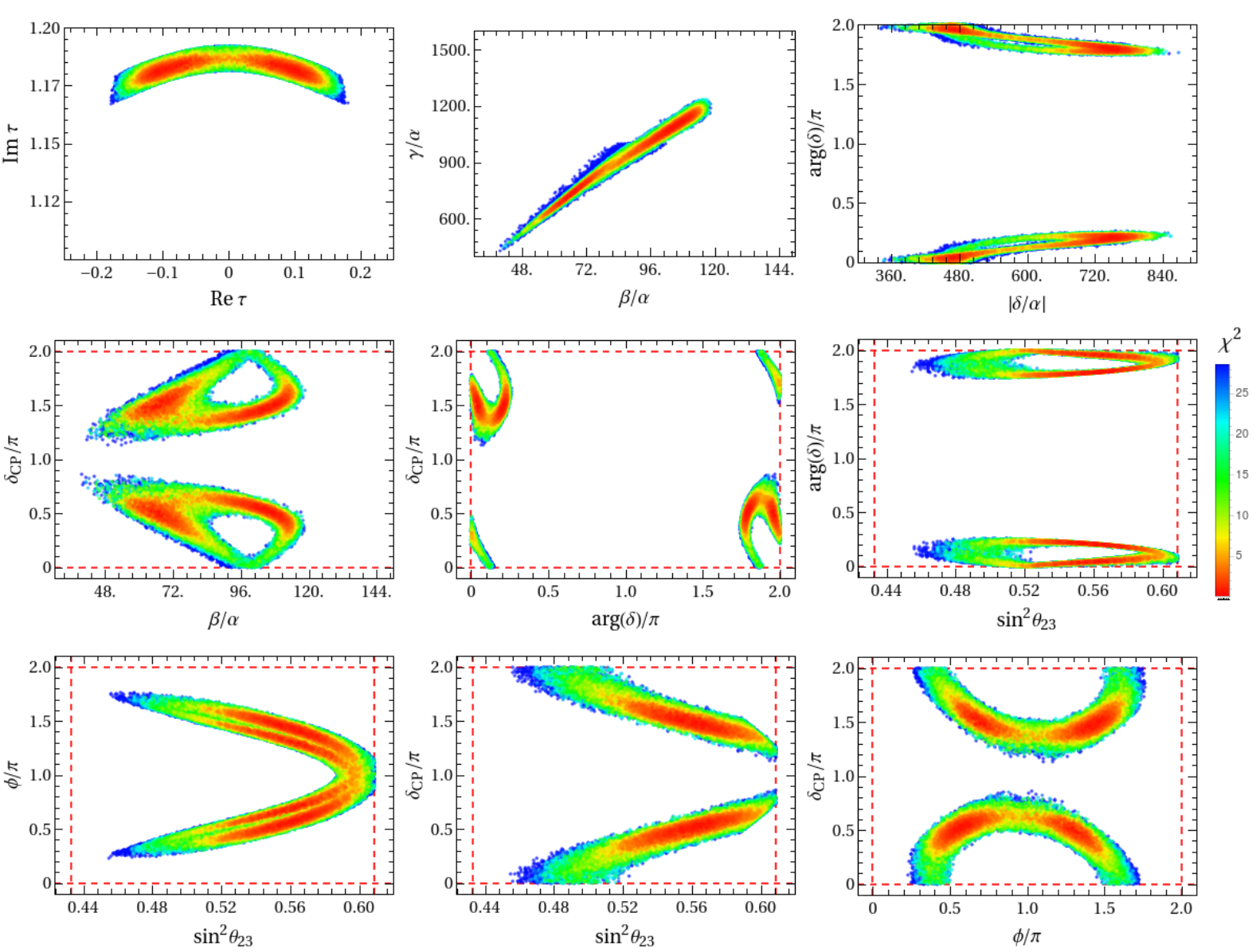}
\caption{The predicted correlations among the input free parameters, neutrino mixing angles and CP violating phases in the \textbf{Model III}.
The plot displays only the points corresponding to choices of
the parameters reproducing $\Delta m^2_{21}$, $\Delta m^2_{31}$
and all the three mixing angles within the $3\sigma$ regions~\cite{Esteban:2018azc}. }
\label{fig:model4}
\end{figure}

The neutrino masses are described by the minimal seesaw model with two right-handed neutrinos. We assign the left-handed leptons $L$ to a triplet $\mathbf{3}$ of $\widetilde{S}_4$, only two right-handed neutrinos are introduced and they are assumed to transform as a doublet $\mathbf{\widetilde{2}'}$ under $\widetilde{S}_4$, while the right-handed charged leptons $E^c_{1}$, $E^c_{2}$ and $E^c_{3}$ transform as singlets $\mathbf{1}$, $\mathbf{1}$ and $\mathbf{\widehat{1}}$ respectively. We choose the modular weights of lepton fields are
\begin{equation}
\label{eq:wtIV}k_L/2=1,\quad  k_{N^{c}}/2=\frac{3}{2},\quad  k_{E_{1}^{c}}/2=1,\quad  k_{E_{2}^{c}}/2=3,\quad  k_{E_{3}^{c}}/2=4\,.
\end{equation}
The masses of the charged leptons and neutrino are described by the following superpotential,
\begin{eqnarray}
\nonumber&&\hskip-0.2in \mathcal{W}_{e}=\alpha ( E^{c}_{1}LY_{\mathbf{3}}^{(2)} )_{\mathbf{1}}H_{d}+\beta (E^{c}_{2}L Y_{\mathbf{3}}^{(4)})_{\mathbf{1}}H_{d}+\gamma ( E^{c}_{3}LY_{\mathbf{\widehat{3}'},I}^{(5)})_{\mathbf{1}}H_{d}+\delta (E^{c}_{3}L Y_{\mathbf{\widehat{3}'},II}^{(5)})_{\mathbf{1}}H_{d}\,, \\
&&\hskip-0.2in  \mathcal{W}_{\nu}=g(N^{c}L Y_{\mathbf{4}}^{\left(\frac{5}{2}\right)})_{\mathbf{1}}H_{u}+\Lambda(N^{c}N^{c} Y_{\widehat{3}'}^{(3)})_{\mathbf{1}}\,.
\label{eq:We4}
\end{eqnarray}
Notice that the term $(N^{c}N^{c} Y_{\mathbf{\widehat{1}'}}^{(3)})_{\mathbf{1}}$ is allowed by symmetries of the model  but give a vanishing contribution because of the antisymmetric CG coefficient for the contraction $\mathbf{\widetilde{2}'}\otimes\mathbf{\widetilde{2}'}\to\mathbf{\widehat{1}}$.
We find the charged lepton and neutrino mass matrices are given by
\begin{eqnarray}
\nonumber M_{e}&=&\begin{pmatrix}
\alpha  Y_{\mathbf{3},1}^{(2)} ~& \alpha  Y_{\mathbf{3},3}^{(2)} ~& \alpha  Y_{\mathbf{3},2}^{(2)} \\
\beta  Y_{\mathbf{3},1}^{(4)} ~& \beta  Y_{\mathbf{3},3}^{(4)} ~& \beta Y_{\mathbf{3},2}^{(4)} \\
\gamma  Y_{\mathbf{\widehat{3}'}I,1}^{(5)}+\delta  Y_{\mathbf{\widehat{3}'}II,1}^{(5)} ~& \gamma  Y_{\mathbf{\widehat{3}'}I,3}^{(5)}+\delta  Y_{\mathbf{\widehat{3}'}II,3}^{(5)} ~& \gamma  Y_{\mathbf{\widehat{3}'}I,2}^{(5)}+\delta  Y_{\mathbf{\widehat{3}'}II,2}^{(5)} \\
\end{pmatrix}v_{d}\,,\\
\label{eq:MIV} M_{N}&=&\begin{pmatrix}
 \sqrt{2} Y_{\mathbf{\widehat{3}'},2}^{(3)} ~& Y_{\mathbf{\widehat{3}'},1}^{(3)} \\
 Y_{\mathbf{\widehat{3}'},1}^{(3)} ~& -\sqrt{2} Y_{\mathbf{\widehat{3}'},3}^{(3)} \\
\end{pmatrix}\Lambda\,,~~~
 M_{D}=g\begin{pmatrix}
 -\sqrt{2} Y_{\mathbf{4},4}^{\left(\frac{5}{2}\right)} ~& -Y_{\mathbf{4},2}^{\left(\frac{5}{2}\right)} ~& \sqrt{3} Y_{\mathbf{4},1}^{\left(\frac{5}{2}\right)} \\
 \sqrt{2} Y_{\mathbf{4},2}^{\left(\frac{5}{2}\right)} ~& \sqrt{3} Y_{\mathbf{4},3}^{\left(\frac{5}{2}\right)} ~& -Y_{\mathbf{4},4}^{\left(\frac{5}{2}\right)}
\end{pmatrix}v_{u}\,.
\end{eqnarray}
The parameters $\alpha$, $\beta$, and $\gamma$ are taken to be real and positive by rephasing right-handed charged lepton fields without loss of generality while the phase of $\delta$ can not be removed. The light neutrino mass matrix only depends on the complex modulus $\tau$ besides the overall mass scale $g^2v^2_u/\Lambda$. We numerically scan over the parameter space, the parameters $\alpha$, $\beta$, $\gamma$ and $|\delta|$ are treated as random numbers between 0 and 2000, the phase of $\delta$ freely varies in the range of 0 and $2\pi$. Since normal ordering neutrino mass spectrum is slightly favored over the inverted ordering spectrum, we will focus on normal ordering case. The best fit values of the input parameters are determined to be
\begin{equation}
\begin{gathered}
\langle\tau\rangle=0.071297+ 1.18399i\,,\quad \beta/\alpha=66.2359\,,\quad \gamma/\alpha=729.8477\,,\\
|\delta|/\alpha=470.1402\,,~ \arg(\delta)=0.043426\pi\,,~ \alpha v_d=1.0589~\text{MeV}\,,~ \frac{gv_u^2}{\Lambda}=22.040~\text{meV}\,.
\end{gathered}
\end{equation}
The lepton mixing parameters and neutrino masses are predicted to be
\begin{eqnarray}
\nonumber&&\sin^2 \theta_{12}=0.31000\,,\quad \sin^2 \theta_{13}=0.022370\,,\quad\sin^2 \theta_{23}=0.56300\,,\\
\nonumber&&\delta_{CP}=1.49646\pi\,,\quad \phi=0.701615\pi\,,\quad m_{1}=0~\text{eV}\,,~~m_{2}=8.5965~\text{meV}\,,\\
&& m_{3}=50.279~\text{meV}\,,\quad ~m_\beta=8.8851~\text{meV}\,,\quad ~ m_{\beta\beta}=3.3787~\text{meV}\,,
\end{eqnarray}
which are in the experimentally preferred $1\sigma$ range~\cite{Esteban:2018azc}, and both effective neutrino masses $m_{\beta}$ and $m_{\beta\beta}$ are far below the sensitivity of forthcoming experiments. Notice that the lightest neutrino is always massless with $m_1=0$ because only two right-handed neutrinos are introduced. Consequently the bound on neutrino mass sum from Planck is fulfilled.
We require all the three lepton mixing angles and the neutrino mass squared splittings $\Delta m^2_{21}$ and $\Delta m^2_{31}$ are in the experimentally preferred $3\sigma$ ranges~\cite{Esteban:2018azc}.
The correlations between the input parameters and observables are shown in figure~\ref{fig:model4}. We find that the mixing angle $\theta_{23}$ and the Dirac CP violation phase $\delta_{CP}$ and the Majorana phase $\phi$ are strongly correlated with each other, and the values of $\delta_{CP}$ around $\pm0.5\pi$ is preferred.

\section{\label{sec:summary}Summary and conclusions}

In the present work, we have extended the modular invariance approach to include the half-integral weight modular forms. It is highly nontrivial to
generalize integral weight modular forms to non-integral case, and
a multiplier system is generally necessary for the consistency definition of non-integral weight modular forms. In order to discuss the action of the full modular group on the half-integral modular forms, one should extend the modular group $\text{SL}_2(\mathbb{Z})$ to the metaplectic group $\text{Mp}_2(\mathbb{Z})$ which is the double covering of $\text{SL}_2(\mathbb{Z})$. As a result, we need to generalize the framework of modular invariant theory to the metaplectic modular invariant theory. Each modular multiplet is specified by its modula weight and the transformation under the finite metaplectic group. We show that the half-integral weight modular forms for the congruence subgroup $\Gamma(4N)$ can be arranged into irreducible multiplets of finite metaplectic group $\widetilde{\Gamma}_{4N}$,
which is the double covering of the homogeneous finite modular group $\Gamma'_{4N}$. We have considered the simplest case of level $4N=4$ in the context of metaplectic modular invariance approach. The half-integral weight modular forms up to weight 6 are constructed in terms of the Jacobi theta functions, and they are decomposed into different irreducible multiplets of $\widetilde{\Gamma}_4$. It is notable that the odd integral weight modular forms are in the representations $\widehat{\mathbf{1}}$, $\mathbf{\widehat{1}'}$, $\mathbf{2'}$, $\widehat{\mathbf{3}}$ and $\mathbf{\widehat{3}'}$, the even integral weight modular forms are in the representations $\mathbf{1}$, $\mathbf{1'}$, $\mathbf{2}$, $\mathbf{3}$ and $\mathbf{3'}$, while the modular forms of weight $n+1/2$ with $n$ a generic non-negative integer are in the representations $\mathbf{\widehat{2}}$, $\mathbf{\widehat{2}'}$, $\mathbf{\widetilde{2}}$, $\mathbf{\widetilde{2}'}$, $\mathbf{4}$ and $\mathbf{4'}$. It is worth noting that in the top-down approach from string theory, the wavefunctions of zero modes and massive modes on the magnetized torus behave as modular forms of weight $1/2$~\cite{Kikuchi:2020frp}.

We present three typical models based on the finite metaplectic group $\widetilde{\Gamma}_4\equiv \widetilde{S}_4$. The neutrino masses are described by the effective Weinberg operator in the \textbf{Model I}, and neutrino masses arise from type I seesaw mechanism in \textbf{Model II} and \textbf{Model III}, and three right-handed neutrinos and two right-handed neutrinos are introduced in \textbf{Model II} and \textbf{Model III} respectively. The structure of these models are rather simple, and there are no additional flavons except the complex modulus $\tau$. The half-integral weight modular forms are involved in either neutrino Yukawa couplings or charged lepton Yukawa couplings. Each model is analyzed numerically, the predictions are in excellent agreement with the experimental data on neutrino oscillation, beta decay, neutrinoless double decay and cosmology. Finally, we perform a comprehensive numerical scan over the parameter space of \textbf{Model III}, some interesting correlations between free parameters and observables are shown in figure~\ref{fig:model4}. Since the $\widetilde{S}_4$ modular symmetry can describe the lepton sector very well, it is interesting to apply $\widetilde{S}_4$ to explain the hierarchical quark masses and CKM mixing matrix.

In a similar fashion, other rational weight modular forms can be studied, and one needs to determine the corresponding metaplectic covering group
to remove the ambiguity of multi-valued branches induced by rational powers. It is very lucky that the rational weight modular forms for the principal congruence subgroup of level odd $N\geq5$ have been constructed by mathematicians~\cite{ibukiyama2000modular,ibukiyama2020graded}, as summarized in table~\ref{tab:MFspace and finite group}. There are no conceptual and mathematical obstacles to perform an analysis similar to the present work, although the group order of the corresponding metaplectic finite group is large. It is promising that the real weight modular forms can also be discussed in an analogous manner~\cite{aoki2017jacobi}. It is
fascinating to use the simplest nontrivial case of real weight modular forms to understand the standard model flavor puzzle in future.

It is known that the K\"ahler potential is not completely fixed by the modular symmetry~\cite{Chen:2019ewa}, nevertheless it could be strongly constrained in the top-down approach combining the modular flavor symmetry with traditional flavor symmetry\cite{Baur:2019kwi,Baur:2019iai,Nilles:2020nnc,Nilles:2020kgo}. The K\"ahler potential is also less constrained the metaplectic modular symmetry. We expect that the K\"ahler potential as well as the structure of the model should also be severely restricted if the metaplectic flavor symmetry is combined with traditional flavor symmetry. Moreover, it is well established that the CP transformation consistent with modular group $\text{SL}_2(\mathbb{Z})$ is uniquely $\tau\rightarrow-\tau^{*}$ up to modular transformations\cite{Novichkov:2019sqv,Baur:2019kwi,Acharya:1995ag,Dent:2001cc,Giedt:2002ns}, and CP is conserved if the value of modulus $\tau$ is pure imaginary or at the border of the fundamental domain. The modular group  $\text{SL}_2(\mathbb{Z})$ is extended to the metaplectic group $\text{Mp}_2(\mathbb{Z})$ in this work. It is interesting to investigate the CP transformation consistent with $\text{Mp}_2(\mathbb{Z})$, the CP conserved values of $\tau$ and the implications for modular models. All these are left for future projects. We conclude that half-integral weight modular forms as well as more general rational weight modular forms provide new opportunities and possibilities for modular model building, and there are many relevant aspects which deserve studying further.

\section*{Acknowledgements}

XGL, BYQ and GJD are supported by the National Natural Science Foundation of China under Grant Nos 11975224, 11835013 and 11947301. CYY acknowledges the China Post-doctoral Science Foundation under Grant No. 2018M641621.

\newpage
\section*{Appendix}

\setcounter{equation}{0}
\renewcommand{\theequation}{\thesection.\arabic{equation}}

\begin{appendix}

\section{\label{sec:app-1} Kronecker symbol and two-cocycle $\zeta_{1/2}(\gamma_1,\gamma_2)$  }

The Kronecker symbol is a multiplicative function with values 1, $-1$, 0. Let $m$ be an odd integer and $a$ be an integer, both $m$ and $a$ can be positive or negative, the Kronecker symbol denoted as $\left(\dfrac{a}{m}\right)$ or $(a|m)$ for is defined by
\begin{equation}
\left(\dfrac{a}{m}\right)=\left(\dfrac{a}{u}\right)\prod_{i=1}^{k}\left(\dfrac{a}{p_i}\right)^{\sigma_i}\,,
\end{equation}
where
\begin{equation}
m=u\cdot p_1^{\sigma_1}\ldots p^{\sigma_k}_k\,,
\end{equation}
is the prime factorization of $m$ with $u=\pm1$ and consequently the $p_i$ are primes. The notation $\left(\dfrac{a}{p_i}\right)$ is the usual Legendre symbol~\cite{stromberg2013weil,cohen2017modular},
\begin{equation}
\left(\dfrac{a}{p_i}\right)=\begin{cases}
0, \qquad &{}a= 0\quad(\mathrm{mod}\ p_i)\,, \\
1,&\gcd(a,p_i)=1,\exists ~n\in\mathbb{Z}, n^2= 1\quad(\mathrm{mod}\ p_i)\,,\\
-1,&\gcd(a,p_i)=1,\nexists~ n\in\mathbb{Z}, n^2= 1\quad(\mathrm{mod}\ p_i)\,.
\end{cases}
\end{equation}
Legendre symbol can also be equivalently defined by means of the explicit formula as follows
\begin{equation}
\left(\dfrac{a}{p}\right)=a^{\frac{p-1}{2}}~~(\mathrm{mod}\ p)\,.
\end{equation}
The quantity $\left(\dfrac{a}{u}\right)$ is simply equal to 1 for $u=1$ and any integer value of $a$. When $u=-1$, it is given by
\begin{equation}
\left(\dfrac{a}{-1}\right)=\begin{cases}
-1,&  \text{for}~a<0\,,\\
1,&  \text{for}~a\geq0\,.
\end{cases}
\end{equation}
The Kronecker symbol has the following basic properties
\begin{eqnarray}
\nonumber&&\left(\dfrac{0}{\pm1}\right)=1,~~~~\left(\dfrac{-1}{m}\right)=(-1)^{\frac{m-1}{2}}~~\text{for~all}~~m=1(\mathrm{mod}~ 2)\,,\\
\nonumber&&\left(\dfrac{2}{m}\right)=(-1)^{\frac{m^2-1}{8}}~~\text{for~all}~~m=1(\mathrm{mod}~ 2)\,,\\
\nonumber&&\left(\dfrac{a}{m}\right)=0~~\text{if}~~\gcd(a,m)>1\,, \\
\nonumber&&\left(\dfrac{a}{m}\right)\left(\dfrac{b}{m}\right)=\left(\dfrac{ab}{m}\right)~~\text{if}~~ab\neq 0\,, \\
\nonumber&&\left(\dfrac{a}{m}\right)\left(\dfrac{a}{n}\right)=\left(\dfrac{a}{mn}\right)~~\text{if}~~mn\neq 0\,, \\
&&\left(\dfrac{a}{m}\right)=\begin{cases}
\left(\dfrac{a}{m}\right),\qquad &{}\text{$m>0$}\,, \\
\left(\dfrac{a}{|m|}\right), &\text{$m<0$ and $a>0$}\,, \\
-\left(\dfrac{a}{|m|}\right), &\text{$m<0$ and $a<0$}\,.
\end{cases}
\end{eqnarray}

The two-cocycle $\zeta_{1/2}(\gamma_1,\gamma_2)$ can be expressed as a formula as follow~\cite{stromberg2013weil},
\begin{equation}
\label{eq:zeta-value}\zeta_{1/2}(\gamma_1,\gamma_2)=\mu(\gamma_1,\gamma_2)s(\gamma_1)s(\gamma_2)s^{-1}(\gamma_1\gamma_2)\,,
\end{equation}
with
\begin{eqnarray}
\nonumber&&\mu(\gamma_1,\gamma_2)=\left(\sigma(\gamma_1)\sigma(\gamma_1\gamma_2),~\sigma(\gamma_2)\sigma(\gamma_1\gamma_2)\right)_{\infty}\,, \\
&&s(\gamma)=\begin{cases}
1,\qquad &{}c\neq 0\\
\mathrm{sign}(d), &c=0
\end{cases}\,,~\qquad~
\sigma(\gamma)=\begin{cases}
c,\qquad&{}c\neq 0\\
d, &c=0
\end{cases}\,,
\end{eqnarray}
and $(x,y)_{\infty}$ is the Hilbert symbol at infinite~\cite{stromberg2013weil,cohen2017modular},
\begin{equation}
(x,y)_{\infty}=\begin{cases}
-1,\qquad &{}\text{$x<0$ and $y<0$}\\
1, &\text{other cases}
\end{cases}\,.
\end{equation}
Hence $\zeta_{1/2}(\gamma_1,\gamma_2)$ can only take values $+1$ and $-1$. If either $\gamma_1$ or $\gamma_2$ is equal to the generator $S$ and $T$, we have
\begin{eqnarray}
\nonumber&&\zeta_{1/2}(\gamma,T)=\zeta_{1/2}(T,\gamma)=1\,,\\
\nonumber&&\zeta_{1/2}(\gamma,S)=\begin{cases}
-1, \quad &{}\text{$c<0$ and $d\leqslant 0$}\,, \\
1, \quad &\text{others}\,,
\end{cases}\\
&&\zeta_{1/2}(S,\gamma)=\begin{cases}
-1, \quad &{}\text{$a\leqslant 0$ and $c<0$} \,, \\
1, \quad &\text{others}\,.
\end{cases}
\end{eqnarray}

\section{\label{sec:app-multiplier}Multiplier system of rational weight modular forms
}

\cleqn

As a general principle, when we discuss the general non-integral weight $r$ modular forms, it is necessary to introduce the so-called multiplier system $v(\gamma)$ to ensure the existence of well-defined automorphy factors $j_r(\gamma,\tau)=v(\gamma)(c\tau+d)^r$, namely $j_r(\gamma_1\gamma_2,\tau)=j_r(\gamma_1,\gamma_2\tau)j_r(\gamma_2,\tau)$ for any $\gamma_1,\gamma_2 \in \Gamma'$, where $\Gamma'$ is a subgroup of $\text{SL}_2(\mathbb{Z})$. Thus the definition of a non-integral weight $r$ modular form for the subgroup $\Gamma'$ is
\begin{equation}
f(\gamma\tau)=v(\gamma)(c\tau+d)^r f(\tau), \quad \gamma \in \Gamma' \,.
\end{equation}
The multiplier system $v(\gamma)$ heavily depends on $\Gamma'$. For principal congruence subgroup $\Gamma(4N)$, the multiplier system is the Kronecker symbol mentioned above, it is used to define the half-integral weight modular forms for $\Gamma(4N)$. For other principal congruence subgroup $\Gamma(N)$ of level odd integer $N\geq5$, a unified construction of multiplier systems denote by $v_N(\gamma)$ is given in~\cite{ibukiyama2000modular}, and the corresponding modular forms are of weight $(N-3)/(2N)$.
as already mentioned in the section~\ref{subsec:rational-weight-MF}. Specifically, $v_N$ is given by the following formula
\begin{equation}
v_N(\gamma)=
\begin{cases}
1 ~~~~~~~~~~~~~~~~~~~~~~~~~~~~~~~~~~~~~~~~~~~~~~~~~~~~~~~~~~~~~~~~~~~~~\qquad\text{if}~~ c=0 \,,\\[0.1in]
\exp\left(-2\pi i\dfrac{3\,\text{sign}(c)(N^2-1)}{8N}\right)\exp\left(2\pi i\dfrac{N^2-1}{8N}\Phi(\gamma)\right) \qquad\text{if}~~ c \neq 0 \,,
\end{cases}
\end{equation}
where $\gamma=\begin{pmatrix}
a&b\\c& d \end{pmatrix} \in \Gamma(N)$, and $\Phi(\gamma)$ is a integer valued defined as
\begin{equation}
\Phi(\gamma)=
\begin{cases}
\dfrac{b}{d} ~~~~~~~~~~~~~~~~~~~\qquad\qquad\qquad\text{if}~~ c=0 \,,\\[0.2in] \dfrac{a+d}{c}-12\, \text{sign}(c)\, s(d,|c|) \qquad\text{if}~~ c \neq 0 \,,
\end{cases}
\end{equation}
where $s(d,|c|)$ is the Dedekind sum with
\begin{equation}
s(h,k)=\sum^k_{\mu=1} \left(\left(\frac{h\mu}{k}\right)\right)\left(\left(\frac{\mu}{k}\right)\right)\,,
\end{equation}
for integers $h, k(k\neq 0)$.
Here $((x))$ is the sawtooth function defined by
\begin{equation}
((x)) =\begin{cases}  x- [x]-\frac{1}{2} ~\quad~ \text{if}~ x \notin \mathbb{Z}\,, \\
0  ~~~~~~~~~~~~~\quad~~ \text{if}~ x \in \mathbb{Z}\,,
\end{cases}
\end{equation}
with $[x]$ the floor function. Note that the multiplier system $v_N(\gamma)$ is an $N$-th root of unity, consequently $v_N(\gamma)^N=1$ for
all $\gamma \in \Gamma(N)$. In short,  $v_N(\gamma)(c\tau+d)^{(N-3)/2N}$ is the automorphy factor for the modular form of weight $(N-3)/(2N)$ at level odd integer $N\geq5$.

\section{\label{sec:app-S4QC}Group theory of $\widetilde{S}_4$}
\cleqn

The double covering group of $\widetilde{S}_4$ has 96 elements, and it can be generated by three generators $\widetilde{S},\widetilde{T}$ and $\widetilde{R}$ obeying the rules:
\begin{equation}
  \widetilde{S}^2=\widetilde{R},\quad (\widetilde{S}\widetilde{T})^3=\widetilde{T}^4=\widetilde{R}^4=1,\quad \widetilde{S}\widetilde{R}=\widetilde{R}\widetilde{S},\quad \widetilde{T}\widetilde{R}=\widetilde{R}\widetilde{T}\,.
\end{equation}
After we input these multiplication rules in \texttt{GAP}~\cite{GAP}, its group ID can be determined as [96, 67]. Notice that $S_4$ is not a subgroup of $\widetilde{S}_4$, it is isomorphic to the quotient group of $\widetilde{S}_4$ over $Z_4^{\widetilde{R}}$, i.e. $S_4\cong \widetilde{S}_4/Z_4^{\widetilde{R}}$, where $Z_4^{\widetilde{R}}=\{1,\widetilde{R},\widetilde{R}^2,\widetilde{R}^3\}$ is the center and a normal subgroup of $\widetilde{S}_4$.
The finite metaplectic group $\widetilde{S}_4$ is a quadruple cover of $S_4$ or double cover of $S'_4$. All the elements of $\widetilde{S}_4$ group can be divided into 16 conjugacy classes:

\begin{eqnarray}
\nonumber 1C_1&=&\{1\} \,, \\
\nonumber 1C_2&=&\left\{\widetilde{R}^2\right\} = (1C_1)\cdot \widetilde{R}^2\,, \\
\nonumber 6C_2&=&\left\{\widetilde{T}^2,\widetilde{T}^2\widetilde{R}^2,\left(\widetilde{S}\widetilde{T}^2\right)^2,\widetilde{S}\widetilde{T}^2\widetilde{S}^3,\left(\widetilde{S}\widetilde{T}^2\right)^2\widetilde{R}^2,
\widetilde{S}\widetilde{T}^2\widetilde{S}^3\widetilde{R}^2\right\} \,, \\
\nonumber 8C_3&=&\left\{\widetilde{S}\widetilde{T},\widetilde{T}\widetilde{S},(\widetilde{S}\widetilde{T})^2,(\widetilde{T}\widetilde{S})^2,\widetilde{T}^2\widetilde{S}\widetilde{T}^3,\widetilde{T}^3\widetilde{S}\widetilde{T}^2,\widetilde{T}^2\widetilde{S}^3\widetilde{T}\widetilde{R}^2,
\widetilde{T}\widetilde{S}^3\widetilde{T}^2\widetilde{R}^2\right\} \,, \\
\nonumber 1C_4&=&\{\widetilde{R}\} = (1C_1)\cdot \widetilde{R} \,, \\
\nonumber 1C_4'&=&\left\{\widetilde{R}^3\right\} = (1C_1)\cdot \widetilde{R}^3 \,, \\
\nonumber 6C_4&=&\left\{\widetilde{T},\widetilde{T}^3\widetilde{S}^2,\widetilde{T}^2\widetilde{S}\widetilde{R}^2,\widetilde{S}\widetilde{T}^2\widetilde{R}^2,\widetilde{T}\widetilde{S}\widetilde{T}\widetilde{R}^2,
\widetilde{S}\widetilde{T}\widetilde{S}^3\widetilde{R}^2\right\} \,, \\
\nonumber 6C_4'&=&\left\{\widetilde{T}\widetilde{R},\widetilde{T}^3\widetilde{S}^2\widetilde{R},\widetilde{T}^2\widetilde{S}\widetilde{R}^3,\widetilde{S}\widetilde{T}^2\widetilde{R}^3,\widetilde{T}\widetilde{S}\widetilde{T}\widetilde{R}^3,\widetilde{S}\widetilde{T}\widetilde{S}^3\widetilde{R}^3\right\} =(6C_4)\cdot \widetilde{R} \,, \\
\nonumber 6C_4''&=&\left\{\widetilde{T}\widetilde{R}^2,\widetilde{T}^3\widetilde{S}^2\widetilde{R}^2,\widetilde{T}^2\widetilde{S},\widetilde{S}\widetilde{T}^2,\widetilde{T}\widetilde{S}\widetilde{T},\widetilde{S}\widetilde{T}\widetilde{S}^3\right\} =(6C_4)\cdot \widetilde{R}^2 \,, \\
\nonumber 6C_4'''&=&\left\{\widetilde{T}\widetilde{R}^3,\widetilde{T}^3\widetilde{S}^2\widetilde{R}^3,\widetilde{T}^2\widetilde{S}\widetilde{R},\widetilde{S}\widetilde{T}^2\widetilde{R},\widetilde{T}\widetilde{S}\widetilde{T}\widetilde{R},\widetilde{S}\widetilde{T}\widetilde{S}^3\widetilde{R}\right\} =(6C_4)\cdot \widetilde{R}^3 \,, \\
\nonumber 6C_4''''&=&\left\{\widetilde{T}^2\widetilde{R},\widetilde{T}^2\widetilde{R}^3,\left(\widetilde{S}\widetilde{T}^2\right)^2\widetilde{R},\widetilde{S}\widetilde{T}^2\widetilde{S}^3\widetilde{R},\left(\widetilde{S}\widetilde{T}^2\right)^2\widetilde{R}^3,\widetilde{S}\widetilde{T}^2\widetilde{S}^3\widetilde{R}^3\right\} = (6C_2)\cdot \widetilde{R} \,, \\
\nonumber 8C_6&=&\left\{\widetilde{S}\widetilde{T}\widetilde{R}^2,\widetilde{T}\widetilde{S}\widetilde{R}^2,(\widetilde{S}\widetilde{T})^2\widetilde{R}^2,(\widetilde{T}\widetilde{S})^2\widetilde{R}^2,
\widetilde{T}^2\widetilde{S}\widetilde{T}^3\widetilde{R}^2,\widetilde{T}^3\widetilde{S}\widetilde{T}^2\widetilde{R}^2,\right.\\
\nonumber &&\left. \widetilde{T}^2\widetilde{S}^3\widetilde{T},\widetilde{T}\widetilde{S}^3\widetilde{T}^2\right\} = (8C_3)\cdot \widetilde{R}^2 \,, \\
\nonumber 12C_8&=&\left\{\widetilde{S},\widetilde{S}\widetilde{R}^2,\widetilde{T}^2\widetilde{S}\widetilde{T}^2,\widetilde{T}^3\widetilde{S}\widetilde{T},\widetilde{T}\widetilde{S}\widetilde{T}^3,
\widetilde{T}^2\widetilde{S}\widetilde{T}^2\widetilde{R}^2,\widetilde{T}^3\widetilde{S}\widetilde{T}\widetilde{R}^2,\widetilde{S}\widetilde{T}^2\widetilde{S}^3\widetilde{T},
\widetilde{T}\widetilde{S}\widetilde{T}^2\widetilde{S}^3,\right.\\
\nonumber &&\left.\widetilde{T}\widetilde{S}\widetilde{T}^3\widetilde{R}^2,\widetilde{S}\widetilde{T}^2\widetilde{S}^3\widetilde{T}\widetilde{R}^2,\widetilde{T}\widetilde{S}\widetilde{T}^2\widetilde{S}^3\widetilde{R}^2\right\} \,, \\
\nonumber 12C_8'&=&\left\{\widetilde{S}\widetilde{R},\widetilde{S}\widetilde{R}^3,\widetilde{T}^2\widetilde{S}\widetilde{T}^2\widetilde{R},\widetilde{T}^3\widetilde{S}\widetilde{T}\widetilde{R},
\widetilde{T}\widetilde{S}\widetilde{T}^3\widetilde{R},\widetilde{T}^2\widetilde{S}\widetilde{T}^2\widetilde{R}^3,\widetilde{T}^3\widetilde{S}\widetilde{T}\widetilde{R}^3,
\widetilde{S}\widetilde{T}^2\widetilde{S}^3\widetilde{T}\widetilde{R},\right.\\
\nonumber &&\left.\widetilde{T}\widetilde{S}\widetilde{T}^2\widetilde{S}^3\widetilde{R},\widetilde{T}\widetilde{S}\widetilde{T}^3\widetilde{R}^3,\widetilde{S}\widetilde{T}^2\widetilde{S}^3\widetilde{T}\widetilde{R}^3,\widetilde{T}\widetilde{S}\widetilde{T}^2\widetilde{S}^3\widetilde{R}^3\right\} = (12C_8)\cdot \widetilde{R} \,, \\
\nonumber 8C_{12}&=&\left\{\widetilde{S}\widetilde{T}\widetilde{R},\widetilde{T}\widetilde{S}\widetilde{R},(\widetilde{S}\widetilde{T})^2\widetilde{R},(\widetilde{T}\widetilde{S})^2\widetilde{R},
\widetilde{T}^2\widetilde{S}\widetilde{T}^3\widetilde{R},\widetilde{T}^3\widetilde{S}\widetilde{T}^2\widetilde{R},\widetilde{T}^2\widetilde{S}^3\widetilde{T}\widetilde{R}^3,\right.\\
\nonumber &&\left. \widetilde{T}\widetilde{S}^3\widetilde{T}^2\widetilde{R}^3\right\} = (8C_3)\cdot \widetilde{R} \,, \\
\nonumber 8C_{12}'&=&\left\{\widetilde{S}\widetilde{T}\widetilde{R}^3,\widetilde{T}\widetilde{S}\widetilde{R}^3,(\widetilde{S}\widetilde{T})^2\widetilde{R}^3,(\widetilde{T}\widetilde{S})^2\widetilde{R}^3,
\widetilde{T}^2\widetilde{S}\widetilde{T}^3\widetilde{R}^3,\widetilde{T}^3\widetilde{S}\widetilde{T}^2\widetilde{R}^3,\right.\\
&& \left.\widetilde{T}^2\widetilde{S}^3\widetilde{T}\widetilde{R},\widetilde{T}\widetilde{S}^3\widetilde{T}^2\widetilde{R}\right\}  =(8C_3)\cdot \widetilde{R}^3\,.
\end{eqnarray}
where $kC_n$ denotes a conjugacy class with $k$ elements of order $n$. Note that some of these conjugacy classes can be written as the product of the others with $\widetilde{R}, \widetilde{R}^2$ or $\widetilde{R}^3$. There are four one-dimensional irreducible representations $\mathbf{1},\mathbf{1}^{\prime},\mathbf{\widehat{1}}$ and $\mathbf{\widehat{1}}^{\prime}$, six two-dimensional irreducible representations $\mathbf{2}, \mathbf{2}',\mathbf{\widehat{2}},\mathbf{\widehat{2}}',\mathbf{\widetilde{2}}$ and $\mathbf{\widetilde{2}}'$, four three-dimensional irreducible representations $\mathbf{3},\mathbf{3}^{\prime},\mathbf{\widehat{3}}$ and $\mathbf{\widehat{3}}^{\prime}$, and two four-dimensional irreducible representations $\mathbf{4},\mathbf{4}'$. We have summarized the explicit matrix representations in table~\ref{tab:Rep_baseB}. In the representations $\mathbf{1}$, $\mathbf{1}^{\prime}$, $\mathbf{2}$, $\mathbf{3}$ and $\mathbf{3}^{\prime}$, the generator $\widetilde{R}=1$ is an identity matrix, the representation matrices of $\widetilde{S}$ and $\widetilde{T}$ coincide with those of $S_4$, consequently $\widetilde{S}_4$ can not be distinguished from $S_4$ in these representations since they are represented by the same set of matrices. In the representations $\mathbf{\widehat{1}}$, $\mathbf{\widehat{1}}^{\prime}$, $\mathbf{2'}$, $\mathbf{\widehat{3}}$ and $\mathbf{\widehat{3}}^{\prime}$, the generator $\widetilde{R}=-1$. The character table of $\widetilde{S}_4$ can be obtained by taking the trace of the representation matrices of the representative elements, and it is shown in table~\ref{tab:character}. Moreover, the Kronecker products between all irreducible representations are given as follows:

\begin{table}[hptb]
\centering
\begin{adjustbox}{width=\textwidth}
\setlength{\tabcolsep}{1.5pt}
\begin{tabular}{|c|c|c|c|c|c|c|c|c|c|c|c|c|c|c|c|c|}
\hline\hline
Classes & $1C_1$ & $1C_2$ & $6C_2$ & $8C_3$ & $1C_4$ & $1C_4'$ & $6C_4$ & $6C_4'$ & $6C_4''$ & $6C_4'''$ & $6C_4''''$ & $8C_6$ & $12C_8$ & $12C_8'$ & $8C_{12}$ & $8C_{12}'$\\ \hline
 &  &  &  &  &  &  &  &  &  &  &  & &  &  & & \\ [-0.15in]
$G$ & $1$ & $\widetilde{R}$ & $\widetilde{T}^2$ & $\widetilde{S}\widetilde{T}$ & $\widetilde{R}$ & $\widetilde{R}^3$ & $\widetilde{T}$ & $\widetilde{T}\widetilde{R}$ & $\widetilde{T}\widetilde{R}^2$ & $\widetilde{T}\widetilde{R}^3$ & $\widetilde{T}^2\widetilde{R}$ & $\widetilde{S}\widetilde{T}\widetilde{R}^2$ & $\widetilde{S}$ & $\widetilde{S}\widetilde{R}$ & $\widetilde{S}\widetilde{T}\widetilde{R}$ & $\widetilde{S}\widetilde{T}\widetilde{R}^3$\\ \hline
$\mathbf{1}$ & $1$ & $1$ & $1$ & $1$ & $1$ & $1$ & $1$ & $1$ & $1$ & $1$ & $1$ & $1$ & $1$ & $1$ & $1$ & $1$ \\
 $\mathbf{1}'$ & $1$ & $1$ & $1$ & $1$ & $1$ & $1$ & $-1$ & $-1$ & $-1$ & $-1$ & $1$ & $1$ & $-1$ & $-1$ & $1$ & $1$ \\
 $\widehat{\mathbf{1}}$ & $1$ & $1$ & $-1$ & $1$ & $-1$ & $-1$ & $-i$ & $i$ & $-i$ & $i$ & $1$ & $1$ & $i$ & $-i$ & $-1$ & $-1$ \\
 $\widehat{\mathbf{1}}'$ & $1$ & $1$ & $-1$ & $1$ & $-1$ & $-1$ & $i$ & $-i$ & $i$ & $-i$ & $1$ & $1$ & $-i$ & $i$ & $-1$ & $-1$ \\
 $\mathbf{2}$ & $2$ & $2$ & $2$ & $-1$ & $2$ & $2$ & $0$ & $0$ & $0$ & $0$ & $2$ & $-1$ & $0$ & $0$ & $-1$ & $-1$ \\
 $\mathbf{2}'$ & $2$ & $2$ & $-2$ & $-1$ & $-2$ & $-2$ & $0$ & $0$ & $0$ & $0$ & $2$ & $-1$ & $0$ & $0$ & $1$ & $1$ \\
 $\widehat{\mathbf{2}}$ & $2$ & $-2$ & $0$ & $-1$ & $2 i$ & $-2 i$ & $1+i$ & $-1+i$ & $-1-i$ & $1-i$ & $0$ & $1$ & $0$ & $0$ & $-i$ & $i$ \\
 $\widehat{\mathbf{2}}'$ & $2$ & $-2$ & $0$ & $-1$ & $2 i$ & $-2 i$ & $-1-i$ & $1-i$ & $1+i$ & $-1+i$ & $0$ & $1$ & $0$ & $0$ & $-i$ & $i$ \\
 $\widetilde{\mathbf{2}}$ & $2$ & $-2$ & $0$ & $-1$ & $-2 i$ & $2 i$ & $1-i$ & $-1-i$ & $-1+i$ & $1+i$ & $0$ & $1$ & $0$ & $0$ & $i$ & $-i$ \\
 $\widetilde{\mathbf{2}}'$ & $2$ & $-2$ & $0$ & $-1$ & $-2 i$ & $2 i$ & $-1+i$ & $1+i$ & $1-i$ & $-1-i$ & $0$ & $1$ & $0$ & $0$ & $i$ & $-i$ \\
 $\mathbf{3}$ & $3$ & $3$ & $-1$ & $0$ & $3$ & $3$ & $1$ & $1$ & $1$ & $1$ & $-1$ & $0$ & $-1$ & $-1$ & $0$ & $0$ \\
 $\mathbf{3}'$ & $3$ & $3$ & $-1$ & $0$ & $3$ & $3$ & $-1$ & $-1$ & $-1$ & $-1$ & $-1$ & $0$ & $1$ & $1$ & $0$ & $0$ \\
 $\widehat{\mathbf{3}}$ & $3$ & $3$ & $1$ & $0$ & $-3$ & $-3$ & $-i$ & $i$ & $-i$ & $i$ & $-1$ & $0$ & $-i$ & $i$ & $0$ & $0$ \\
 $\widehat{\mathbf{3}}'$ & $3$ & $3$ & $1$ & $0$ & $-3$ & $-3$ & $i$ & $-i$ & $i$ & $-i$ & $-1$ & $0$ & $i$ & $-i$ & $0$ & $0$ \\
 $\mathbf{4}$ & $4$ & $-4$ & $0$ & $1$ & $4 i$ & $-4 i$ & $0$ & $0$ & $0$ & $0$ & $0$ & $-1$ & $0$ & $0$ & $i$ & $-i$ \\
    $\mathbf{4}'$ & $4$ & $-4$ & $0$ & $1$ & $-4 i$ & $4 i$ & $0$ & $0$ & $0$ & $0$ & $0$ & $-1$ & $0$ & $0$ & $-i$ & $i$ \\\hline\hline
  \end{tabular}
\end{adjustbox}
\caption{\label{tab:character}Character table of $\widetilde{S}_4$, and we give a representative element for each conjugacy class in the second row.}
\end{table}

%%%%%%%%%%%%%%%%%%%%%%%%%%%%%%%%%%%%%%%%%%%%%%%%%%%%%%%%

\begin{eqnarray}
\nonumber&&\mathbf{1} \otimes \mathbf{1} = \mathbf{1'} \otimes \mathbf{1'} = \mathbf{\widehat{1}} \otimes \mathbf{\widehat{1}'} = \mathbf{1},\quad \mathbf{1} \otimes \mathbf{1'} = \mathbf{\widehat{1}} \otimes \mathbf{\widehat{1}} = \mathbf{\widehat{1}'} \otimes \mathbf{\widehat{1}'} = \mathbf{1'}, \\
\nonumber&&\mathbf{1} \otimes \mathbf{\widehat{1}} = \mathbf{1'} \otimes \mathbf{\widehat{1}'} = \mathbf{\widehat{1}},\quad \mathbf{1} \otimes \mathbf{\widehat{1}'} = \mathbf{1'} \otimes \mathbf{\widehat{1}} = \mathbf{\widehat{1}'}, \\[2ex]
\nonumber&&\mathbf{1} \otimes \mathbf{2} = \mathbf{1'} \otimes \mathbf{2} = \mathbf{\widehat{1}} \otimes \mathbf{2'} = \mathbf{\widehat{1}'} \otimes \mathbf{2'} = \mathbf{2},\quad \mathbf{1} \otimes \mathbf{2'} = \mathbf{1'} \otimes \mathbf{2'} = \mathbf{\widehat{1}} \otimes \mathbf{2} = \mathbf{\widehat{1}'} \otimes \mathbf{2} = \mathbf{2'}, \\
\nonumber&&\mathbf{1} \otimes \mathbf{\widehat{2}} = \mathbf{1'} \otimes \mathbf{\widehat{2}'} = \mathbf{\widehat{1}} \otimes \mathbf{\widetilde{2}'} = \mathbf{\widehat{1}'} \otimes \mathbf{\widetilde{2}} = \mathbf{\widehat{2}},\quad \mathbf{1} \otimes \mathbf{\widehat{2}'} = \mathbf{1'} \otimes \mathbf{\widehat{2}} = \mathbf{\widehat{1}} \otimes \mathbf{\widetilde{2}} = \mathbf{\widehat{1}'} \otimes \mathbf{\widetilde{2}'} = \mathbf{\widehat{2}'}, \\
\nonumber&&\mathbf{1} \otimes \mathbf{\widetilde{2}} = \mathbf{1'} \otimes \mathbf{\widetilde{2}'} = \mathbf{\widehat{1}} \otimes \mathbf{\widehat{2}} = \mathbf{\widehat{1}'} \otimes \mathbf{\widehat{2}'} = \mathbf{\widetilde{2}},\quad \mathbf{1} \otimes \mathbf{\widetilde{2}'} = \mathbf{1'} \otimes \mathbf{\widetilde{2}} = \mathbf{\widehat{1}} \otimes \mathbf{\widehat{2}'} = \mathbf{\widehat{1}'} \otimes \mathbf{\widehat{2}} = \mathbf{\widetilde{2}'}, \\[2ex]
\nonumber&&\mathbf{1} \otimes \mathbf{3} = \mathbf{1'} \otimes \mathbf{3'} = \mathbf{\widehat{1}} \otimes \mathbf{\widehat{3}'} = \mathbf{\widehat{1}'} \otimes \mathbf{\widehat{3}} = \mathbf{3},\quad \mathbf{1} \otimes \mathbf{3'} = \mathbf{1'} \otimes \mathbf{3} = \mathbf{\widehat{1}} \otimes \mathbf{\widehat{3}} = \mathbf{\widehat{1}'} \otimes \mathbf{\widehat{3}'} = \mathbf{3'}, \\
\nonumber&&\mathbf{1} \otimes \mathbf{\widehat{3}} = \mathbf{1'} \otimes \mathbf{\widehat{3}'} = \mathbf{\widehat{1}} \otimes \mathbf{3} = \mathbf{\widehat{1}'} \otimes \mathbf{3'} = \mathbf{\widehat{3}},\quad \mathbf{1} \otimes \mathbf{\widehat{3}'} = \mathbf{1'} \otimes \mathbf{\widehat{3}} = \mathbf{\widehat{1}} \otimes \mathbf{3'} = \mathbf{\widehat{1}'} \otimes \mathbf{3} = \mathbf{\widehat{3}'}, \\[2ex]
\nonumber&&\mathbf{1} \otimes \mathbf{4} = \mathbf{1'} \otimes \mathbf{4} = \mathbf{\widehat{1}} \otimes \mathbf{4'} = \mathbf{\widehat{1}'} \otimes \mathbf{4'} = \mathbf{2} \otimes \mathbf{\widehat{2}} = \mathbf{2} \otimes \mathbf{\widehat{2}'} = \mathbf{2'} \otimes \mathbf{\widetilde{2}} = \mathbf{2'} \otimes \mathbf{\widetilde{2}'} = \mathbf{4},\\
\nonumber&&\mathbf{1} \otimes \mathbf{4'} = \mathbf{1'} \otimes \mathbf{4'} = \mathbf{\widehat{1}} \otimes \mathbf{4} = \mathbf{\widehat{1}'} \otimes \mathbf{4} = \mathbf{2} \otimes \mathbf{\widetilde{2}} = \mathbf{2} \otimes \mathbf{\widetilde{2}'} = \mathbf{2'} \otimes \mathbf{\widehat{2}} = \mathbf{2'} \otimes \mathbf{\widehat{2}'} = \mathbf{4'}, \\[2ex]
\nonumber&&\mathbf{2} \otimes \mathbf{2} = \mathbf{2'} \otimes \mathbf{2'} = \mathbf{1} \oplus \mathbf{1'} \oplus \mathbf{2},\quad \mathbf{2} \otimes \mathbf{2'} = \mathbf{\widehat{1}} \oplus \mathbf{\widehat{1}'} \oplus \mathbf{2'}, \\
\nonumber&&\mathbf{\widehat{2}} \otimes \mathbf{\widehat{2}} = \mathbf{\widehat{2}'} \otimes \mathbf{\widehat{2}'} = \mathbf{\widetilde{2}} \otimes \mathbf{\widetilde{2}'} = \mathbf{\widehat{1}'} \oplus \mathbf{\widehat{3}'}, \quad \mathbf{\widehat{2}} \otimes \mathbf{\widehat{2}'} = \mathbf{\widetilde{2}} \otimes \mathbf{\widetilde{2}} = \mathbf{\widetilde{2}'} \otimes \mathbf{\widetilde{2}'} = \mathbf{\widehat{1}} \oplus \mathbf{\widehat{3}},\\
\nonumber&&\mathbf{\widehat{2}} \otimes \mathbf{\widetilde{2}} = \mathbf{\widehat{2}'} \otimes \mathbf{\widetilde{2}'} = \mathbf{1} \oplus \mathbf{3},\quad \mathbf{\widehat{2}} \otimes \mathbf{\widetilde{2}'} = \mathbf{\widehat{2}'} \otimes \mathbf{\widetilde{2}} = \mathbf{1'} \oplus \mathbf{3'} \\[2ex]
\nonumber&&\mathbf{2} \otimes \mathbf{3} = \mathbf{2} \otimes \mathbf{3'} = \mathbf{2'} \otimes \mathbf{\widehat{3}} = \mathbf{2'} \otimes \mathbf{\widehat{3}'} = \mathbf{3} \oplus \mathbf{3'}, \quad \mathbf{2} \otimes \mathbf{\widehat{3}} = \mathbf{2} \otimes \mathbf{\widehat{3}'} = \mathbf{2'} \otimes \mathbf{3} = \mathbf{2'} \otimes \mathbf{3'} = \mathbf{\widehat{3}} \oplus \mathbf{\widehat{3}'}, \\
\nonumber&&\mathbf{\widehat{2}} \otimes \mathbf{3} = \mathbf{\widehat{2}'} \otimes \mathbf{3'} = \mathbf{\widetilde{2}} \otimes \mathbf{\widehat{3}'} = \mathbf{\widetilde{2}'} \otimes \mathbf{\widehat{3}} = \mathbf{\widehat{2}} \oplus \mathbf{4}, \quad \mathbf{\widehat{2}} \otimes \mathbf{3'} = \mathbf{\widehat{2}'} \otimes \mathbf{3} = \mathbf{\widetilde{2}} \otimes \mathbf{\widehat{3}} = \mathbf{\widetilde{2}'} \otimes \mathbf{\widehat{3}'} = \mathbf{\widehat{2}'} \oplus \mathbf{4}, \\
\nonumber&&\mathbf{\widehat{2}} \otimes \mathbf{\widehat{3}} = \mathbf{\widehat{2}'} \otimes \mathbf{\widehat{3}'} = \mathbf{\widetilde{2}} \otimes \mathbf{3} = \mathbf{\widetilde{2}'} \otimes \mathbf{3'} = \mathbf{\widetilde{2}} \oplus \mathbf{4'}, \quad \mathbf{\widehat{2}} \otimes \mathbf{\widehat{3}'} = \mathbf{\widehat{2}'} \otimes \mathbf{\widehat{3}} = \mathbf{\widetilde{2}} \otimes \mathbf{3'} = \mathbf{\widetilde{2}'} \otimes \mathbf{3} = \mathbf{\widetilde{2}'} \oplus \mathbf{4'}, \\[2ex]
\nonumber&&\mathbf{2} \otimes \mathbf{4} = \mathbf{2'} \otimes \mathbf{4'} = \mathbf{\widehat{2}} \oplus \mathbf{\widehat{2}'} \oplus \mathbf{4},\quad \mathbf{2} \otimes \mathbf{4'} = \mathbf{2'} \otimes \mathbf{4} = \mathbf{\widetilde{2}} \oplus \mathbf{\widetilde{2}'} \oplus \mathbf{4'}, \\
\nonumber&&\mathbf{\widehat{2}} \otimes \mathbf{4} = \mathbf{\widehat{2}'} \otimes \mathbf{4} = \mathbf{\widetilde{2}} \otimes \mathbf{4'} = \mathbf{\widetilde{2}'} \otimes \mathbf{4'} = \mathbf{2'} \oplus \mathbf{\widehat{3}} \oplus \mathbf{\widehat{3}'}, \\
\nonumber&&\mathbf{\widehat{2}} \otimes \mathbf{4'} = \mathbf{\widehat{2}'} \otimes \mathbf{4'} = \mathbf{\widetilde{2}} \otimes \mathbf{4} = \mathbf{\widetilde{2}'} \otimes \mathbf{4} = \mathbf{2} \oplus \mathbf{3} \oplus \mathbf{3'}, \\[2ex]
\nonumber&&\mathbf{3} \otimes \mathbf{3} = \mathbf{3'} \otimes \mathbf{3'} = \mathbf{\widehat{3}} \otimes \mathbf{\widehat{3}'} = \mathbf{1} \oplus \mathbf{2} \oplus \mathbf{3} \oplus \mathbf{3'}, \quad \mathbf{3} \otimes \mathbf{3'} = \mathbf{\widehat{3}} \otimes \mathbf{\widehat{3}} = \mathbf{\widehat{3}'} \otimes \mathbf{\widehat{3}'} = \mathbf{1'} \oplus \mathbf{2} \oplus \mathbf{3} \oplus \mathbf{3'}, \\
\nonumber&&\mathbf{3} \otimes \mathbf{\widehat{3}} = \mathbf{3'} \otimes \mathbf{\widehat{3}'} = \mathbf{\widehat{1}} \oplus \mathbf{2'} \oplus \mathbf{\widehat{3}} \oplus \mathbf{\widehat{3}'}, \quad \mathbf{3} \otimes \mathbf{\widehat{3}'} = \mathbf{3'} \otimes \mathbf{\widehat{3}} = \mathbf{\widehat{1}'} \oplus \mathbf{2'} \oplus \mathbf{\widehat{3}} \oplus \mathbf{\widehat{3}'}, \\[3ex]
\nonumber&&\mathbf{3} \otimes \mathbf{4} = \mathbf{3'} \otimes \mathbf{4} = \mathbf{\widehat{3}} \otimes \mathbf{4'} = \mathbf{\widehat{3}'} \otimes \mathbf{4'} = \mathbf{\widehat{2}} \oplus \mathbf{\widehat{2}'} \oplus \mathbf{4} \oplus \mathbf{4}, \\
\nonumber&&\mathbf{3} \otimes \mathbf{4'} = \mathbf{3'} \otimes \mathbf{4'} = \mathbf{\widehat{3}} \otimes \mathbf{4} = \mathbf{\widehat{3}'} \otimes \mathbf{4} = \mathbf{\widetilde{2}} \oplus \mathbf{\widetilde{2}'} \oplus \mathbf{4'} \oplus \mathbf{4'}, \\[2ex]
\nonumber&&\mathbf{4} \otimes \mathbf{4} = \mathbf{4'} \otimes \mathbf{4'} = \mathbf{\widehat{1}} \oplus \mathbf{\widehat{1}'} \oplus \mathbf{2'} \oplus \mathbf{\widehat{3}} \oplus \mathbf{\widehat{3}} \oplus \mathbf{\widehat{3}'} \oplus \mathbf{\widehat{3}'}, \\
&&\mathbf{4} \otimes \mathbf{4'} = \mathbf{1} \oplus \mathbf{1'} \oplus \mathbf{2} \oplus \mathbf{3} \oplus \mathbf{3} \oplus \mathbf{3'} \oplus \mathbf{3'}.
\end{eqnarray}

%%%%%%%%%%%%%%%%%%%%%%%%%%%%%%%%%%%%%%%%%%%%%%%%%%%%%%%%

\begin{table}[!t]
\centering
\begin{tabular}{|c|c|c|c|}\hline\hline
 & $\widetilde{S}$ & $\widetilde{T}$ & $\widetilde{R}$\\ \hline
    $\mathbf{1},\mathbf{1^{\prime}}$ & $\pm 1$ & $\pm 1$ & $1$ \\ \hline
    $\mathbf{\widehat{1}},\mathbf{\widehat{1}^{\prime}}$ & $\pm i$ & $\mp i$ & $-1$ \\ \hline
    $\mathbf{2}$ & $\frac{1}{2}\begin{pmatrix}
 -1 & \sqrt{3} \\
 \sqrt{3} & 1 \\
\end{pmatrix}$ & $\begin{pmatrix}
 1 & 0 \\
 0 & -1 \\
\end{pmatrix}$ & $\begin{pmatrix}{1} & {0} \\ {0} & {1}\end{pmatrix}$ \\ \hline
  $\mathbf{2}'$ & $\frac{i}{2}\begin{pmatrix}
 -1 & \sqrt{3} \\
 \sqrt{3} & 1 \\
\end{pmatrix}$ & $-i\begin{pmatrix}
 1 & 0 \\
 0 & -1 \\
\end{pmatrix}$ & $-\begin{pmatrix}{1} & {0} \\ {0} & {1}\end{pmatrix}$ \\ \hline
  $\widehat{\mathbf{2}},\widehat{\mathbf{2}}'$ & $\pm\dfrac{e^{\pi i/4}}{\sqrt{2}}\begin{pmatrix}
 -1 & 1 \\
 1 & 1 \\
\end{pmatrix}$ & $\pm\begin{pmatrix}
 1 & 0 \\
 0 & i \\
\end{pmatrix}$ & $i\begin{pmatrix}{1} & {0} \\ {0} & {1}\end{pmatrix}$ \\ \hline
 $\widetilde{\mathbf{2}},\widetilde{\mathbf{2}}'$ & $\pm\dfrac{i e^{\pi i/4} }{\sqrt{2}}\begin{pmatrix}
 -1 & 1 \\
 1 & 1 \\
\end{pmatrix}$ & $\mp i\begin{pmatrix}
 1 & 0 \\
 0 & i \\
\end{pmatrix}$ & $-i\begin{pmatrix}{1} & {0} \\ {0} & {1}\end{pmatrix}$ \\ \hline
    $\mathbf{3},\mathbf{3^{\prime}}$ & $\pm\dfrac{1}{2}\begin{pmatrix}
 0 & \sqrt{2} & \sqrt{2} \\
 \sqrt{2} & -1 & 1 \\
 \sqrt{2} & 1 & -1 \\
\end{pmatrix}$ & $\pm\begin{pmatrix}
 1 & 0 & 0 \\
 0 & i & 0 \\
 0 & 0 & -i \\
\end{pmatrix}$ & $\begin{pmatrix}{1} & {0} & {0} \\ {0} & {1} & {0} \\ {0} & {0} & {1}\end{pmatrix}$ \\ \hline
  $\mathbf{\widehat{3}}, \mathbf{\widehat{3}^{\prime}}$ & $\pm\dfrac{i}{2}\begin{pmatrix}
 0 & \sqrt{2} & \sqrt{2} \\
 \sqrt{2} & -1 & 1 \\
 \sqrt{2} & 1 & -1 \\
\end{pmatrix}$ & $\mp i\begin{pmatrix}
 1 & 0 & 0 \\
 0 & i & 0 \\
 0 & 0 & -i \\
\end{pmatrix}$ & $-\begin{pmatrix}{1} & {0} & {0} \\ {0} & {1} & {0} \\ {0} & {0} & {1}\end{pmatrix}$ \\ \hline
  $\mathbf{4}$ & $\dfrac{e^{\pi i/4} }{2 \sqrt{2}}\begin{pmatrix}
 1 & \sqrt{3} & 1 & \sqrt{3} \\
 \sqrt{3} & -1 & \sqrt{3} & -1 \\
 1 & \sqrt{3} & -1 & -\sqrt{3} \\
 \sqrt{3} & -1 & -\sqrt{3} & 1 \\
\end{pmatrix}$ & $\begin{pmatrix}
 1 & 0 & 0 & 0 \\
 0 & -1 & 0 & 0 \\
 0 & 0 & i & 0 \\
 0 & 0 & 0 & -i \\
\end{pmatrix}$ & $i\begin{pmatrix}{1} & {0} & {0} & {0} \\ {0} & {1} & {0} & {0} \\ {0} & {0} & {1} & {0} \\ {0} & {0} & {0} & {1}\end{pmatrix}$ \\ \hline
 $\mathbf{4}'$ & $\dfrac{i e^{\pi i/4}}{2 \sqrt{2}}\begin{pmatrix}
 1 & \sqrt{3} & 1 & \sqrt{3} \\
 \sqrt{3} & -1 & \sqrt{3} & -1 \\
 1 & \sqrt{3} & -1 & -\sqrt{3} \\
 \sqrt{3} & -1 & -\sqrt{3} & 1 \\
\end{pmatrix}$ & $-i\begin{pmatrix}
 1 & 0 & 0 & 0 \\
 0 & -1 & 0 & 0 \\
 0 & 0 & i & 0 \\
 0 & 0 & 0 & -i \\
\end{pmatrix}$ & $-i\begin{pmatrix}{1} & {0} & {0} & {0} \\ {0} & {1} & {0} & {0} \\ {0} & {0} & {1} & {0} \\ {0} & {0} & {0} & {1}\end{pmatrix}$ \\ \hline\hline
\end{tabular}
\caption{The representation matrices of the generators $\widetilde{S}, \widetilde{T}$ and $\widetilde{R}$ for different irreducible representations of $\widetilde{S}_{4}$ in the $\widetilde{T}$-diagonal basis. }
\label{tab:Rep_baseB}
\end{table}
We list the Clebsch-Gordan coefficients of $\widetilde{S}_4$ in following.
All CG coefficients are expressed in the form of $\alpha\otimes\beta$, we use $\alpha_i(\beta_i)$ to denote the component of the left (right) basis vector $\alpha (\beta)$. The notations $\mathbf{I}$, $\mathbf{II}$,  $\mathbf{III}$ and $\mathbf{IV}$ stand for singlet, doublet, triplet and quartet representations of $\widetilde{S}_4$ respectively.

\begin{itemize}
\item $\mathbf{I}\otimes\mathbf{I}\to\mathbf{I}$\,,
\begin{equation}
\begin{array}{lll}
\begin{array}{c}
    ~\\ n=0 \\
  \end{array} ~~~~&
  \left.\begin{array}{l}
\mathbf{1}\otimes\mathbf{1}\to\mathbf{1_{s}},~~
\mathbf{1}\otimes\mathbf{1'}\to\mathbf{1'}\\
\mathbf{1}\otimes\mathbf{\widehat{1}}\to\mathbf{\widehat{1}},~~~
\mathbf{1}\otimes\mathbf{\widehat{1}'}\to\mathbf{\widehat{1}'}\\
\mathbf{1'}\otimes\mathbf{1'}\to\mathbf{1_{s}},~
\mathbf{1'}\otimes\mathbf{\widehat{1}}\to\mathbf{\widehat{1}'}\\
\mathbf{1'}\otimes\mathbf{\widehat{1}'}\to\mathbf{\widehat{1}},~~
\mathbf{\widehat{1}}\otimes\mathbf{\widehat{1}}\to\mathbf{1'_{s}}\\
\mathbf{\widehat{1}}\otimes\mathbf{\widehat{1}'}\to\mathbf{1},~~~
\mathbf{\widehat{1}'}\otimes\mathbf{\widehat{1}'}\to\mathbf{1'_{s}}
  \end{array}\right\} &~~~
\begin{array}{l}
\mathbf{I} \sim \alpha\beta
\end{array}
\end{array}\nonumber
\end{equation}
\end{itemize}

\begin{itemize}
\item $\mathbf{I}\otimes\mathbf{II}\to\mathbf{II}$\,,
\begin{equation}
\begin{array}{lll}
\begin{array}{c}
    ~\\ \\  \\ \\ [-1ex]n=0 \\ \\ \\ \\  \\ \\ \\[2.1ex]n=1 \\
  \end{array} ~~~~&
  \left.\begin{array}{l}
 \mathbf{1}\otimes\mathbf{2}\to\mathbf{2},~~~
\mathbf{1}\otimes\mathbf{2'}\to\mathbf{2'}\\
\mathbf{1}\otimes\mathbf{\widehat{2}}\to\mathbf{\widehat{2}},~~~
\mathbf{1}\otimes\mathbf{\widehat{2}'}\to\mathbf{\widehat{2}'}\\
\mathbf{1}\otimes\mathbf{\widetilde{2}}\to\mathbf{\widetilde{2}},~~~
\mathbf{1}\otimes\mathbf{\widetilde{2}'}\to\mathbf{\widetilde{2}'}\\
\mathbf{1'}\otimes\mathbf{\widehat{2}}\to\mathbf{\widehat{2}'},~~
\mathbf{1'}\otimes\mathbf{\widehat{2}'}\to\mathbf{\widehat{2}}\\
\mathbf{1'}\otimes\mathbf{\widetilde{2}}\to\mathbf{\widetilde{2}'},~~
\mathbf{1'}\otimes\mathbf{\widetilde{2}'}\to\mathbf{\widetilde{2}}\\
\mathbf{\widehat{1}}\otimes\mathbf{2}\to\mathbf{2'},~~~
\mathbf{\widehat{1}}\otimes\mathbf{\widehat{2}}\to\mathbf{\widetilde{2}}\\
\mathbf{\widehat{1}}\otimes\mathbf{\widehat{2}'}\to\mathbf{\widetilde{2}'},~~
\mathbf{\widehat{1}}\otimes\mathbf{\widetilde{2}}\to\mathbf{\widehat{2}'}\\
\mathbf{\widehat{1}}\otimes\mathbf{\widetilde{2}'}\to\mathbf{\widehat{2}},~~~
\mathbf{\widehat{1}'}\otimes\mathbf{2'}\to\mathbf{2}\\
\mathbf{\widehat{1}'}\otimes\mathbf{\widehat{2}}\to\mathbf{\widetilde{2}'},~~
\mathbf{\widehat{1}'}\otimes\mathbf{\widehat{2}'}\to\mathbf{\widetilde{2}}\\
\mathbf{\widehat{1}'}\otimes\mathbf{\widetilde{2}}\to\mathbf{\widehat{2}},~~~
\mathbf{\widehat{1}'}\otimes\mathbf{\widetilde{2}'}\to\mathbf{\widehat{2}'}\\ \\
\mathbf{1'}\otimes\mathbf{2}\to\mathbf{2},~~~
\mathbf{1'}\otimes\mathbf{2'}\to\mathbf{2'}\\
\mathbf{\widehat{1}}\otimes\mathbf{2'}\to\mathbf{2},~~~
\mathbf{\widehat{1}'}\otimes\mathbf{2}\to\mathbf{2'}
  \end{array}\right\} &~~~
\begin{array}{l}
\mathbf{II} \sim  \alpha M^{(n)}\begin{pmatrix}
\beta_1\\
\beta_2
\end{pmatrix}\\
\end{array}
\end{array}\nonumber
\end{equation}
\end{itemize}

where $M^{(0)}=
\begin{pmatrix}
  1 & 0\\0 & 1
\end{pmatrix}
$, $M^{(1)}=
\begin{pmatrix}
  0 & 1\\-1 & 0
\end{pmatrix}
$, it's the same below.

\begin{itemize}
\item $\mathbf{I}\otimes\mathbf{III}\to\mathbf{III}$\,,
\begin{equation}
\begin{array}{lll}
\begin{array}{c}
    ~\\ n=0 \\
  \end{array} ~~~~&
  \left.\begin{array}{l}
\mathbf{1}\otimes\mathbf{3}\to\mathbf{3},~~~
\mathbf{1}\otimes\mathbf{3'}\to\mathbf{3'}\\
\mathbf{1}\otimes\mathbf{\widehat{3}}\to\mathbf{\widehat{3}},~~~
\mathbf{1}\otimes\mathbf{\widehat{3}'}\to\mathbf{\widehat{3}'}\\
\mathbf{1'}\otimes\mathbf{3}\to\mathbf{3'},~~
\mathbf{1'}\otimes\mathbf{3'}\to\mathbf{3}\\
\mathbf{1'}\otimes\mathbf{\widehat{3}}\to\mathbf{\widehat{3}'},~~
\mathbf{1'}\otimes\mathbf{\widehat{3}'}\to\mathbf{\widehat{3}}\\
\mathbf{\widehat{1}}\otimes\mathbf{3}\to\mathbf{\widehat{3}},~~~\,
\mathbf{\widehat{1}}\otimes\mathbf{3'}\to\mathbf{\widehat{3}'}\\
\mathbf{\widehat{1}}\otimes\mathbf{\widehat{3}}\to\mathbf{3'},~~~
\mathbf{\widehat{1}}\otimes\mathbf{\widehat{3}'}\to\mathbf{3}\\
\mathbf{\widehat{1}'}\otimes\mathbf{3}\to\mathbf{\widehat{3}'},~~\,
\mathbf{\widehat{1}'}\otimes\mathbf{3'}\to\mathbf{\widehat{3}}\\
\mathbf{\widehat{1}'}\otimes\mathbf{\widehat{3}}\to\mathbf{3},~~~
\mathbf{\widehat{1}'}\otimes\mathbf{\widehat{3}'}\to\mathbf{3'}
  \end{array}\right\} &~~~
\begin{array}{l}
\mathbf{III} \sim \alpha\begin{pmatrix}
\beta_1\\
\beta_2\\
\beta_3
\end{pmatrix}
\end{array}
\end{array}\nonumber
\end{equation}
\end{itemize}

\begin{itemize}
\item $\mathbf{I}\otimes\mathbf{IV}\to\mathbf{IV}$\,,
\begin{equation}
\begin{array}{lll}
\begin{array}{c}
    ~\\ [-4ex]n=0  \\ \\ \\ \\[3.5ex]n=1 \\
  \end{array} ~~~~&
  \left.\begin{array}{l}
\mathbf{1}\otimes\mathbf{4}\to\mathbf{4}\\
\mathbf{1}\otimes\mathbf{4'}\to\mathbf{4'}\\
\mathbf{\widehat{1}}\otimes\mathbf{4}\to\mathbf{4'}\\
\mathbf{\widehat{1}'}\otimes\mathbf{4'}\to\mathbf{4}\\ \\
\mathbf{1'}\otimes\mathbf{4}\to\mathbf{4}\\
\mathbf{1'}\otimes\mathbf{4'}\to\mathbf{4'}\\
\mathbf{\widehat{1}}\otimes\mathbf{4'}\to\mathbf{4}\\
\mathbf{\widehat{1}'}\otimes\mathbf{4}\to\mathbf{4'}
  \end{array}\right\} &~~~
\begin{array}{l}
\mathbf{IV} \sim \alpha\begin{pmatrix}
  M^{(n)}\begin{pmatrix}
\beta_1\\
\beta_2
\end{pmatrix}\\[3ex]
  M^{(n)}\begin{pmatrix}
\beta_3\\
\beta_4
\end{pmatrix}
\end{pmatrix}
\end{array}
\end{array}\nonumber
\end{equation}
\end{itemize}

\begin{itemize}
\item $\mathbf{II}\otimes\mathbf{II}\to\mathbf{I}_1\oplus\mathbf{I}_2\oplus\mathbf{II}$\,,
\begin{equation}
\begin{array}{lll}
\begin{array}{c}
    ~\\ [-1.1ex]n=0  \\ \\[1ex]n=1 \\
  \end{array} ~~~~&
  \left.\begin{array}{l}
\mathbf{2}\otimes\mathbf{2}\to\mathbf{1'_{a}}\oplus\mathbf{1_{s}}\oplus\mathbf{2_{s}}\\
\mathbf{2}\otimes\mathbf{2'}\to\mathbf{\widehat{1}'}\oplus\mathbf{\widehat{1}}\oplus\mathbf{2'}\\\\
\mathbf{2'}\otimes\mathbf{2'}\to\mathbf{1_{a}}\oplus\mathbf{1'_{s}}\oplus\mathbf{2_{s}}
  \end{array}\right\} &~~~
\begin{array}{l}
\mathbf{I}_1 \sim \alpha_1\beta_2-\alpha_2\beta_1\\[1ex]
\mathbf{I}_2 \sim \alpha_1\beta_1+\alpha_2\beta_2\\[1ex]
\mathbf{II} \sim M^{(n)}\begin{pmatrix}
-\alpha_1\beta_1+\alpha_2\beta_2\\
\alpha_1\beta_2+\alpha_2\beta_1
\end{pmatrix}
\end{array}
\end{array}\nonumber
\end{equation}
\end{itemize}

\begin{itemize}
\item $\mathbf{II}\otimes\mathbf{II}\to\mathbf{IV}$\,,
\begin{equation}
\begin{array}{lll}
\begin{array}{c}
    ~\\ [-4ex]n=0  \\ \\ \\ \\[3.5ex]n=1 \\
  \end{array} ~~~~&
  \left.\begin{array}{l}
\mathbf{2}\otimes\mathbf{\widehat{2}}\to\mathbf{4}\\
\mathbf{2}\otimes\mathbf{\widetilde{2}}\to\mathbf{4'}\\
\mathbf{2'}\otimes\mathbf{\widehat{2}}\to\mathbf{4'}\\
\mathbf{2'}\otimes\mathbf{\widetilde{2}'}\to\mathbf{4}\\\\
\mathbf{2}\otimes\mathbf{\widehat{2}'}\to\mathbf{4}\\
\mathbf{2}\otimes\mathbf{\widetilde{2}'}\to\mathbf{4'}\\
\mathbf{2'}\otimes\mathbf{\widehat{2}'}\to\mathbf{4'}\\
\mathbf{2'}\otimes\mathbf{\widetilde{2}}\to\mathbf{4}
  \end{array}\right\} &~~~
\begin{array}{l}
\mathbf{IV} \sim \begin{pmatrix}
  M^{(n)}\begin{pmatrix}
    \alpha_1\\
   -\alpha_2\\
 \end{pmatrix}\beta_1\\[3ex]
 M^{(n)}\begin{pmatrix}
  -\alpha_1\\
  \alpha_2
 \end{pmatrix}\beta_2
\end{pmatrix}
\end{array}
\end{array}\nonumber
\end{equation}
\end{itemize}

\begin{itemize}
\item $\mathbf{II}\otimes\mathbf{II}\to\mathbf{I}\oplus\mathbf{III}$\,,
\begin{equation}
\begin{array}{lll}
\begin{array}{c}
    ~\\ n=0 \\
  \end{array} ~~~~&
  \left.\begin{array}{l}
\mathbf{\widehat{2}}\otimes\mathbf{\widehat{2}}\to\mathbf{\widehat{1}'_{a}}\oplus\mathbf{\widehat{3}'_{s}}\\
\mathbf{\widehat{2}}\otimes\mathbf{\widehat{2}'}\to\mathbf{\widehat{1}}\oplus\mathbf{\widehat{3}}\\
\mathbf{\widehat{2}}\otimes\mathbf{\widetilde{2}}\to\mathbf{1}\oplus\mathbf{3}\\
\mathbf{\widehat{2}}\otimes\mathbf{\widetilde{2}'}\to\mathbf{1'}\oplus\mathbf{3'}\\
\mathbf{\widehat{2}'}\otimes\mathbf{\widehat{2}'}\to\mathbf{\widehat{1}'_{a}}\oplus\mathbf{\widehat{3}'_{s}}\\
\mathbf{\widehat{2}'}\otimes\mathbf{\widetilde{2}}\to\mathbf{1'}\oplus\mathbf{3'}\\
\mathbf{\widehat{2}'}\otimes\mathbf{\widetilde{2}'}\to\mathbf{1}\oplus\mathbf{3}\\
\mathbf{\widetilde{2}}\otimes\mathbf{\widetilde{2}}\to\mathbf{\widehat{1}_{a}}\oplus\mathbf{\widehat{3}_{s}}\\
\mathbf{\widetilde{2}}\otimes\mathbf{\widetilde{2}'}\to\mathbf{\widehat{1}'}\oplus\mathbf{\widehat{3}'}\\
\mathbf{\widetilde{2}'}\otimes\mathbf{\widetilde{2}'}\to\mathbf{\widehat{1}_{a}}\oplus\mathbf{\widehat{3}_{s}}
  \end{array}\right\} &~~~
\begin{array}{l}
\mathbf{I} \sim \alpha_1\beta_2-\alpha_2\beta_1\\[1ex]
\mathbf{III} \sim \begin{pmatrix}
\alpha_1\beta_2+\alpha_2\beta_1\\
-\sqrt{2}\alpha_2\beta_2\\
\sqrt{2}\alpha_1\beta_1
\end{pmatrix}
\end{array}
\end{array}\nonumber
\end{equation}
\end{itemize}

\begin{itemize}
\item $\mathbf{II}\otimes\mathbf{III}\to\mathbf{III}_1\oplus\mathbf{III}_2$\,,
\begin{equation}
\begin{array}{lll}
\begin{array}{c}
    ~\\ n=0 \\
  \end{array} ~~~~&
  \left.\begin{array}{l}
\mathbf{2}\otimes\mathbf{3}\to\mathbf{3}\oplus\mathbf{3'}\\
\mathbf{2}\otimes\mathbf{3'}\to\mathbf{3'}\oplus\mathbf{3}\\
\mathbf{2}\otimes\mathbf{\widehat{3}}\to\mathbf{\widehat{3}}\oplus\mathbf{\widehat{3}'}\\
\mathbf{2}\otimes\mathbf{\widehat{3}'}\to\mathbf{\widehat{3}'}\oplus\mathbf{\widehat{3}}\\
\mathbf{2'}\otimes\mathbf{3}\to\mathbf{\widehat{3}}\oplus\mathbf{\widehat{3}'}\\
\mathbf{2'}\otimes\mathbf{3'}\to\mathbf{\widehat{3}'}\oplus\mathbf{\widehat{3}}\\
\mathbf{2'}\otimes\mathbf{\widehat{3}}\to\mathbf{3'}\oplus\mathbf{3}\\
\mathbf{2'}\otimes\mathbf{\widehat{3}'}\to\mathbf{3}\oplus\mathbf{3'}
  \end{array}\right\} &~~~
\begin{array}{l}
\mathbf{III}_1 \sim \begin{pmatrix}
2\alpha_1\beta_1\\
-\alpha_1\beta_2+\sqrt{3}\alpha_2\beta_3\\
-\alpha_1\beta_3+\sqrt{3}\alpha_2\beta_2
\end{pmatrix}\\[5ex]
\mathbf{III}_2 \sim \begin{pmatrix}
-2\alpha_2\beta_1\\
\sqrt{3}\alpha_1\beta_3+\alpha_2\beta_2\\
\sqrt{3}\alpha_1\beta_2+\alpha_2\beta_3
\end{pmatrix}
\end{array}
\end{array}\nonumber
\end{equation}
\end{itemize}

\begin{itemize}
\item $\mathbf{II}\otimes\mathbf{III}\to\mathbf{II}\oplus\mathbf{IV}$\,,
\begin{equation}
\begin{array}{lll}
\begin{array}{c}
    ~\\ [-3ex]n=0  \\ \\ \\ \\ \\ \\ \\ \\[3ex]n=1 \\
  \end{array} ~~~~&
  \left.\begin{array}{l}
\mathbf{\widehat{2}}\otimes\mathbf{3'}\to\mathbf{\widehat{2}'}\oplus\mathbf{4}\\
\mathbf{\widehat{2}}\otimes\mathbf{\widehat{3}'}\to\mathbf{\widetilde{2}'}\oplus\mathbf{4'}\\
\mathbf{\widehat{2}'}\otimes\mathbf{3}\to\mathbf{\widehat{2}'}\oplus\mathbf{4}\\
\mathbf{\widehat{2}'}\otimes\mathbf{\widehat{3}}\to\mathbf{\widetilde{2}'}\oplus\mathbf{4'}\\
\mathbf{\widetilde{2}}\otimes\mathbf{3'}\to\mathbf{\widetilde{2}'}\oplus\mathbf{4'}\\
\mathbf{\widetilde{2}}\otimes\mathbf{\widehat{3}}\to\mathbf{\widehat{2}'}\oplus\mathbf{4}\\
\mathbf{\widetilde{2}'}\otimes\mathbf{3}\to\mathbf{\widetilde{2}'}\oplus\mathbf{4'}\\
\mathbf{\widetilde{2}'}\otimes\mathbf{\widehat{3}'}\to\mathbf{\widehat{2}'}\oplus\mathbf{4}\\\\
\mathbf{\widehat{2}}\otimes\mathbf{3}\to\mathbf{\widehat{2}}\oplus\mathbf{4}\\
\mathbf{\widehat{2}}\otimes\mathbf{\widehat{3}}\to\mathbf{\widetilde{2}}\oplus\mathbf{4'}\\
\mathbf{\widehat{2}'}\otimes\mathbf{3'}\to\mathbf{\widehat{2}}\oplus\mathbf{4}\\
\mathbf{\widehat{2}'}\otimes\mathbf{\widehat{3}'}\to\mathbf{\widetilde{2}}\oplus\mathbf{4'}\\
\mathbf{\widetilde{2}}\otimes\mathbf{3}\to\mathbf{\widetilde{2}}\oplus\mathbf{4'}\\
\mathbf{\widetilde{2}}\otimes\mathbf{\widehat{3}'}\to\mathbf{\widehat{2}}\oplus\mathbf{4}\\
\mathbf{\widetilde{2}'}\otimes\mathbf{3'}\to\mathbf{\widetilde{2}}\oplus\mathbf{4'}\\
\mathbf{\widetilde{2}'}\otimes\mathbf{\widehat{3}}\to\mathbf{\widehat{2}}\oplus\mathbf{4}
  \end{array}\right\} &~~~
\begin{array}{l}
\mathbf{II} \sim \begin{pmatrix}
-\alpha_1\beta_1+\sqrt{2}\alpha_2\beta_3\\
\sqrt{2}\alpha_1\beta_2+\alpha_2\beta_1
\end{pmatrix}\\[5ex]
\mathbf{IV} \sim
   \begin{pmatrix}
     M^{(n)}\begin{pmatrix}
       -\sqrt{3}\alpha_2\beta_2\\
       \sqrt{2}\alpha_1\beta_1+\alpha_2\beta_3\\
     \end{pmatrix}\\[2ex]
     M^{(n)}\begin{pmatrix}
       \sqrt{3}\alpha_1\beta_3\\
       -\alpha_1\beta_2+\sqrt{2}\alpha_2\beta_1
     \end{pmatrix}
   \end{pmatrix}
\end{array}
\end{array}\nonumber
\end{equation}
\end{itemize}

\begin{itemize}
\item $\mathbf{II}\otimes\mathbf{IV}\to\mathbf{II}_1\oplus\mathbf{II}_2\oplus\mathbf{IV}$\,,
\begin{equation}
\begin{array}{lll}
\begin{array}{c}
    ~\\[-5ex] n=0  \\[1ex] \\\\ [-1ex]n=1 \\
  \end{array} ~~~~&
  \left.\begin{array}{l}
   \mathbf{2'}\otimes\mathbf{4'}\to\mathbf{\widehat{2}}\oplus\mathbf{\widehat{2}'}\oplus\mathbf{4}\\\\
   \mathbf{2}\otimes\mathbf{4}\to\mathbf{\widehat{2}'}\oplus\mathbf{\widehat{2}}\oplus\mathbf{4}\\
\mathbf{2}\otimes\mathbf{4'}\to\mathbf{\widetilde{2}'}\oplus\mathbf{\widetilde{2}}\oplus\mathbf{4'}\\
\mathbf{2'}\otimes\mathbf{4}\to\mathbf{\widetilde{2}'}\oplus\mathbf{\widetilde{2}}\oplus\mathbf{4'}
  \end{array}\right\} &~~~
\begin{array}{l}
\mathbf{II}_1 \sim \begin{pmatrix}
\alpha_1\beta_2+\alpha_2\beta_1\\
-\alpha_1\beta_4-\alpha_2\beta_3
\end{pmatrix}\\[3ex]
\mathbf{II}_2 \sim \begin{pmatrix}
\alpha_1\beta_1-\alpha_2\beta_2\\
-\alpha_1\beta_3+\alpha_2\beta_4
\end{pmatrix}\\[3ex]
\mathbf{IV} \sim \begin{pmatrix}
     M^{(n)}\begin{pmatrix}
       \alpha_1\beta_2-\alpha_2\beta_1\\
       \alpha_1\beta_1+\alpha_2\beta_2\\
     \end{pmatrix}\\[2ex]
     M^{(n)}\begin{pmatrix}
       \alpha_1\beta_4-\alpha_2\beta_3\\
       \alpha_1\beta_3+\alpha_2\beta_4
     \end{pmatrix}
   \end{pmatrix}
\end{array}
\end{array}\nonumber
\end{equation}
\end{itemize}

\begin{itemize}
\item $\mathbf{II}\otimes\mathbf{IV}\to\mathbf{II}\oplus\mathbf{III}_1\oplus\mathbf{III}_2$\,,
\begin{equation}
\begin{array}{lll}
\begin{array}{c}
    ~\\ [-3ex]n=0  \\ \\ \\ \\[3ex]n=1 \\
  \end{array} ~~~~&
  \left.\begin{array}{l}
\mathbf{\widehat{2}}\otimes\mathbf{4}\to\mathbf{2'}\oplus\mathbf{\widehat{3}'}\oplus\mathbf{\widehat{3}}\\
\mathbf{\widehat{2}'}\otimes\mathbf{4'}\to\mathbf{2}\oplus\mathbf{3'}\oplus\mathbf{3}\\
\mathbf{\widetilde{2}'}\otimes\mathbf{4}\to\mathbf{2}\oplus\mathbf{3'}\oplus\mathbf{3}\\
\mathbf{\widetilde{2}'}\otimes\mathbf{4'}\to\mathbf{2'}\oplus\mathbf{\widehat{3}'}\oplus\mathbf{\widehat{3}}\\\\
\mathbf{\widehat{2}}\otimes\mathbf{4'}\to\mathbf{2}\oplus\mathbf{3}\oplus\mathbf{3'}\\
\mathbf{\widehat{2}'}\otimes\mathbf{4}\to\mathbf{2'}\oplus\mathbf{\widehat{3}}\oplus\mathbf{\widehat{3}'}\\
\mathbf{\widetilde{2}}\otimes\mathbf{4}\to\mathbf{2}\oplus\mathbf{3}\oplus\mathbf{3'}\\
\mathbf{\widetilde{2}}\otimes\mathbf{4'}\to\mathbf{2'}\oplus\mathbf{\widehat{3}}\oplus\mathbf{\widehat{3}'}
  \end{array}\right\} &~~~
\begin{array}{l}
\mathbf{II} \sim M^{(n)}\begin{pmatrix}
\alpha_1\beta_4+\alpha_2\beta_2\\
\alpha_1\beta_3+\alpha_2\beta_1
\end{pmatrix}\\[5ex]
\mathbf{III}_1 \sim \begin{pmatrix}
\sqrt{2}\alpha_1\beta_3-\sqrt{2}\alpha_2\beta_1\\
\sqrt{3}\alpha_1\beta_2+\alpha_2\beta_3\\
\alpha_1\beta_1+\sqrt{3}\alpha_2\beta_4
\end{pmatrix}\\[5ex]
\mathbf{III}_2 \sim \begin{pmatrix}
\sqrt{2}\alpha_1\beta_4-\sqrt{2}\alpha_2\beta_2\\
-\sqrt{3}\alpha_1\beta_1+\alpha_2\beta_4\\
\alpha_1\beta_2-\sqrt{3}\alpha_2\beta_3
\end{pmatrix}
\end{array}
\end{array}\nonumber
\end{equation}
\end{itemize}

\begin{itemize}
\item $\mathbf{III}\otimes\mathbf{III}\to\mathbf{I}\oplus\mathbf{II}\oplus\mathbf{III}_1\oplus\mathbf{III}_2$\,,
\begin{equation}
\begin{array}{lll}
\begin{array}{c}
    ~\\ [-3ex]n=0  \\ \\ \\ \\[3ex]n=1 \\
  \end{array} ~~~~&
  \left.\begin{array}{l}
\mathbf{3}\otimes\mathbf{3}\to\mathbf{1_{s}}\oplus\mathbf{2_{s}}\oplus\mathbf{3_{a}}\oplus\mathbf{3'_{s}}\\
\mathbf{3}\otimes\mathbf{\widehat{3}}\to\mathbf{\widehat{1}}\oplus\mathbf{2'}\oplus\mathbf{\widehat{3}}\oplus\mathbf{\widehat{3}'}\\
\mathbf{3'}\otimes\mathbf{3'}\to\mathbf{1_{s}}\oplus\mathbf{2_{s}}\oplus\mathbf{3_{a}}\oplus\mathbf{3'_{s}}\\
\mathbf{3'}\otimes\mathbf{\widehat{3}'}\to\mathbf{\widehat{1}}\oplus\mathbf{2'}\oplus\mathbf{\widehat{3}}\oplus\mathbf{\widehat{3}'}\\
\mathbf{\widehat{3}}\otimes\mathbf{\widehat{3}'}\to\mathbf{1}\oplus\mathbf{2}\oplus\mathbf{3}\oplus\mathbf{3'}\\\\
\mathbf{3}\otimes\mathbf{3'}\to\mathbf{1'}\oplus\mathbf{2}\oplus\mathbf{3'}\oplus\mathbf{3}\\
\mathbf{3}\otimes\mathbf{\widehat{3}'}\to\mathbf{\widehat{1}'}\oplus\mathbf{2'}\oplus\mathbf{\widehat{3}'}\oplus\mathbf{\widehat{3}}\\
\mathbf{3'}\otimes\mathbf{\widehat{3}}\to\mathbf{\widehat{1}'}\oplus\mathbf{2'}\oplus\mathbf{\widehat{3}'}\oplus\mathbf{\widehat{3}}\\
\mathbf{\widehat{3}}\otimes\mathbf{\widehat{3}}\to\mathbf{1'_{s}}\oplus\mathbf{2_{s}}\oplus\mathbf{3'_{a}}\oplus\mathbf{3_{s}}\\
\mathbf{\widehat{3}'}\otimes\mathbf{\widehat{3}'}\to\mathbf{1'_{s}}\oplus\mathbf{2_{s}}\oplus\mathbf{3'_{a}}\oplus\mathbf{3_{s}}
  \end{array}\right\} &~~~
\begin{array}{l}
\mathbf{I} \sim \alpha_1\beta_1+\alpha_2\beta_3+\alpha_3\beta_2\\[3ex]
\mathbf{II} \sim M^{(n)}\begin{pmatrix}
2\alpha_1\beta_1-\alpha_2\beta_3-\alpha_3\beta_2\\
\sqrt{3}\alpha_2\beta_2+\sqrt{3}\alpha_3\beta_3
\end{pmatrix}\\[3ex]
\mathbf{III}_1 \sim \begin{pmatrix}
\alpha_2\beta_3-\alpha_3\beta_2\\
\alpha_1\beta_2-\alpha_2\beta_1\\
-\alpha_1\beta_3+\alpha_3\beta_1
\end{pmatrix}\\[5ex]
\mathbf{III}_2 \sim \begin{pmatrix}
\alpha_2\beta_2-\alpha_3\beta_3\\
-\alpha_1\beta_3-\alpha_3\beta_1\\
\alpha_1\beta_2+\alpha_2\beta_1
\end{pmatrix}
\end{array}
\end{array}\nonumber
\end{equation}
\end{itemize}

\begin{itemize}
\item $\mathbf{III}\otimes\mathbf{IV}\to\mathbf{II}_1\oplus\mathbf{II}_2\oplus\mathbf{IV}_1\oplus\mathbf{IV}_2$\,,
\begin{equation}
\begin{array}{lll}
\begin{array}{c}
    ~\\ [-3ex]n=0  \\ \\ \\ \\[3ex]n=1 \\
  \end{array} ~~~~&
  \left.\begin{array}{l}
\mathbf{3'}\otimes\mathbf{4}\to\mathbf{\widehat{2}'}\oplus\mathbf{\widehat{2}}\oplus\mathbf{4_{I}}\oplus\mathbf{4_{II}}\\
\mathbf{3'}\otimes\mathbf{4'}\to\mathbf{\widetilde{2}'}\oplus\mathbf{\widetilde{2}}\oplus\mathbf{4'_{I}}\oplus\mathbf{4'_{II}}\\
\mathbf{\widehat{3}}\otimes\mathbf{4'}\to\mathbf{\widehat{2}'}\oplus\mathbf{\widehat{2}}\oplus\mathbf{4_{I}}\oplus\mathbf{4_{II}}\\
\mathbf{\widehat{3}'}\otimes\mathbf{4}\to\mathbf{\widetilde{2}'}\oplus\mathbf{\widetilde{2}}\oplus\mathbf{4'_{I}}\oplus\mathbf{4'_{II}}\\\\
\mathbf{3}\otimes\mathbf{4}\to\mathbf{\widehat{2}}\oplus\mathbf{\widehat{2}'}\oplus\mathbf{4_{I}}\oplus\mathbf{4_{II}}\\
\mathbf{3}\otimes\mathbf{4'}\to\mathbf{\widetilde{2}}\oplus\mathbf{\widetilde{2}'}\oplus\mathbf{4'_{I}}\oplus\mathbf{4'_{II}}\\
\mathbf{\widehat{3}}\otimes\mathbf{4}\to\mathbf{\widetilde{2}}\oplus\mathbf{\widetilde{2}'}\oplus\mathbf{4'_{I}}\oplus\mathbf{4'_{II}}\\
\mathbf{\widehat{3}'}\otimes\mathbf{4'}\to\mathbf{\widehat{2}}\oplus\mathbf{\widehat{2}'}\oplus\mathbf{4_{I}}\oplus\mathbf{4_{II}}
  \end{array}\right\} & \begin{array}{l}
\mathbf{II}_1 \sim \begin{pmatrix}
\sqrt{2}\alpha_1\beta_1-\sqrt{3}\alpha_2\beta_4-\alpha_3\beta_3\\
\sqrt{2}\alpha_1\beta_3+\alpha_2\beta_1+\sqrt{3}\alpha_3\beta_2
\end{pmatrix}\\[3ex]
\mathbf{II}_2 \sim \begin{pmatrix}
\sqrt{2}\alpha_1\beta_2+\sqrt{3}\alpha_2\beta_3-\alpha_3\beta_4\\
\sqrt{2}\alpha_1\beta_4+\alpha_2\beta_2-\sqrt{3}\alpha_3\beta_1
\end{pmatrix}\\[4ex]
\mathbf{IV}_1 \sim \begin{pmatrix}
     M^{(n)}\begin{pmatrix}
       \alpha_1\beta_2-\sqrt{6}\alpha_2\beta_3-2\sqrt{2}\alpha_3\beta_4\\
       3\alpha_1\beta_1+\sqrt{6}\alpha_2\beta_4\\
     \end{pmatrix}\\[2ex]
     M^{(n)}\begin{pmatrix}
       -\alpha_1\beta_4-2\sqrt{2}\alpha_2\beta_2-\sqrt{6}\alpha_3\beta_1\\
       -3\alpha_1\beta_3+\sqrt{6}\alpha_3\beta_2
     \end{pmatrix}
   \end{pmatrix}\\[7ex]
\mathbf{IV}_2 \sim \begin{pmatrix}
     M^{(n)}\begin{pmatrix}
       3\alpha_1\beta_2-\sqrt{6}\alpha_2\beta_3\\
       \alpha_1\beta_1+\sqrt{6}\alpha_2\beta_4-2\sqrt{2}\alpha_3\beta_3\\
     \end{pmatrix}\\[2ex]
     M^{(n)}\begin{pmatrix}
       -3\alpha_1\beta_4-\sqrt{6}\alpha_3\beta_1\\
       -\alpha_1\beta_3-2\sqrt{2}\alpha_2\beta_1+\sqrt{6}\alpha_3\beta_2
     \end{pmatrix}
   \end{pmatrix}
\end{array}
\end{array}\nonumber
\end{equation}
\end{itemize}

\begin{itemize}
\item $\mathbf{IV}\otimes\mathbf{IV}\to\mathbf{I}_1\oplus\mathbf{I}_2\oplus\mathbf{II}\oplus\mathbf{III}_1\oplus\mathbf{III}_2\oplus\mathbf{III}_3\oplus\mathbf{III}_4$\,,
\begin{equation}
\begin{array}{lll}
\begin{array}{c}
    \\[-3.5ex] n=0  \\[1.5ex] \\  [0ex]n=1 \\
  \end{array} ~~&
  \left.\begin{array}{l}
\mathbf{4}\otimes\mathbf{4}\to\mathbf{\widehat{1}_{s}}\oplus\mathbf{\widehat{1}'_{a}}\oplus\mathbf{2'_{a}}\oplus\mathbf{\widehat{3}_{a}}\oplus\mathbf{\widehat{3}_{s}}\oplus\mathbf{\widehat{3}'_{sI}}\oplus\mathbf{\widehat{3}'_{sII}}\\\\
\mathbf{4}\otimes\mathbf{4'}\to\mathbf{1'}\oplus\mathbf{1}\oplus\mathbf{2}\oplus\mathbf{3'_{II}}\oplus\mathbf{3'_{I}}\oplus\mathbf{3_{I}}\oplus\mathbf{3_{II}}\\
\mathbf{4'}\otimes\mathbf{4'}\to\mathbf{\widehat{1}'_{s}}\oplus\mathbf{\widehat{1}_{a}}\oplus\mathbf{2'_{a}}\oplus\mathbf{\widehat{3}'_{a}}\oplus\mathbf{\widehat{3}'_{s}}\oplus\mathbf{\widehat{3}_{sI}}\oplus\mathbf{\widehat{3}_{sII}}
  \end{array}\right\} & \\ [10ex]
& \begin{array}{l}
\mathbf{I}_1 \sim \alpha_1\beta_4-\alpha_2\beta_3-\alpha_3\beta_2+\alpha_4\beta_1\\[4ex]
\mathbf{I}_2 \sim \alpha_1\beta_3+\alpha_2\beta_4-\alpha_3\beta_1-\alpha_4\beta_2\\[4ex]
\mathbf{II} \sim M^{(n)}\begin{pmatrix}
-\alpha_1\beta_4-\alpha_2\beta_3+\alpha_3\beta_2+\alpha_4\beta_1\\
\alpha_1\beta_3-\alpha_2\beta_4-\alpha_3\beta_1+\alpha_4\beta_2
\end{pmatrix}\\[4ex]
\mathbf{III}_1 \sim \begin{pmatrix}
\alpha_1\beta_4-\alpha_2\beta_3+\alpha_3\beta_2-\alpha_4\beta_1\\
\sqrt{2}\alpha_3\beta_4-\sqrt{2}\alpha_4\beta_3\\
-\sqrt{2}\alpha_1\beta_2+\sqrt{2}\alpha_2\beta_1
\end{pmatrix}\\[5ex]
\mathbf{III}_2 \sim \begin{pmatrix}
\sqrt{2}\alpha_1\beta_4+\sqrt{2}\alpha_2\beta_3+\sqrt{2}\alpha_3\beta_2+\sqrt{2}\alpha_4\beta_1\\
\sqrt{3}\alpha_1\beta_1-\sqrt{3}\alpha_2\beta_2-\alpha_3\beta_4-\alpha_4\beta_3\\
\alpha_1\beta_2+\alpha_2\beta_1-\sqrt{3}\alpha_3\beta_3+\sqrt{3}\alpha_4\beta_4
\end{pmatrix}\\[5ex]
\mathbf{III}_3 \sim \begin{pmatrix}
\alpha_1\beta_3-3\alpha_2\beta_4+\alpha_3\beta_1-3\alpha_4\beta_2\\
-\sqrt{6}\alpha_1\beta_2-\sqrt{6}\alpha_2\beta_1-2\sqrt{2}\alpha_3\beta_3\\
2\sqrt{2}\alpha_1\beta_1+\sqrt{6}\alpha_3\beta_4+\sqrt{6}\alpha_4\beta_3
\end{pmatrix}\\[5ex]
\mathbf{III}_4 \sim \begin{pmatrix}
\alpha_1\beta_3+\alpha_2\beta_4+\alpha_3\beta_1+\alpha_4\beta_2\\
\sqrt{2}\alpha_3\beta_3+\sqrt{2}\alpha_4\beta_4\\
-\sqrt{2}\alpha_1\beta_1-\sqrt{2}\alpha_2\beta_2
\end{pmatrix}
\end{array}
\end{array}\nonumber
\end{equation}
\end{itemize}

\section{\label{sec:app-another-basis}CG coefficients and modular forms in another basis }

\begin{table}[!t]
\centering
\begin{tabular}{|c|c|c|c|} \hline\hline
 & $\widetilde{S}$ & $\widetilde{T}$ & $\widetilde{R}$\\ \hline
    $\mathbf{1},\mathbf{1^{\prime}}$ & $\pm 1$ & $\pm 1$ & $1$ \\ \hline
    $\mathbf{\widehat{1}},\mathbf{\widehat{1}^{\prime}}$ & $\pm i$ & $\mp i$ & $-1$ \\ \hline
    $\mathbf{2}$ & $\begin{pmatrix}{0} & {1} \\ {1} & {0}\end{pmatrix}$ & $\begin{pmatrix}{0} & {\omega^{2}} \\ {\omega} & {0}\end{pmatrix}$ & $\begin{pmatrix}{1} & {0} \\ {0} & {1}\end{pmatrix}$ \\ \hline
  $\mathbf{2}'$ & $i\begin{pmatrix}{0} & {1} \\ {1} & {0}\end{pmatrix}$ & $-i\begin{pmatrix}{0} & {\omega^{2}} \\ {\omega} & {0}\end{pmatrix}$ & $-\begin{pmatrix}{1} & {0} \\ {0} & {1}\end{pmatrix}$ \\ \hline
  $\widehat{\mathbf{2}},\widehat{\mathbf{2}}'$ & $\pm\dfrac{e^{\pi i/4}}{\sqrt{3}}\begin{pmatrix} \sqrt{2} & 1 \\ 1 & -\sqrt{2} \end{pmatrix}$ & $\pm\dfrac{ie^{\pi i/4}}{\sqrt{3}}\begin{pmatrix} -\sqrt{2} \omega & -\omega^2 \\ -\omega & \sqrt{2} \omega^2 \end{pmatrix}$ & $i\begin{pmatrix}{1} & {0} \\ {0} & {1}\end{pmatrix}$ \\ \hline
 $\widetilde{\mathbf{2}},\mathbf{\widetilde{2}'}$ & $\pm\dfrac{ie^{\pi i/4}}{\sqrt{3}}\begin{pmatrix} \sqrt{2} & 1 \\ 1 & -\sqrt{2} \end{pmatrix}$ & $\pm\dfrac{e^{\pi i/4}}{\sqrt{3}}\begin{pmatrix} -\sqrt{2} \omega & -\omega^2 \\ -\omega & \sqrt{2} \omega^2 \end{pmatrix}$ & $-i\begin{pmatrix}{1} & {0} \\ {0} & {1}\end{pmatrix}$ \\ \hline
    $\mathbf{3},\mathbf{3^{\prime}}$ & $\pm\dfrac{1}{3}\begin{pmatrix}{1} & {-2} & {-2} \\ {-2} & {-2} & {1} \\ {-2} & {1} & {-2}\end{pmatrix}$ & $\pm\dfrac{1}{3}\begin{pmatrix}{1} & {-2 \omega^{2}} & {-2 \omega} \\ {-2} & {-2 \omega^{2}} & {\omega} \\ {-2} & {\omega^{2}} & {-2 \omega}\end{pmatrix}$ & $\begin{pmatrix}{1} & {0} & {0} \\ {0} & {1} & {0} \\ {0} & {0} & {1}\end{pmatrix}$ \\ \hline
  $\mathbf{\widehat{3}}, \mathbf{\widehat{3}^{\prime}}$ & $\pm\dfrac{i}{3}\begin{pmatrix}{1} & {-2} & {-2} \\ {-2} & {-2} & {1} \\ {-2} & {1} & {-2}\end{pmatrix}$ & $\mp\dfrac{i}{3}\begin{pmatrix}{1} & {-2 \omega^{2}} & {-2 \omega} \\ {-2} & {-2 \omega^{2}} & {\omega} \\ {-2} & {\omega^{2}} & {-2 \omega}\end{pmatrix}$ & $-\begin{pmatrix}{1} & {0} & {0} \\ {0} & {1} & {0} \\ {0} & {0} & {1}\end{pmatrix}$ \\ \hline
  $\mathbf{4}$ & $\dfrac{e^{\pi i/4}}{\sqrt{3}}\begin{pmatrix} 0 & 1 & 0 & \sqrt{2} \\ 1 & 0 & \sqrt{2} & 0 \\ 0 & \sqrt{2} & 0 & -1 \\ \sqrt{2} & 0 & -1 & 0 \end{pmatrix}$ & $-\dfrac{ie^{\pi i/4}}{\sqrt{3}}\begin{pmatrix} 0 & 1 & 0 & \sqrt{2} \omega^2 \\ 1 & 0 & \sqrt{2} \omega & 0 \\ 0 & \sqrt{2} & 0 & -\omega^2 \\ \sqrt{2} & 0 & -\omega & 0 \end{pmatrix}$ & $i\begin{pmatrix}{1} & {0} & {0} & {0} \\ {0} & {1} & {0} & {0} \\ {0} & {0} & {1} & {0} \\ {0} & {0} & {0} & {1}\end{pmatrix}$ \\ \hline
 $\mathbf{4}'$ & $\dfrac{ie^{\pi i/4}}{\sqrt{3}}\begin{pmatrix} 0 & 1 & 0 & \sqrt{2} \\ 1 & 0 & \sqrt{2} & 0 \\ 0 & \sqrt{2} & 0 & -1 \\ \sqrt{2} & 0 & -1 & 0 \end{pmatrix}$ & $-\dfrac{e^{\pi i/4}}{\sqrt{3}}\begin{pmatrix} 0 & 1 & 0 & \sqrt{2} \omega^2 \\ 1 & 0 & \sqrt{2} \omega & 0 \\ 0 & \sqrt{2} & 0 & -\omega^2 \\ \sqrt{2} & 0 & -\omega & 0 \end{pmatrix}$ & $-i\begin{pmatrix}{1} & {0} & {0} & {0} \\ {0} & {1} & {0} & {0} \\ {0} & {0} & {1} & {0} \\ {0} & {0} & {0} & {1}\end{pmatrix}$ \\ \hline\hline
  \end{tabular}
\caption{The representation matrices of the generators $\widetilde{S}, \widetilde{T}$ and $\widetilde{R}$ for the irreducible representations of $\widetilde{S}_{4}$ in another basis, where $\omega=e^{2 \pi i/3}$.}
\label{tab:Rep_baseA}
\end{table}

In the following, we present another basis of $\widetilde{S}_4$, and the representation matrices of the generators $\widetilde{S}$, $\widetilde{T}$, $\widetilde{R}$ are listed in table~\ref{tab:Rep_baseA}.  In the irreducible representations $\mathbf{1}$, $\mathbf{1}^{\prime}$, $\mathbf{\widehat{1}}$, $\mathbf{\widehat{1}}^{\prime}$, $\mathbf{2}$, $\mathbf{2'}$, $\mathbf{3}$, $\mathbf{3}^{\prime}$, $\mathbf{\widehat{3}}$ and $\mathbf{\widehat{3}}^{\prime}$, the generators $\widetilde{S}$, $\widetilde{T}$ and $\widetilde{R}$ are represented by the same matrices as those of $S'_4$ in our previous work~\cite{Liu:2020akv}. Therefore $\widetilde{S}_4$ can not be distinguished from $S'_4$ in these representations, and this basis is more convenient to compare $\widetilde{S}_4$ models with $S'_4$ models of~\cite{Liu:2020akv}. Moreover, the CG coefficients are simpler in this basis although the $q-$expansions of the modular forms are more complex. In the following, we present all the CG coefficients of $\widetilde{S}_4$ group in this basis, and the components of the multiplets in the tensor product are denoted by $\alpha_i$ and $\beta_i$.

\begin{itemize}
\item $\mathbf{I}\otimes\mathbf{I}\to\mathbf{I}$\,,
\begin{equation}
\begin{array}{lll}
\begin{array}{c}
    ~\\ p=0 \\
  \end{array} ~~~~&
  \left.\begin{array}{l}
\mathbf{1}\otimes\mathbf{1}\to\mathbf{1_{s}},~~
\mathbf{1}\otimes\mathbf{1'}\to\mathbf{1'}\\
\mathbf{1}\otimes\mathbf{\widehat{1}}\to\mathbf{\widehat{1}},~~~
\mathbf{1}\otimes\mathbf{\widehat{1}'}\to\mathbf{\widehat{1}'}\\
\mathbf{1'}\otimes\mathbf{1'}\to\mathbf{1_{s}},~
\mathbf{1'}\otimes\mathbf{\widehat{1}}\to\mathbf{\widehat{1}'}\\
\mathbf{1'}\otimes\mathbf{\widehat{1}'}\to\mathbf{\widehat{1}},~~
\mathbf{\widehat{1}}\otimes\mathbf{\widehat{1}}\to\mathbf{1'_{s}}\\
\mathbf{\widehat{1}}\otimes\mathbf{\widehat{1}'}\to\mathbf{1},~~~
\mathbf{\widehat{1}'}\otimes\mathbf{\widehat{1}'}\to\mathbf{1'_{s}}
  \end{array}\right\} &~~~
\begin{array}{l}
\mathbf{I} \sim \alpha\beta
\end{array}
\end{array}\nonumber
\end{equation}
\end{itemize}

\begin{itemize}
\item $\mathbf{I}\otimes\mathbf{II}\to\mathbf{II}$\,,
\begin{equation}
\begin{array}{lll}
\begin{array}{c}
    ~\\ \\  \\ \\ [-1ex]p=0 \\ \\ \\ \\  \\ \\ \\[0.1ex]p=1 \\
  \end{array} ~~~~&
  \left.\begin{array}{l}
\mathbf{1}\otimes\mathbf{2}\to\mathbf{2},~~~\,
\mathbf{1}\otimes\mathbf{2'}\to\mathbf{2'}\\
\mathbf{1}\otimes\mathbf{\widehat{2}}\to\mathbf{\widehat{2}},~~~\,
\mathbf{1}\otimes\mathbf{\widehat{2}'}\to\mathbf{\widehat{2}'}\\
\mathbf{1}\otimes\mathbf{\tilde{2}}\to\mathbf{\tilde{2}},~~~\,
\mathbf{1}\otimes\mathbf{\tilde{2}'}\to\mathbf{\tilde{2}'}\\
\mathbf{1'}\otimes\mathbf{\widehat{2}}\to\mathbf{\widehat{2}'},~~
\mathbf{1'}\otimes\mathbf{\widehat{2}'}\to\mathbf{\widehat{2}}\\
\mathbf{1'}\otimes\mathbf{\tilde{2}}\to\mathbf{\tilde{2}'},~~
\mathbf{1'}\otimes\mathbf{\tilde{2}'}\to\mathbf{\tilde{2}}\\
\mathbf{\widehat{1}}\otimes\mathbf{2}\to\mathbf{2'},~~~
\mathbf{\widehat{1}}\otimes\mathbf{\widehat{2}}\to\mathbf{\tilde{2}}\\
\mathbf{\widehat{1}}\otimes\mathbf{\widehat{2}'}\to\mathbf{\tilde{2}'},~~
\mathbf{\widehat{1}}\otimes\mathbf{\tilde{2}}\to\mathbf{\widehat{2}'}\\
\mathbf{\widehat{1}}\otimes\mathbf{\tilde{2}'}\to\mathbf{\widehat{2}},~~~
\mathbf{\widehat{1}'}\otimes\mathbf{2'}\to\mathbf{2}\\
\mathbf{\widehat{1}'}\otimes\mathbf{\widehat{2}}\to\mathbf{\tilde{2}'},~~
\mathbf{\widehat{1}'}\otimes\mathbf{\widehat{2}'}\to\mathbf{\tilde{2}}\\
\mathbf{\widehat{1}'}\otimes\mathbf{\tilde{2}}\to\mathbf{\widehat{2}},~~~
\mathbf{\widehat{1}'}\otimes\mathbf{\tilde{2}'}\to\mathbf{\widehat{2}'}\\\\
\mathbf{1'}\otimes\mathbf{2}\to\mathbf{2},~~~
\mathbf{1'}\otimes\mathbf{2'}\to\mathbf{2'}\\
\mathbf{\widehat{1}}\otimes\mathbf{2'}\to\mathbf{2},~~~
\mathbf{\widehat{1}'}\otimes\mathbf{2}\to\mathbf{2'}
  \end{array}\right\} &~~~
\begin{array}{l}
\mathbf{II} \sim \alpha\begin{pmatrix}
\beta_1\\
(-1)^p\beta_2
\end{pmatrix}\\
\end{array}
\end{array}\nonumber
\end{equation}
\end{itemize}

\begin{itemize}
\item $\mathbf{I}\otimes\mathbf{III}\to\mathbf{III}$\,,
\begin{equation}
\begin{array}{lll}
\begin{array}{c}
    ~\\ p=0 \\
  \end{array} ~~~~&
  \left.\begin{array}{l}
\mathbf{1}\otimes\mathbf{3}\to\mathbf{3},~~~
\mathbf{1}\otimes\mathbf{3'}\to\mathbf{3'}\\
\mathbf{1}\otimes\mathbf{\widehat{3}}\to\mathbf{\widehat{3}},~~~
\mathbf{1}\otimes\mathbf{\widehat{3}'}\to\mathbf{\widehat{3}'}\\
\mathbf{1'}\otimes\mathbf{3}\to\mathbf{3'},~~
\mathbf{1'}\otimes\mathbf{3'}\to\mathbf{3}\\
\mathbf{1'}\otimes\mathbf{\widehat{3}}\to\mathbf{\widehat{3}'},~~
\mathbf{1'}\otimes\mathbf{\widehat{3}'}\to\mathbf{\widehat{3}}\\
\mathbf{\widehat{1}}\otimes\mathbf{3}\to\mathbf{\widehat{3}},~~~\,
\mathbf{\widehat{1}}\otimes\mathbf{3'}\to\mathbf{\widehat{3}'}\\
\mathbf{\widehat{1}}\otimes\mathbf{\widehat{3}}\to\mathbf{3'},~~~
\mathbf{\widehat{1}}\otimes\mathbf{\widehat{3}'}\to\mathbf{3}\\
\mathbf{\widehat{1}'}\otimes\mathbf{3}\to\mathbf{\widehat{3}'},~~
\mathbf{\widehat{1}'}\otimes\mathbf{3'}\to\mathbf{\widehat{3}}\\
\mathbf{\widehat{1}'}\otimes\mathbf{\widehat{3}}\to\mathbf{3},~~~
\mathbf{\widehat{1}'}\otimes\mathbf{\widehat{3}'}\to\mathbf{3'}
  \end{array}\right\} &~~~
\begin{array}{l}
\mathbf{III} \sim \alpha\begin{pmatrix}
\beta_1\\
\beta_2\\
\beta_3
\end{pmatrix}
\end{array}
\end{array}\nonumber
\end{equation}
\end{itemize}

\begin{itemize}
\item $\mathbf{I}\otimes\mathbf{IV}\to\mathbf{IV}$\,,
\begin{equation}
\begin{array}{lll}
\begin{array}{c}
    ~\\ [-4ex]p=0  \\ \\ \\ \\[3.5ex]p=1 \\
  \end{array} ~~~~&
  \left.\begin{array}{l}
\mathbf{1}\otimes\mathbf{4}\to\mathbf{4}\\
\mathbf{1}\otimes\mathbf{4'}\to\mathbf{4'}\\
\mathbf{\widehat{1}}\otimes\mathbf{4}\to\mathbf{4'}\\
\mathbf{\widehat{1}'}\otimes\mathbf{4'}\to\mathbf{4}\\\\
\mathbf{1'}\otimes\mathbf{4}\to\mathbf{4}\\
\mathbf{1'}\otimes\mathbf{4'}\to\mathbf{4'}\\
\mathbf{\widehat{1}}\otimes\mathbf{4'}\to\mathbf{4}\\
\mathbf{\widehat{1}'}\otimes\mathbf{4}\to\mathbf{4'}
  \end{array}\right\} &~~~
\begin{array}{l}
\mathbf{IV} \sim \alpha\begin{pmatrix}
\beta_1\\
(-1)^p\beta_2\\
\beta_3\\
(-1)^p\beta_4
\end{pmatrix}
\end{array}
\end{array}\nonumber
\end{equation}
\end{itemize}

\begin{itemize}
\item $\mathbf{II}\otimes\mathbf{II}\to\mathbf{I}_1\oplus\mathbf{I}_2\oplus\mathbf{II}$\,,
\begin{equation}
\begin{array}{lll}
\begin{array}{c}
    ~\\ [-1.1ex]p=0  \\ \\[1ex]p=1 \\
  \end{array} ~~~~&
  \left.\begin{array}{l}
\mathbf{2}\otimes\mathbf{2}\to\mathbf{1'_{a}}\oplus\mathbf{1_{s}}\oplus\mathbf{2_{s}}\\
\mathbf{2}\otimes\mathbf{2'}\to\mathbf{\widehat{1}'}\oplus\mathbf{\widehat{1}}\oplus\mathbf{2'}\\\\
\mathbf{2'}\otimes\mathbf{2'}\to\mathbf{1_{a}}\oplus\mathbf{1'_{s}}\oplus\mathbf{2_{s}}
  \end{array}\right\} &~~~
\begin{array}{l}
\mathbf{I}_1 \sim \alpha_1\beta_2-\alpha_2\beta_1\\[1ex]
\mathbf{I}_2 \sim \alpha_1\beta_2+\alpha_2\beta_1\\[1ex]
\mathbf{II} \sim \begin{pmatrix}
\alpha_2\beta_2\\
(-1)^p\alpha_1\beta_1
\end{pmatrix}
\end{array}
\end{array}\nonumber
\end{equation}
\end{itemize}

\begin{itemize}
\item $\mathbf{II}\otimes\mathbf{II}\to\mathbf{IV}$\,,
\begin{equation}
\begin{array}{lll}
\begin{array}{c}
    ~\\ [-4ex]p=0  \\ \\ \\ \\[3.5ex]p=1 \\
  \end{array} ~~~~&
  \left.\begin{array}{l}
\mathbf{2}\otimes\mathbf{\widehat{2}}\to\mathbf{4}\\
\mathbf{2}\otimes\mathbf{\tilde{2}}\to\mathbf{4'}\\
\mathbf{2'}\otimes\mathbf{\widehat{2}}\to\mathbf{4'}\\
\mathbf{2'}\otimes\mathbf{\tilde{2}'}\to\mathbf{4}\\\\
\mathbf{2}\otimes\mathbf{\widehat{2}'}\to\mathbf{4}\\
\mathbf{2}\otimes\mathbf{\tilde{2}'}\to\mathbf{4'}\\
\mathbf{2'}\otimes\mathbf{\widehat{2}'}\to\mathbf{4'}\\
\mathbf{2'}\otimes\mathbf{\tilde{2}}\to\mathbf{4}
  \end{array}\right\} &~~~
\begin{array}{l}
\mathbf{IV} \sim \begin{pmatrix}
(-1)^p\alpha_2\beta_1\\
\alpha_1\beta_2\\
-(-1)^p\alpha_2\beta_2\\
\alpha_1\beta_1
\end{pmatrix}
\end{array}
\end{array}\nonumber
\end{equation}
\end{itemize}

\begin{itemize}
\item $\mathbf{II}\otimes\mathbf{II}\to\mathbf{I}\oplus\mathbf{III}$\,,
\begin{equation}
\begin{array}{lll}
\begin{array}{c}
    ~\\ p=0 \\ \\
  \end{array} ~~~~&
  \left.\begin{array}{l}
\mathbf{\widehat{2}}\otimes\mathbf{\widehat{2}}\to\mathbf{\widehat{1}'_{a}}\oplus\mathbf{\widehat{3}'_{s}}\\
\mathbf{\widehat{2}}\otimes\mathbf{\widehat{2}'}\to\mathbf{\widehat{1}}\oplus\mathbf{\widehat{3}}\\
\mathbf{\widehat{2}}\otimes\mathbf{\tilde{2}}\to\mathbf{1}\oplus\mathbf{3}\\
\mathbf{\widehat{2}}\otimes\mathbf{\tilde{2}'}\to\mathbf{1'}\oplus\mathbf{3'}\\
\mathbf{\widehat{2}'}\otimes\mathbf{\widehat{2}'}\to\mathbf{\widehat{1}'_{a}}\oplus\mathbf{\widehat{3}'_{s}}\\
\mathbf{\widehat{2}'}\otimes\mathbf{\tilde{2}}\to\mathbf{1'}\oplus\mathbf{3'}\\
\mathbf{\widehat{2}'}\otimes\mathbf{\tilde{2}'}\to\mathbf{1}\oplus\mathbf{3}\\
\mathbf{\tilde{2}}\otimes\mathbf{\tilde{2}}\to\mathbf{\widehat{1}_{a}}\oplus\mathbf{\widehat{3}_{s}}\\
\mathbf{\tilde{2}}\otimes\mathbf{\tilde{2}'}\to\mathbf{\widehat{1}'}\oplus\mathbf{\widehat{3}'}\\
\mathbf{\tilde{2}'}\otimes\mathbf{\tilde{2}'}\to\mathbf{\widehat{1}_{a}}\oplus\mathbf{\widehat{3}_{s}}
  \end{array}\right\} &~~~
\begin{array}{l}
\mathbf{I} \sim \alpha_1\beta_2-\alpha_2\beta_1\\[5ex]
\mathbf{III} \sim \begin{pmatrix}
\alpha_1\beta_2+\alpha_2\beta_1\\
\sqrt{2}\alpha_1\beta_1\\
-\sqrt{2}\alpha_2\beta_2
\end{pmatrix}
\end{array}
\end{array}\nonumber
\end{equation}
\end{itemize}

\begin{itemize}
\item $\mathbf{II}\otimes\mathbf{III}\to\mathbf{III}_1\oplus\mathbf{III}_2$\,,
\begin{equation}
\begin{array}{lll}
\begin{array}{c}
    ~\\ p=0 \\
  \end{array} ~~~~&
  \left.\begin{array}{l}
\mathbf{2}\otimes\mathbf{3}\to\mathbf{3'}\oplus\mathbf{3}\\
\mathbf{2}\otimes\mathbf{3'}\to\mathbf{3}\oplus\mathbf{3'}\\
\mathbf{2}\otimes\mathbf{\widehat{3}}\to\mathbf{\widehat{3}'}\oplus\mathbf{\widehat{3}}\\
\mathbf{2}\otimes\mathbf{\widehat{3}'}\to\mathbf{\widehat{3}}\oplus\mathbf{\widehat{3}'}\\
\mathbf{2'}\otimes\mathbf{3}\to\mathbf{\widehat{3}'}\oplus\mathbf{\widehat{3}}\\
\mathbf{2'}\otimes\mathbf{3'}\to\mathbf{\widehat{3}}\oplus\mathbf{\widehat{3}'}\\
\mathbf{2'}\otimes\mathbf{\widehat{3}}\to\mathbf{3}\oplus\mathbf{3'}\\
\mathbf{2'}\otimes\mathbf{\widehat{3}'}\to\mathbf{3'}\oplus\mathbf{3}
  \end{array}\right\} &~~~
\begin{array}{l}
\mathbf{III}_1 \sim \begin{pmatrix}
\alpha_1\beta_2-\alpha_2\beta_3\\
\alpha_1\beta_3-\alpha_2\beta_1\\
\alpha_1\beta_1-\alpha_2\beta_2
\end{pmatrix}\\[5ex]
\mathbf{III}_2 \sim \begin{pmatrix}
\alpha_1\beta_2+\alpha_2\beta_3\\
\alpha_1\beta_3+\alpha_2\beta_1\\
\alpha_1\beta_1+\alpha_2\beta_2
\end{pmatrix}
\end{array}
\end{array}\nonumber
\end{equation}
\end{itemize}

\begin{itemize}
\item $\mathbf{II}\otimes\mathbf{III}\to\mathbf{II}\oplus\mathbf{IV}$\,,
\begin{equation}
\begin{array}{lll}
\begin{array}{c}
    ~\\ [-3ex]p=0  \\ \\ \\ \\ \\ \\ \\ \\[3ex]p=1 \\
  \end{array} ~~~~&
  \left.\begin{array}{l}
\mathbf{\widehat{2}}\otimes\mathbf{3}\to\mathbf{\widehat{2}}\oplus\mathbf{4}\\
\mathbf{\widehat{2}}\otimes\mathbf{\widehat{3}}\to\mathbf{\tilde{2}}\oplus\mathbf{4'}\\
\mathbf{\widehat{2}'}\otimes\mathbf{3'}\to\mathbf{\widehat{2}}\oplus\mathbf{4}\\
\mathbf{\widehat{2}'}\otimes\mathbf{\widehat{3}'}\to\mathbf{\tilde{2}}\oplus\mathbf{4'}\\
\mathbf{\tilde{2}}\otimes\mathbf{3}\to\mathbf{\tilde{2}}\oplus\mathbf{4'}\\
\mathbf{\tilde{2}}\otimes\mathbf{\widehat{3}'}\to\mathbf{\widehat{2}}\oplus\mathbf{4}\\
\mathbf{\tilde{2}'}\otimes\mathbf{3'}\to\mathbf{\tilde{2}}\oplus\mathbf{4'}\\
\mathbf{\tilde{2}'}\otimes\mathbf{\widehat{3}}\to\mathbf{\widehat{2}}\oplus\mathbf{4}\\\\
\mathbf{\widehat{2}}\otimes\mathbf{3'}\to\mathbf{\widehat{2}'}\oplus\mathbf{4}\\
\mathbf{\widehat{2}}\otimes\mathbf{\widehat{3}'}\to\mathbf{\tilde{2}'}\oplus\mathbf{4'}\\
\mathbf{\widehat{2}'}\otimes\mathbf{3}\to\mathbf{\widehat{2}'}\oplus\mathbf{4}\\
\mathbf{\widehat{2}'}\otimes\mathbf{\widehat{3}}\to\mathbf{\tilde{2}'}\oplus\mathbf{4'}\\
\mathbf{\tilde{2}}\otimes\mathbf{3'}\to\mathbf{\tilde{2}'}\oplus\mathbf{4'}\\
\mathbf{\tilde{2}}\otimes\mathbf{\widehat{3}}\to\mathbf{\widehat{2}'}\oplus\mathbf{4}\\
\mathbf{\tilde{2}'}\otimes\mathbf{3}\to\mathbf{\tilde{2}'}\oplus\mathbf{4'}\\
\mathbf{\tilde{2}'}\otimes\mathbf{\widehat{3}'}\to\mathbf{\widehat{2}'}\oplus\mathbf{4}
  \end{array}\right\} &~~~
\begin{array}{l}
\mathbf{II} \sim \begin{pmatrix}
\alpha_1\beta_1-\sqrt{2}\alpha_2\beta_2\\
-\sqrt{2}\alpha_1\beta_3-\alpha_2\beta_1
\end{pmatrix}\\[5ex]
\mathbf{IV} \sim \begin{pmatrix}
\alpha_1\beta_2-\sqrt{2}\alpha_2\beta_3\\
-(-1)^p(\sqrt{2}\alpha_1\beta_2-\alpha_2\beta_3)\\
\sqrt{2}\alpha_1\beta_1+\alpha_2\beta_2\\
(-1)^p(\alpha_1\beta_3-\sqrt{2}\alpha_2\beta_1)
\end{pmatrix}
\end{array}
\end{array}\nonumber
\end{equation}
\end{itemize}

\begin{itemize}
\item $\mathbf{II}\otimes\mathbf{IV}\to\mathbf{II}_1\oplus\mathbf{II}_2\oplus\mathbf{IV}$\,,
\begin{equation}
\begin{array}{lll}
\begin{array}{c}
    ~\\ [-0.1ex]p=0  \\ \\[3ex]p=1 \\
  \end{array} ~~~~&
  \left.\begin{array}{l}
\mathbf{2}\otimes\mathbf{4}\to\mathbf{\widehat{2}'}\oplus\mathbf{\widehat{2}}\oplus\mathbf{4}\\
\mathbf{2}\otimes\mathbf{4'}\to\mathbf{\tilde{2}'}\oplus\mathbf{\tilde{2}}\oplus\mathbf{4'}\\
\mathbf{2'}\otimes\mathbf{4}\to\mathbf{\tilde{2}'}\oplus\mathbf{\tilde{2}}\oplus\mathbf{4'}\\\\
\mathbf{2'}\otimes\mathbf{4'}\to\mathbf{\widehat{2}}\oplus\mathbf{\widehat{2}'}\oplus\mathbf{4}
  \end{array}\right\} &~~~
\begin{array}{l}
\mathbf{II}_1 \sim \begin{pmatrix}
\alpha_1\beta_1-\alpha_2\beta_4\\
-\alpha_1\beta_3-\alpha_2\beta_2
\end{pmatrix}\\[3ex]
\mathbf{II}_2 \sim \begin{pmatrix}
\alpha_1\beta_1+\alpha_2\beta_4\\
-\alpha_1\beta_3+\alpha_2\beta_2
\end{pmatrix}\\[3ex]
\mathbf{IV} \sim \begin{pmatrix}
\alpha_1\beta_4\\
-(-1)^p\alpha_2\beta_3\\
-\alpha_1\beta_2\\
(-1)^p\alpha_2\beta_1
\end{pmatrix}
\end{array}
\end{array}\nonumber
\end{equation}
\end{itemize}

\begin{itemize}
\item $\mathbf{II}\otimes\mathbf{IV}\to\mathbf{II}\oplus\mathbf{III}_1\oplus\mathbf{III}_2$\,,
\begin{equation}
\begin{array}{lll}
\begin{array}{c}
    ~\\ [-3ex]p=0  \\ \\ \\ \\[3ex]p=1 \\
  \end{array} ~~~~&
  \left.\begin{array}{l}
\mathbf{\widehat{2}}\otimes\mathbf{4}\to\mathbf{2'}\oplus\mathbf{\widehat{3}}\oplus\mathbf{\widehat{3}'}\\
\mathbf{\widehat{2}'}\otimes\mathbf{4'}\to\mathbf{2}\oplus\mathbf{3}\oplus\mathbf{3'}\\
\mathbf{\tilde{2}'}\otimes\mathbf{4}\to\mathbf{2}\oplus\mathbf{3}\oplus\mathbf{3'}\\
\mathbf{\tilde{2}'}\otimes\mathbf{4'}\to\mathbf{2'}\oplus\mathbf{\widehat{3}}\oplus\mathbf{\widehat{3}'}\\\\
\mathbf{\widehat{2}}\otimes\mathbf{4'}\to\mathbf{2}\oplus\mathbf{3'}\oplus\mathbf{3}\\
\mathbf{\widehat{2}'}\otimes\mathbf{4}\to\mathbf{2'}\oplus\mathbf{\widehat{3}'}\oplus\mathbf{\widehat{3}}\\
\mathbf{\tilde{2}}\otimes\mathbf{4}\to\mathbf{2}\oplus\mathbf{3'}\oplus\mathbf{3}\\
\mathbf{\tilde{2}}\otimes\mathbf{4'}\to\mathbf{2'}\oplus\mathbf{\widehat{3}'}\oplus\mathbf{\widehat{3}}
  \end{array}\right\} &~~~
\begin{array}{l}
\mathbf{II} \sim \begin{pmatrix}
\alpha_1\beta_2-\alpha_2\beta_4\\
(-1)^p(\alpha_1\beta_3+\alpha_2\beta_1)
\end{pmatrix}\\[5ex]
\mathbf{III}_1 \sim \begin{pmatrix}
\sqrt{2}\alpha_1\beta_4-\sqrt{2}\alpha_2\beta_3\\
\alpha_1\beta_3-\alpha_2\beta_1-\sqrt{2}\alpha_2\beta_2\\
-\sqrt{2}\alpha_1\beta_1+\alpha_1\beta_2+\alpha_2\beta_4
\end{pmatrix}\\[5ex]
\mathbf{III}_2 \sim \begin{pmatrix}
\sqrt{2}\alpha_1\beta_4+\sqrt{2}\alpha_2\beta_3\\
-\alpha_1\beta_3+\alpha_2\beta_1-\sqrt{2}\alpha_2\beta_2\\
\sqrt{2}\alpha_1\beta_1+\alpha_1\beta_2+\alpha_2\beta_4
\end{pmatrix}
\end{array}
\end{array}\nonumber
\end{equation}
\end{itemize}

\begin{itemize}
\item $\mathbf{III}\otimes\mathbf{III}\to\mathbf{I}\oplus\mathbf{II}\oplus\mathbf{III}_1\oplus\mathbf{III}_2$\,,
\begin{equation}
\begin{array}{lll}
\begin{array}{c}
    ~\\ [-3ex]p=0  \\ \\ \\ \\[3ex]p=1 \\
  \end{array} ~~~~&
  \left.\begin{array}{l}
\mathbf{3}\otimes\mathbf{3}\to\mathbf{1_{s}}\oplus\mathbf{2_{s}}\oplus\mathbf{3'_{s}}\oplus\mathbf{3_{a}}\\
\mathbf{3}\otimes\mathbf{\widehat{3}}\to\mathbf{\widehat{1}}\oplus\mathbf{2'}\oplus\mathbf{\widehat{3}'}\oplus\mathbf{\widehat{3}}\\
\mathbf{3'}\otimes\mathbf{3'}\to\mathbf{1_{s}}\oplus\mathbf{2_{s}}\oplus\mathbf{3'_{s}}\oplus\mathbf{3_{a}}\\
\mathbf{3'}\otimes\mathbf{\widehat{3}'}\to\mathbf{\widehat{1}}\oplus\mathbf{2'}\oplus\mathbf{\widehat{3}'}\oplus\mathbf{\widehat{3}}\\
\mathbf{\widehat{3}}\otimes\mathbf{\widehat{3}'}\to\mathbf{1}\oplus\mathbf{2}\oplus\mathbf{3'}\oplus\mathbf{3}\\\\
\mathbf{3}\otimes\mathbf{3'}\to\mathbf{1'}\oplus\mathbf{2}\oplus\mathbf{3}\oplus\mathbf{3'}\\
\mathbf{3}\otimes\mathbf{\widehat{3}'}\to\mathbf{\widehat{1}'}\oplus\mathbf{2'}\oplus\mathbf{\widehat{3}}\oplus\mathbf{\widehat{3}'}\\
\mathbf{3'}\otimes\mathbf{\widehat{3}}\to\mathbf{\widehat{1}'}\oplus\mathbf{2'}\oplus\mathbf{\widehat{3}}\oplus\mathbf{\widehat{3}'}\\
\mathbf{\widehat{3}}\otimes\mathbf{\widehat{3}}\to\mathbf{1'_{s}}\oplus\mathbf{2_{s}}\oplus\mathbf{3_{s}}\oplus\mathbf{3'_{a}}\\
\mathbf{\widehat{3}'}\otimes\mathbf{\widehat{3}'}\to\mathbf{1'_{s}}\oplus\mathbf{2_{s}}\oplus\mathbf{3_{s}}\oplus\mathbf{3'_{a}}
  \end{array}\right\} &~~~
\begin{array}{l}
\mathbf{I} \sim \alpha_1\beta_1+\alpha_2\beta_3+\alpha_3\beta_2\\[3ex]
\mathbf{II} \sim \begin{pmatrix}
\alpha_1\beta_3+\alpha_2\beta_2+\alpha_3\beta_1\\
(-1)^p(\alpha_1\beta_2+\alpha_2\beta_1+\alpha_3\beta_3)
\end{pmatrix}\\[5ex]
\mathbf{III}_1 \sim \begin{pmatrix}
2\alpha_1\beta_1-\alpha_2\beta_3-\alpha_3\beta_2\\
-\alpha_1\beta_2-\alpha_2\beta_1+2\alpha_3\beta_3\\
-\alpha_1\beta_3+2\alpha_2\beta_2-\alpha_3\beta_1
\end{pmatrix}\\[5ex]
\mathbf{III}_2 \sim \begin{pmatrix}
\alpha_2\beta_3-\alpha_3\beta_2\\
\alpha_1\beta_2-\alpha_2\beta_1\\
-\alpha_1\beta_3+\alpha_3\beta_1
\end{pmatrix}
\end{array}
\end{array}\nonumber
\end{equation}
\end{itemize}

\begin{itemize}
\item $\mathbf{III}\otimes\mathbf{IV}\to\mathbf{II}_1\oplus\mathbf{II}_2\oplus\mathbf{IV}_1\oplus\mathbf{IV}_2$\,,
\begin{equation}
\begin{array}{lll}
\begin{array}{c}
    ~\\ [-3ex]p=0  \\ \\ \\ \\[3ex]p=1 \\
  \end{array} ~~~~&
  \left.\begin{array}{l}
\mathbf{3}\otimes\mathbf{4}\to\mathbf{\widehat{2}'}\oplus\mathbf{\widehat{2}}\oplus\mathbf{4_{I}}\oplus\mathbf{4_{II}}\\
\mathbf{3}\otimes\mathbf{4'}\to\mathbf{\tilde{2}'}\oplus\mathbf{\tilde{2}}\oplus\mathbf{4'_{I}}\oplus\mathbf{4'_{II}}\\
\mathbf{\widehat{3}}\otimes\mathbf{4}\to\mathbf{\tilde{2}'}\oplus\mathbf{\tilde{2}}\oplus\mathbf{4'_{I}}\oplus\mathbf{4'_{II}}\\
\mathbf{\widehat{3}'}\otimes\mathbf{4'}\to\mathbf{\widehat{2}'}\oplus\mathbf{\widehat{2}}\oplus\mathbf{4_{I}}\oplus\mathbf{4_{II}}\\\\
\mathbf{3'}\otimes\mathbf{4}\to\mathbf{\widehat{2}}\oplus\mathbf{\widehat{2}'}\oplus\mathbf{4_{I}}\oplus\mathbf{4_{II}}\\
\mathbf{3'}\otimes\mathbf{4'}\to\mathbf{\tilde{2}}\oplus\mathbf{\tilde{2}'}\oplus\mathbf{4'_{I}}\oplus\mathbf{4'_{II}}\\
\mathbf{\widehat{3}}\otimes\mathbf{4'}\to\mathbf{\widehat{2}}\oplus\mathbf{\widehat{2}'}\oplus\mathbf{4_{I}}\oplus\mathbf{4_{II}}\\
\mathbf{\widehat{3}'}\otimes\mathbf{4}\to\mathbf{\tilde{2}}\oplus\mathbf{\tilde{2}'}\oplus\mathbf{4'_{I}}\oplus\mathbf{4'_{II}}
  \end{array}\right\} & \begin{array}{l}
\mathbf{II}_1 \sim \begin{pmatrix}
\sqrt{2}\alpha_1\beta_3-\alpha_2\beta_4+\alpha_3\beta_1+\sqrt{2}\alpha_3\beta_2\\
\sqrt{2}\alpha_1\beta_4-\sqrt{2}\alpha_2\beta_1+\alpha_2\beta_2+\alpha_3\beta_3
\end{pmatrix}\\[4ex]
\mathbf{II}_2 \sim \begin{pmatrix}
\sqrt{2}\alpha_1\beta_3+\alpha_2\beta_4+\alpha_3\beta_1-\sqrt{2}\alpha_3\beta_2\\
-\sqrt{2}\alpha_1\beta_4-\sqrt{2}\alpha_2\beta_1-\alpha_2\beta_2+\alpha_3\beta_3
\end{pmatrix}\\[4ex]
\mathbf{IV}_1 \sim \begin{pmatrix}
\sqrt{2}\alpha_1\beta_2-\alpha_3\beta_4\\
(-1)^p(\sqrt{2}\alpha_1\beta_1-\alpha_2\beta_3)\\
-\sqrt{2}\alpha_2\beta_4-\alpha_3\beta_2\\
-(-1)^p(\alpha_2\beta_1+\sqrt{2}\alpha_3\beta_3)
\end{pmatrix}\\[6ex]
\mathbf{IV}_2 \sim \begin{pmatrix}
\alpha_1\beta_1+\sqrt{2}\alpha_2\beta_3\\
-(-1)^p(\alpha_1\beta_2+\sqrt{2}\alpha_3\beta_4)\\
-\alpha_1\beta_3+\sqrt{2}\alpha_3\beta_1\\
(-1)^p(\alpha_1\beta_4-\sqrt{2}\alpha_2\beta_2)
\end{pmatrix}
\end{array}
\end{array}\nonumber
\end{equation}
\end{itemize}

\begin{itemize}
\item $\mathbf{IV}\otimes\mathbf{IV}\to\mathbf{I}_1\oplus\mathbf{I}_2\oplus\mathbf{II}\oplus\mathbf{III}_1\oplus\mathbf{III}_2\oplus\mathbf{III}_3\oplus\mathbf{III}_4$\,,
\begin{equation}
\begin{array}{lll}
\begin{array}{c}
     p=0  \\[1ex] \\  [0ex]p=1 \\
  \end{array} ~~&
  \left.\begin{array}{l}
\mathbf{4}\otimes\mathbf{4}\to\mathbf{\widehat{1}'_{a}}\oplus\mathbf{\widehat{1}_{s}}\oplus\mathbf{2'_{a}}\oplus\mathbf{\widehat{3}'_{sI}}\oplus\mathbf{\widehat{3}_{s}}\oplus\mathbf{\widehat{3}'_{sII}}\oplus\mathbf{\widehat{3}_{a}}\\\\
\mathbf{4}\otimes\mathbf{4'}\to\mathbf{1}\oplus\mathbf{1'}\oplus\mathbf{2}\oplus\mathbf{3_{I}}\oplus\mathbf{3'_{I}}\oplus\mathbf{3_{II}}\oplus\mathbf{3'_{II}}\\
\mathbf{4'}\otimes\mathbf{4'}\to\mathbf{\widehat{1}_{a}}\oplus\mathbf{\widehat{1}'_{s}}\oplus\mathbf{2'_{a}}\oplus\mathbf{\widehat{3}_{sI}}\oplus\mathbf{\widehat{3}'_{s}}\oplus\mathbf{\widehat{3}_{sII}}\oplus\mathbf{\widehat{3}'_{a}}
  \end{array}\right\} & \\ [12ex]
& \begin{array}{l}
\mathbf{I}_1 \sim \alpha_1\beta_2-\alpha_2\beta_1+\alpha_3\beta_4-\alpha_4\beta_3\\[3ex]
\mathbf{I}_2 \sim \alpha_1\beta_2+\alpha_2\beta_1+\alpha_3\beta_4+\alpha_4\beta_3\\[3ex]
\mathbf{II} \sim \begin{pmatrix}
\alpha_1\beta_3-\alpha_3\beta_1\\
(-1)^p(-\alpha_2\beta_4+\alpha_4\beta_2)
\end{pmatrix}\\[3ex]
\mathbf{III}_1 \sim \begin{pmatrix}
\sqrt{2}\alpha_1\beta_1-\sqrt{2}\alpha_2\beta_2\\
\alpha_2\beta_4-\sqrt{2}\alpha_3\beta_3+\alpha_4\beta_2\\
-\alpha_1\beta_3-\alpha_3\beta_1+\sqrt{2}\alpha_4\beta_4
\end{pmatrix}\\[4ex]
\mathbf{III}_2 \sim \begin{pmatrix}
\sqrt{2}\alpha_1\beta_1+\sqrt{2}\alpha_2\beta_2\\
-\alpha_2\beta_4-\sqrt{2}\alpha_3\beta_3-\alpha_4\beta_2\\
-\alpha_1\beta_3-\alpha_3\beta_1-\sqrt{2}\alpha_4\beta_4
\end{pmatrix}\\[4ex]
\mathbf{III}_3 \sim \begin{pmatrix}
\alpha_1\beta_2+\alpha_2\beta_1-\alpha_3\beta_4-\alpha_4\beta_3\\
\sqrt{2}\alpha_1\beta_4+\sqrt{2}\alpha_4\beta_1\\
\sqrt{2}\alpha_2\beta_3+\sqrt{2}\alpha_3\beta_2
\end{pmatrix}\\[4ex]
\mathbf{III}_4 \sim \begin{pmatrix}
\alpha_1\beta_2-\alpha_2\beta_1-\alpha_3\beta_4+\alpha_4\beta_3\\
\sqrt{2}\alpha_1\beta_4-\sqrt{2}\alpha_4\beta_1\\
-\sqrt{2}\alpha_2\beta_3+\sqrt{2}\alpha_3\beta_2
\end{pmatrix}
\end{array}
\end{array}\nonumber
\end{equation}
\end{itemize}

The modular forms in this basis is a little more complicated than before. At weight $1/2$, the modular forms can be arranged into the irreducible representation $\mathbf{\widehat{2}}$  of $\widetilde{S_4}$:
\begin{equation}
 Y_{\mathbf{\widehat{2}}}^{(\frac{1}{2})}(\tau)=\begin{pmatrix}
\vartheta_1(\tau) \\ \vartheta_2(\tau)
\end{pmatrix} = \begin{pmatrix}
\omega^2 ~&~ i+\omega \\ \frac{\sqrt{2}+\sqrt{6}}{2} ~&~ e^{i\pi/4}
\end{pmatrix} \begin{pmatrix}
e_1(\tau) \\ e_2(\tau)
\end{pmatrix}\,,
\end{equation}
where $\omega=e^{2\pi i/3}$, $e_{1}(\tau)$ and $e_{2}(\tau)$ are given in Eq.~\eqref{eq:e12-half-weight}. The $q-$expansion of $\vartheta_1(\tau)$ and $\vartheta_2(\tau)$ can be given by
\begin{align}
\nonumber
\vartheta_1(\tau)&=\omega^2+2(1+\omega)q^{1/4}+2\omega^2 q+2(1+\omega)q^{9/4}+2\omega^2 q^{4}+2(1+\omega)q^{25/4}+2\omega^2 q^{9} +\dots \,,\\
\nonumber
\vartheta_2(\tau)&= \frac{\sqrt{2}+\sqrt{6}}{2}+2e^{i\pi/4}q^{1/4}+(\sqrt{2}+\sqrt{6})q+2e^{i\pi/4}q^{9/4}+(\sqrt{2}+\sqrt{6})q^{4}+2e^{i\pi/4}q^{25/4}\dots\,.
\end{align}
As before, all higher-weight modular forms can be constructed from the tensor products of $Y^{(\frac{1}{2})}_{\mathbf{\widehat{2}}}(\tau)$ by using the CG coefficients in current basis of group $\widetilde{S}_4$. For instance, we find the weight 1 modular forms make up a triplet $\mathbf{\widehat{3}'}$ of $\widetilde{S}_4$
\begin{equation}
Y_{\mathbf{\widehat{3}'}}^{(1)}=\frac{1}{\sqrt{2}}\left(Y_{\mathbf{\widehat{2}}}^{(\frac{1}{2})}Y_{\mathbf{\widehat{2}}}^{(\frac{1}{2})}\right)_{\mathbf{\widehat{3}'_{s}}}=\begin{pmatrix}
\sqrt{2}\vartheta_1\vartheta_2\\
-\vartheta_2^2\\
\vartheta_1^2\\
\end{pmatrix}\,.
\end{equation}
The modular forms of higher weights in the new basis are the same tensor products as before, except that the overall normalization coefficients may be different. The modular forms of weight $3/2$ can be arranged into a quartet representation $\mathbf{4'}$ of $\widetilde{S}_4$:
\begin{equation}
Y_{\mathbf{4'}}^{(\frac{3}{2})}=\left(Y_{\mathbf{\widehat{2}}}^{(\frac{1}{2})}Y_{\mathbf{\widehat{3}'}}^{(1)}\right)_{\mathbf{4'}}=\begin{pmatrix}
\vartheta_1^3+\sqrt{2}\vartheta_2^3\\
\sqrt{2}\vartheta_1^3-\vartheta_2^3\\
3\vartheta_1^2\vartheta_2\\
3\vartheta_1\vartheta_2^2\\
\end{pmatrix}\,.
\end{equation}
At weight $2$, we have five independent modular forms which can be decomposed into a doublet $\mathbf{2}$ and a triplet $\mathbf{3}$,
\begin{eqnarray}
\nonumber &&Y_{\mathbf{2}}^{(2)}=-\frac{1}{\sqrt{2}}\left(Y_{\mathbf{\widehat{2}}}^{(\frac{1}{2})}Y_{\mathbf{4'}}^{(\frac{3}{2})}\right)_{\mathbf{2}}=\begin{pmatrix}
2\sqrt{2}\vartheta_1\vartheta_2^3-\vartheta_1^4\\
\vartheta_2^4+2\sqrt{2}\vartheta_1^3\vartheta_2\\
\end{pmatrix}\,,\\
&&Y_{\mathbf{3}}^{(2)}=\frac{1}{2\sqrt{2}}\left(Y_{\mathbf{\widehat{2}}}^{(\frac{1}{2})}Y_{\mathbf{4'}}^{(\frac{3}{2})}\right)_{\mathbf{3}}=\begin{pmatrix}
3\vartheta_1^2\vartheta_2^2\\
\vartheta_2^4-\sqrt{2}\vartheta_1^3\vartheta_2\\
\vartheta_1^4+\sqrt{2}\vartheta_1\vartheta_2^3\\
\end{pmatrix}\,.
\end{eqnarray}
The weight $5/2$ modular forms  can be arranged into a doublet $\mathbf{\widehat{2}}$ and a quartet $\mathbf{4}$,
\begin{eqnarray}
\nonumber&&Y_{\mathbf{\widehat{2}}}^{(\frac{5}{2})}=-\left(Y_{\mathbf{\widehat{2}}}^{(\frac{1}{2})}Y_{\mathbf{3}}^{(2)}\right)_{\mathbf{\widehat{2}}}=\begin{pmatrix}
\sqrt{2}\vartheta_2^5-5\vartheta_1^3\vartheta_2^2\\
\sqrt{2}\vartheta_1^5+5\vartheta_1^2\vartheta_2^3\\
\end{pmatrix}\,,\\
  &&Y_{\mathbf{4}}^{(\frac{5}{2})}=-\left(Y_{\mathbf{\widehat{2}}}^{(\frac{1}{2})}Y_{\mathbf{2}}^{(2)}\right)_{\mathbf{4}}=\begin{pmatrix}
-\vartheta_1\left(\vartheta_2^4+2\sqrt{2}\vartheta_1^3\vartheta_2\right)\\
\vartheta_2\left(\vartheta_1^4-2\sqrt{2}\vartheta_1\vartheta_2^3\right)\\
\vartheta_2^5+2\sqrt{2}\vartheta_1^3\vartheta_2^2\\
\vartheta_1^5-2\sqrt{2}\vartheta_1^2\vartheta_2^3\\
\end{pmatrix}\,,
\end{eqnarray}
There are seven modular forms of weight 3, which transform as a singlet and two triplets under $\widetilde{S}_4$,
\begin{eqnarray}
  \nonumber&&Y_{\mathbf{\widehat{1}'}}^{(3)}=\frac{1}{\sqrt{2}}\left(Y_{\mathbf{\widehat{2}}}^{(\frac{1}{2})}Y_{\mathbf{\widehat{2}}}^{(\frac{5}{2})}\right)_{\mathbf{\widehat{1}'_{a}}}=\vartheta_1^6+5\sqrt{2}\vartheta_1^3\vartheta_2^3-\vartheta_2^6\,,\\
  \nonumber&&Y_{\mathbf{\widehat{3}}}^{(3)}=\frac{1}{\sqrt{2}}\left(Y_{\mathbf{\widehat{2}}}^{(\frac{1}{2})}Y_{\mathbf{4}}^{(\frac{5}{2})}\right)_{\mathbf{\widehat{3}}}=\begin{pmatrix}
\vartheta_1^6-4\sqrt{2}\vartheta_1^3\vartheta_2^3-\vartheta_2^6\\
3\vartheta_1\vartheta_2^2\left(\vartheta_1^3+\sqrt{2}\vartheta_2^3\right)\\
3\sqrt{2}\vartheta_1^5\vartheta_2-3\vartheta_1^2\vartheta_2^4\\
\end{pmatrix}\,,\\
 &&Y_{\mathbf{\widehat{3}'}}^{(3)}=-\frac{1}{\sqrt{2}}\left(Y_{\mathbf{\widehat{2}}}^{(\frac{1}{2})}Y_{\mathbf{\widehat{2}}}^{(\frac{5}{2})}\right)_{\mathbf{\widehat{3}'_{s}}}=\begin{pmatrix}
-\vartheta_1^6-\vartheta_2^6\\
5\vartheta_1^4\vartheta_2^2-\sqrt{2}\vartheta_1\vartheta_2^5\\
\sqrt{2}\vartheta_1^5\vartheta_2+5\vartheta_1^2\vartheta_2^4\\
\end{pmatrix}\,.
\end{eqnarray}
At weight $7/2$, we have eight modular forms which can be decomposed into three $\widetilde{S}_4$ multiplets $\mathbf{\tilde{2}'}$, $\mathbf{\tilde{2}}$ and $\mathbf{4'}$,
\begin{eqnarray}
\nonumber &&Y_{\mathbf{\tilde{2}'}}^{(\frac{7}{2})}=\left(Y_{\mathbf{\widehat{2}}}^{(\frac{1}{2})}Y_{\mathbf{\widehat{1}'}}^{(3)}\right)_{\mathbf{\tilde{2}'}}=\left(\vartheta_1^6+5\sqrt{2}\vartheta_1^3\vartheta_2^3-\vartheta_2^6\right)\begin{pmatrix}
\vartheta_1\\
\vartheta_2\\
\end{pmatrix}\,,\\
\nonumber &&Y_{\mathbf{\tilde{2}}}^{(\frac{7}{2})}=\left(Y_{\mathbf{\widehat{2}}}^{(\frac{1}{2})}Y_{\mathbf{\widehat{3}}}^{(3)}\right)_{\mathbf{\tilde{2}}}=\begin{pmatrix}
\vartheta_1^7-7\sqrt{2}\vartheta_1^4\vartheta_2^3-7\vartheta_1\vartheta_2^6\\
\vartheta_2^7+7\sqrt{2}\vartheta_1^3\vartheta_2^4-7\vartheta_1^6\vartheta_2\\
\end{pmatrix}\,,\\
&& Y_{\mathbf{4'}}^{(\frac{7}{2})}=\left(Y_{\mathbf{\widehat{2}}}^{(\frac{1}{2})}Y_{\mathbf{\widehat{3}}}^{(3)}\right)_{\mathbf{4'}}=\begin{pmatrix}
6\sqrt{2}\vartheta_1^2\vartheta_2^5-3\vartheta_1^5\vartheta_2^2\\
-3\vartheta_1^2\vartheta_2^2\left(2\sqrt{2}\vartheta_1^3+\vartheta_2^3\right)\\
\sqrt{2}\vartheta_1^7-5\vartheta_1^4\vartheta_2^3+2\sqrt{2}\vartheta_1\vartheta_2^6\\
\sqrt{2}\vartheta_2^7+5\vartheta_1^3\vartheta_2^4+2\sqrt{2}\vartheta_1^6\vartheta_2\\
\end{pmatrix}\,,
\end{eqnarray}
The weight 4 modular forms of level 4 can be arranged into a singlet $\mathbf{1}$, a doublet $\mathbf{2}$, and two triplets $\mathbf{3}$, $\mathbf{3'}$ of $\widetilde{S}_4$
\begin{eqnarray}
\nonumber &&Y_{\mathbf{1}}^{(4)}=-\frac{1}{2\sqrt{2}}\left(Y_{\mathbf{\widehat{2}}}^{(\frac{1}{2})}Y_{\mathbf{\tilde{2}}}^{(\frac{7}{2})}\right)_{\mathbf{1}}=
\vartheta_1\vartheta_2\left(2\sqrt{2}\vartheta_1^6-7\vartheta_1^3\vartheta_2^3-2\sqrt{2}\vartheta_2^6\right)\,,\\
\nonumber && Y_{\mathbf{2}}^{(4)}=-\frac{1}{\sqrt{2}}\left(Y_{\mathbf{\widehat{2}}}^{(\frac{1}{2})}Y_{\mathbf{4'}}^{(\frac{7}{2})}\right)_{\mathbf{2}}=\begin{pmatrix}
\vartheta_2^8+4\sqrt{2}\vartheta_1^3\vartheta_2^5+8\vartheta_1^6\vartheta_2^2\\
\vartheta_1^8-4\sqrt{2}\vartheta_1^5\vartheta_2^3+8\vartheta_1^2\vartheta_2^6\\
\end{pmatrix}\,,\\
\nonumber && Y_{\mathbf{3}}^{(4)}=-\frac{1}{\sqrt{2}}\left(Y_{\mathbf{\widehat{2}}}^{(\frac{1}{2})}Y_{\mathbf{\tilde{2}}}^{(\frac{7}{2})}\right)_{\mathbf{3}}=\begin{pmatrix}
3\sqrt{2}\vartheta_1\vartheta_2\left(\vartheta_1^6+\vartheta_2^6\right)\\
-\vartheta_1^8+7\sqrt{2}\vartheta_1^5\vartheta_2^3+7\vartheta_1^2\vartheta_2^6\\
\vartheta_2^8+7\sqrt{2}\vartheta_1^3\vartheta_2^5-7\vartheta_1^6\vartheta_2^2\\
\end{pmatrix}\,,\\
 && Y_{\mathbf{3'}}^{(4)}=\frac{1}{\sqrt{2}}\left(Y_{\mathbf{\widehat{2}}}^{(\frac{1}{2})}Y_{\mathbf{\tilde{2}'}}^{(\frac{7}{2})}\right)_{\mathbf{3'}}=\begin{pmatrix}
\sqrt{2}\vartheta_1^7\vartheta_2+10\vartheta_1^4\vartheta_2^4-\sqrt{2}\vartheta_1\vartheta_2^7\\
\vartheta_1^8+5\sqrt{2}\vartheta_1^5\vartheta_2^3-\vartheta_1^2\vartheta_2^6\\
\vartheta_2^8-5\sqrt{2}\vartheta_1^3\vartheta_2^5-\vartheta_1^6\vartheta_2^2\\
\end{pmatrix}\,.
\end{eqnarray}
We find 10 modular forms at weight $9/2$,
\begin{eqnarray}
\nonumber && Y_{\mathbf{\widehat{2}}}^{(\frac{9}{2})}=\left(Y_{\mathbf{\widehat{2}}}^{(\frac{1}{2})}Y_{\mathbf{1}}^{(4)}\right)_{\mathbf{\widehat{2}}}=\vartheta_1\vartheta_2\left(2\sqrt{2}\vartheta_1^6-7\vartheta_1^3\vartheta_2^3-2\sqrt{2}\vartheta_2^6\right)\begin{pmatrix}
\vartheta_1\\
\vartheta_2\\
\end{pmatrix}\,,\\
\nonumber && Y_{\mathbf{4}I}^{(\frac{9}{2})}=\left(Y_{\mathbf{\widehat{2}}}^{(\frac{1}{2})}Y_{\mathbf{2}}^{(4)}\right)_{\mathbf{4}}=\begin{pmatrix}
\vartheta_1^9-4\sqrt{2}\vartheta_1^6\vartheta_2^3+8\vartheta_1^3\vartheta_2^6\\
\vartheta_2^9+4\sqrt{2}\vartheta_1^3\vartheta_2^6+8\vartheta_1^6\vartheta_2^3\\
-\vartheta_2\left(\vartheta_1^8-4\sqrt{2}\vartheta_1^5\vartheta_2^3+8\vartheta_1^2\vartheta_2^6\right)\\
\vartheta_1\left(\vartheta_2^8+4\sqrt{2}\vartheta_1^3\vartheta_2^5+8\vartheta_1^6\vartheta_2^2\right)\\
\end{pmatrix}\,,\\
 && Y_{\mathbf{4}II}^{(\frac{9}{2})}=-\left(Y_{\mathbf{\widehat{2}}}^{(\frac{1}{2})}Y_{\mathbf{3}}^{(4)}\right)_{\mathbf{4}}=\begin{pmatrix}
\vartheta_1^9-14\sqrt{2}\vartheta_1^6\vartheta_2^3+7\vartheta_1^3\vartheta_2^6+\sqrt{2}\vartheta_2^9\\
-\sqrt{2}\vartheta_1^9+7\vartheta_1^6\vartheta_2^3+14\sqrt{2}\vartheta_1^3\vartheta_2^6+\vartheta_2^9\\
-\vartheta_1^2\vartheta_2\left(5\vartheta_1^6+7\sqrt{2}\vartheta_1^3\vartheta_2^3+13\vartheta_2^6\right)\\
5\vartheta_1\vartheta_2^8-7\sqrt{2}\vartheta_1^4\vartheta_2^5+13\vartheta_1^7\vartheta_2^2\\
\end{pmatrix}\,.
\end{eqnarray}
There are 11 independent weight 5 modular forms of level 4, which can be decomposed as $\mathbf{2'}\oplus\mathbf{\widehat{3}}\oplus\mathbf{\widehat{3}'}\oplus\mathbf{\widehat{3}'}$ under $\widetilde{S}_4$,
\begin{eqnarray}
\nonumber && Y_{\mathbf{2'}}^{(5)}=-\frac{1}{\sqrt{2}}\left(Y_{\mathbf{\widehat{2}}}^{(\frac{1}{2})}Y_{\mathbf{4}II}^{(\frac{9}{2})}\right)_{\mathbf{2'}}=\begin{pmatrix}
\vartheta_1\left(\vartheta_1^9+3\sqrt{2}\vartheta_1^6\vartheta_2^3-21\vartheta_1^3\vartheta_2^6+2\sqrt{2}\vartheta_2^9\right)\\
\vartheta_2\left(2\sqrt{2}\vartheta_1^9+21\vartheta_1^6\vartheta_2^3+3\sqrt{2}\vartheta_1^3\vartheta_2^6-\vartheta_2^9\right)\\
\end{pmatrix}\,,\\
\nonumber && Y_{\mathbf{\widehat{3}}}^{(5)}=-\frac{1}{\sqrt{2}}\left(Y_{\mathbf{\widehat{2}}}^{(\frac{1}{2})}Y_{\mathbf{4}I}^{(\frac{9}{2})}\right)_{\mathbf{\widehat{3}}}=\begin{pmatrix}
-9\vartheta_1^2\vartheta_2^2\left(\vartheta_1^6+\vartheta_2^6\right)\\
\vartheta_2\left(\sqrt{2}\vartheta_1^9+12\sqrt{2}\vartheta_1^3\vartheta_2^6+\vartheta_2^9\right)\\
\vartheta_1\left(\vartheta_1^9-12\sqrt{2}\vartheta_1^6\vartheta_2^3-\sqrt{2}\vartheta_2^9\right)\\
\end{pmatrix}\,,\\
\nonumber && Y_{\mathbf{\widehat{3}'}I}^{(5)}=\frac{1}{\sqrt{2}}\left(Y_{\mathbf{\widehat{2}}}^{(\frac{1}{2})}Y_{\mathbf{4}I}^{(\frac{9}{2})}\right)_{\mathbf{\widehat{3}'}}=\begin{pmatrix}
7\vartheta_1^8\vartheta_2^2+8\sqrt{2}\vartheta_1^5\vartheta_2^5-7\vartheta_1^2\vartheta_2^8\\
\vartheta_2\left(\sqrt{2}\vartheta_1^9-16\vartheta_1^6\vartheta_2^3+4\sqrt{2}\vartheta_1^3\vartheta_2^6-\vartheta_2^9\right)\\
\vartheta_1\left(\vartheta_1^9+4\sqrt{2}\vartheta_1^6\vartheta_2^3+16\vartheta_1^3\vartheta_2^6+\sqrt{2}\vartheta_2^9\right)\\
\end{pmatrix}\,,\\
   && Y_{\mathbf{\widehat{3}'}II}^{(5)}=\frac{1}{2\sqrt{2}}\left(Y_{\mathbf{\widehat{2}}}^{(\frac{1}{2})}Y_{\mathbf{4}II}^{(\frac{9}{2})}\right)_{\mathbf{\widehat{3}'}}=\vartheta_1\vartheta_2\begin{pmatrix}
4\vartheta_1^7\vartheta_2-7\sqrt{2}\vartheta_1^4\vartheta_2^4-4\vartheta_1\vartheta_2^7\\
2\sqrt{2}\vartheta_1^8-7\vartheta_1^5\vartheta_2^3-2\sqrt{2}\vartheta_1^2\vartheta_2^6\\
2\sqrt{2}\vartheta_2^8+7\vartheta_1^3\vartheta_2^5-2\sqrt{2}\vartheta_1^6\vartheta_2^2\\
\end{pmatrix}\,.
\end{eqnarray}
At weight $11/2$, we find 12 independent modular forms as follows
\begin{eqnarray}
\nonumber && Y_{\mathbf{\tilde{2}}}^{(\frac{11}{2})}=-\left(Y_{\mathbf{\widehat{2}}}^{(\frac{1}{2})}Y_{\mathbf{\widehat{3}}}^{(5)}\right)_{\mathbf{\tilde{2}}}=\begin{pmatrix}
\vartheta_2^2\left(11\vartheta_1^9+33\vartheta_1^3\vartheta_2^6+\sqrt{2}\vartheta_2^9\right)\\
\vartheta_1^2\left(\sqrt{2}\vartheta_1^9-33\vartheta_1^6\vartheta_2^3-11\vartheta_2^9\right)\\
\end{pmatrix}\,,\\
  \nonumber && Y_{\mathbf{\tilde{2}'}}^{(\frac{11}{2})}=\left(Y_{\mathbf{\widehat{2}}}^{(\frac{1}{2})}Y_{\mathbf{\widehat{3}'}I}^{(5)}\right)_{\mathbf{\tilde{2}'}}=\begin{pmatrix}
\vartheta_2^2\left(5\vartheta_1^9+24\sqrt{2}\vartheta_1^6\vartheta_2^3-15\vartheta_1^3\vartheta_2^6+\sqrt{2}\vartheta_2^9\right)\\
-\vartheta_1^2\left(\sqrt{2}\vartheta_1^9+15\vartheta_1^6\vartheta_2^3+24\sqrt{2}\vartheta_1^3\vartheta_2^6-5\vartheta_2^9\right)\\
\end{pmatrix}\,,\\
\nonumber && Y_{\mathbf{4'}I}^{(\frac{11}{2})}=-\left(Y_{\mathbf{\widehat{2}}}^{(\frac{1}{2})}Y_{\mathbf{\widehat{3}'}I}^{(5)}\right)_{\mathbf{4'}}=\begin{pmatrix}
3\vartheta_1\vartheta_2^4\left(8\vartheta_1^6+4\sqrt{2}\vartheta_1^3\vartheta_2^3+\vartheta_2^6\right)\\
-3\vartheta_1^4\vartheta_2\left(\vartheta_1^6-4\sqrt{2}\vartheta_1^3\vartheta_2^3+8\vartheta_2^6\right)\\
\vartheta_2^{11}+3\sqrt{2}\vartheta_1^3\vartheta_2^8-8\sqrt{2}\vartheta_1^9\vartheta_2^2\\
\vartheta_1^2\left(\vartheta_1^9-3\sqrt{2}\vartheta_1^6\vartheta_2^3+8\sqrt{2}\vartheta_2^9\right)\\
\end{pmatrix}\,,\\
 && Y_{\mathbf{4'}II}^{(\frac{11}{2})}=\left(Y_{\mathbf{\widehat{2}}}^{(\frac{1}{2})}Y_{\mathbf{\widehat{3}'}II}^{(5)}\right)_{\mathbf{4'}}=\vartheta_1\vartheta_2\begin{pmatrix}
2\sqrt{2}\vartheta_1^9-3\vartheta_1^6\vartheta_2^3-9\sqrt{2}\vartheta_1^3\vartheta_2^6-4\vartheta_2^9\\
4\vartheta_1^9-9\sqrt{2}\vartheta_1^6\vartheta_2^3+3\vartheta_1^3\vartheta_2^6+2\sqrt{2}\vartheta_2^9\\
6\sqrt{2}\vartheta_1^8\vartheta_2-21\vartheta_1^5\vartheta_2^4-6\sqrt{2}\vartheta_1^2\vartheta_2^7\\
-6\sqrt{2}\vartheta_1\vartheta_2^8-21\vartheta_1^4\vartheta_2^5+6\sqrt{2}\vartheta_1^7\vartheta_2^2\\
\end{pmatrix}\,.
\end{eqnarray}
The weight 6 modular forms can be arranged into two singlets $\mathbf{1}$, $\mathbf{1'}$, a doublet $\mathbf{2}$, and three triplets $\mathbf{3}\oplus\mathbf{3}\oplus\mathbf{3'}$ under $\widetilde{S}_4$:
\begin{eqnarray}
\nonumber && Y_{\mathbf{1}}^{(6)}=\frac{1}{\sqrt{2}}\left(Y_{\mathbf{\widehat{2}}}^{(\frac{1}{2})}Y_{\mathbf{\tilde{2}}}^{(\frac{11}{2})}\right)_{\mathbf{1}}=
\left(\vartheta_1^2+\vartheta_2^2\right)\left(\vartheta_1^4-\vartheta_1^2\vartheta_2^2+\vartheta_2^4\right)\left(\vartheta_1^6-22\sqrt{2}\vartheta_1^3\vartheta_2^3-\vartheta_2^6\right)\,,\\
\nonumber && Y_{\mathbf{1'}}^{(6)}=-\frac{1}{\sqrt{2}}\left(Y_{\mathbf{\widehat{2}}}^{(\frac{1}{2})}Y_{\mathbf{\tilde{2}'}}^{(\frac{11}{2})}\right)_{\mathbf{1'}}=
\vartheta_1^{12}+10\sqrt{2}\vartheta_1^9\vartheta_2^3+48\vartheta_1^6\vartheta_2^6-10\sqrt{2}\vartheta_1^3\vartheta_2^9+\vartheta_2^{12}\,,\\
\nonumber && Y_{\mathbf{2}}^{(6)}=-\frac{1}{\sqrt{2}}\left(Y_{\mathbf{\widehat{2}}}^{(\frac{1}{2})}Y_{\mathbf{4'}I}^{(\frac{11}{2})}\right)_{\mathbf{2}}=\vartheta_1\vartheta_2\begin{pmatrix}
\vartheta_1\left(2\sqrt{2}\vartheta_1^9-15\vartheta_1^6\vartheta_2^3+12\sqrt{2}\vartheta_1^3\vartheta_2^6+8\vartheta_2^9\right)\\
\vartheta_2\left(-8\vartheta_1^9+12\sqrt{2}\vartheta_1^6\vartheta_2^3+15\vartheta_1^3\vartheta_2^6+2\sqrt{2}\vartheta_2^9\right)\\
\end{pmatrix}\,,\\
\nonumber && Y_{\mathbf{3}I}^{(6)}=\frac{1}{\sqrt{2}}\left(Y_{\mathbf{\widehat{2}}}^{(\frac{1}{2})}Y_{\mathbf{4'}I}^{(\frac{11}{2})}\right)_{\mathbf{3}}=\begin{pmatrix}
\vartheta_1^{12}-11\sqrt{2}\vartheta_1^9\vartheta_2^3+11\sqrt{2}\vartheta_1^3\vartheta_2^9+\vartheta_2^{12}\\
\vartheta_1\vartheta_2^2\left(11\vartheta_1^9+33\vartheta_1^3\vartheta_2^6+\sqrt{2}\vartheta_2^9\right)\\
\vartheta_1^2\vartheta_2\left(-\sqrt{2}\vartheta_1^9+33\vartheta_1^6\vartheta_2^3+11\vartheta_2^9\right)\\
\end{pmatrix}\,,\\
\nonumber && Y_{\mathbf{3}II}^{(6)}=-\frac{1}{2\sqrt{2}}\left(Y_{\mathbf{\widehat{2}}}^{(\frac{1}{2})}Y_{\mathbf{4'}II}^{(\frac{11}{2})}\right)_{\mathbf{3}}=\vartheta_1\vartheta_2\begin{pmatrix}
-6\sqrt{2}\vartheta_1^8\vartheta_2^2+21\vartheta_1^5\vartheta_2^5+6\sqrt{2}\vartheta_1^2\vartheta_2^8\\
\vartheta_2\left(4\vartheta_1^9-9\sqrt{2}\vartheta_1^6\vartheta_2^3+3\vartheta_1^3\vartheta_2^6+2\sqrt{2}\vartheta_2^9\right)\\
-2\sqrt{2}\vartheta_1^{10}+3\vartheta_1^7\vartheta_2^3+9\sqrt{2}\vartheta_1^4\vartheta_2^6+4\vartheta_1\vartheta_2^9\\
\end{pmatrix}\,,\\
 && Y_{\mathbf{3'}}^{(6)}=-\frac{1}{\sqrt{2}}\left(Y_{\mathbf{\widehat{2}}}^{(\frac{1}{2})}Y_{\mathbf{4'}I}^{(\frac{11}{2})}\right)_{\mathbf{3'}}=\begin{pmatrix}
-\left(\vartheta_1^6+5\sqrt{2}\vartheta_1^3\vartheta_2^3-\vartheta_2^6\right)\left(\vartheta_1^6+\vartheta_2^6\right)\\
\vartheta_1\vartheta_2^2\left(5\vartheta_1^9+24\sqrt{2}\vartheta_1^6\vartheta_2^3-15\vartheta_1^3\vartheta_2^6+\sqrt{2}\vartheta_2^9\right)\\
\vartheta_1^2\vartheta_2\left(\sqrt{2}\vartheta_1^9+15\vartheta_1^6\vartheta_2^3+24\sqrt{2}\vartheta_1^3\vartheta_2^6-5\vartheta_2^9\right)\\
\end{pmatrix}\,.
\end{eqnarray}
Finally we give the vacuum alignments of half weight modular form $Y^{(\frac{1}{2})}_\mathbf{\widehat{2}}(\tau)$ at the modular transformation fixed points in this basis as follows
\begin{align}
\nonumber
& Y^{(\frac{1}{2})}_\mathbf{\widehat{2}}(\tau_S)=Y_S\begin{pmatrix}
1 \\ -\sqrt{2}-\sqrt{3}
\end{pmatrix}, ~\quad~ Y^{(\frac{1}{2})}_\mathbf{\widehat{2}}(\tau_{ST})=Y_{ST}\begin{pmatrix}
0 \\ 1
\end{pmatrix} \,,\\
& Y^{(\frac{1}{2})}_\mathbf{\widehat{2}}(\tau_T)=Y_T\begin{pmatrix}
\omega \\ -\sqrt{2}-e^{7\pi i/12}\sqrt{3}
\end{pmatrix}, ~\quad~ Y^{(\frac{1}{2})}_\mathbf{\widehat{2}}(\tau_{TS})=Y_{TS}\begin{pmatrix}
1 \\ -\sqrt{2}
\end{pmatrix}\,,
\end{align}
with $Y_S=-0.70975-0.09344i$,~ $Y_{ST}=2.42826$,~ $Y_{TS}=-1.35419-0.36285i$ and $Y_{T}=\omega$.
\end{appendix}

\newpage
\providecommand{\href}[2]{#2}\begingroup\raggedright\endgroup

\end{document}